\documentclass[twocolumn]{aastex631}

\usepackage{amsmath}
\usepackage{amssymb}

\newcommand{\Chandra}{{\em Chandra}}

\newcommand{\NH}{$N_\mathrm{H}$}

\shorttitle{3D thermodynamic structures of the ICM across X-ray surface brightness edges}
\shortauthors{Ueda \& Ichinohe}
\graphicspath{{./}{figures/}}

\begin{document}

\title{
Three-dimensional thermodynamic structures of the intracluster medium across edges in the X-ray surface brightness of massive, bright, dynamically-active galaxy clusters
}

\author[0000-0001-6252-7922]{Shutaro Ueda}
\affiliation{
Academia Sinica Institute of Astronomy and Astrophysics (ASIAA), 11F of AS/NTU Astronomy-Mathematics Building, No.1, Sec. 4, Roosevelt Rd, Taipei 106216, Taiwan
}

\author[0000-0002-6102-1441]{Yuto Ichinohe}
\affiliation{RIKEN Nishina Center for Accelerator-Based Science, 2-1 Hirosawa, Wako, Saitama 351-0198, Japan
}



\begin{abstract}
We present a detailed study of three-dimensional (3D) thermodynamic structures of the intracluster medium (ICM) across edges in the X-ray surface brightness of four massive, bright, dynamically-active galaxy clusters (A3667, A2319, A520, and A2146), with the {\em Chandra} X-ray Observatory. Based on a forward modeling approach developed in previous work, we extend this approach with more generalized ICM density and temperature profiles, allowing us to apply uniformly to the observed X-ray surface brightness profiles to detect edges and measure the 3D thermodynamic profiles of the ICM simultaneously and self-consistently. With the forward modeling analysis, we find, in agreement with previous works, that the obtained 3D thermodynamic structures of the ICM across the edges in A3667 and A2319 are consistent with the characteristics of cold fronts, whereas those in A520 and A2146 are consistent with the nature of shock fronts. We find that the azimuthal distribution of the pressure ratio at the cold front in A3667 shows a different trend from that in A2319. For the shock fronts in A520 and A2146, the observed 3D temperature profiles of the ICM indicate that the temperature is highest at the position of the shock front. In the case of the sector exhibiting $\mathcal{M} = 2.4$ in A520, the ICM temperature appears isothermal with a temperature of $\sim 10$\,keV until $\sim 300$\,kpc away from the shock front in the post-shock region, being consistent with the hypothesis of the instant-equilibration model for shock-heating. 
\end{abstract}



\section{Introduction} 
\label{sec:intro}

Galaxy clusters are the most massive, largest gravitationally bound objects in the Universe, and contain a large amount of baryons trapped in the deep gravitational potential well dominated by dark matter. The greatest majority of the baryons in galaxy clusters resides in the diffuse, hot X-ray emitting gas, known as the intracluster medium (ICM). Galaxy clusters still continue to grow in mass through mergers with small or similar-mass objects, as well as through continuous accretion of material from their surrounding large-scale environments. Cluster merger activities leave an apparent mark on the X-ray surface brightness of the ICM in galaxy clusters.

Observations with the \Chandra\ X-ray Observatory have discovered sharp brightness edges seen in the X-ray surface brightness of the ICM in galaxy clusters, thanks to its high-angular resolution. Such edges have been recognized as cold fronts and shock fronts \citep[e.g.,][]{Markevitch00, Vikhlinin01, Markevitch02, Owers09, Ghizzardi10, Johnson10, Wang16b, Botteon18b, Erdim19}. Cold fronts are frequently found in both merging and relaxed clusters such as Abell\,3667 \citep{Vikhlinin01, Vikhlinin02, Churazov04b, Datta14, Sarazin16, Ichinohe17, Storm18, Omiya24} and Abell\,2142 \citep{Ettori00b, Markevitch00, Owers11, Rossetti13, Liu18, Wang18}. Cold fronts have been used to probe the microphysics of the ICM such as magnetic fields and viscosity \citep[e.g.,][]{Ettori00b, Markevitch07, Ichinohe17, Ichinohe19, Ichinohe21}. However, shock fronts are relatively rare, compared to cold fronts. Major merging clusters such as the Bullet cluster \citep{Markevitch02, Wik14, Di_Mascolo19b}, Abell\,520 \citep{Govoni04, Markevitch05, Wang16b, Wang18b}, and Abell\,2146 \citep{Russell12b, Russell22} exhibit a clear shape of shock fronts. Shock fronts are a useful tool to investigate not only gas dynamics, including turbulence and cosmic-ray (re)acceleration, but also energy equilibration between the ions and the electrons \citep[e.g.,][]{Markevitch06, Inoue16}.

Although both cold fronts and shock fronts appear as edges in the X-ray surface brightness, the thermodynamic profiles of the ICM across these edges are different between these two fronts. For cold fronts, the ICM density in the inner region (i.e., closer to the cluster center) is higher, while the ICM temperature is lower than in the outer region. In some cold fronts, no significant thermal pressure difference between the inner and outer regions is reported \citep[e.g.,][]{Markevitch00, Markevitch01}. On the other hand, for shock fronts, the ICM temperature in the inner region (i.e., the post-shock region) is higher than that in the outer region (i.e., the pre-shock region) because of shock-heating. Therefore, these distinct thermodynamic features of the ICM are crucial to distinguish between cold fronts and shock fronts.

The temperature profile of the ICM in the post-shock region is a key to study the mechanisms of shock-heating. In general, two types of models for the shock-heating mechanism are considered: instant-equilibration and adiabatic compression \citep{Markevitch07}. Since these two models predict different three-dimensional (3D) temperature profiles, particularly, the ICM temperature at the position of shock fronts, extensive studies have been conducted on a variety of shock fronts in merging clusters \citep[e.g.,][]{Markevitch06, Russell12b, Sarazin16, Wang18b, Di_Mascolo19b, Russell22, Sarkar22, Sarkar24}. However, these two models are still under debate. In this context, measuring the positions of shock fronts and ICM temperature at these points accurately is important to explore these two models.

In this paper, we analyze the datasets of four galaxy clusters (Abell\,3667, Abell\,2319, Abell\,520 and Abell\,2146) obtained with the \Chandra\ X-ray Observatory, to perform accurate measurements of the 3D ICM density and temperature profiles and their jumps at the interface. To this end, we extend a forward modeling approach, originally developed in \cite{Umetsu22}, to handle more generalized profiles of the ICM density and temperature. We apply this extended approach uniformly to the observed X-ray surface brightness profiles of our sample, allowing us not only to determine the position of the interface but also to measure the 3D thermodynamic structures of the ICM across the interface, particularly, below the interface. These observed 3D profiles are crucial not only for constraining the mechanisms of shock-heating but also for investigating the mechanisms to generate cold fronts.

This paper is organized as follows. Section~\ref{sec:sample} describes a brief summary of previous studies related to our sample. Section~\ref{sec:obs} summarizes the \Chandra\ observations and data reduction. Section~\ref{sec:ana} describes our forward modeling approach and its application to the cold fronts and shock fronts in our sample. In Section~\ref{sec:discussion}, we discuss the obtained results and their implications. Finally, conclusions of this paper are summarized in Section~\ref{sec:summary}.

Throughout the paper, we assume a spatially flat Lambda cold dark matter ($\Lambda$CDM) model with a matter density parameter of $\Omega_{\rm m} = 0.3$ and a Hubble constant of $H_{0} = 70$\,km\,s$^{-1}$\,Mpc$^{-1}$.

\section{Sample}
\label{sec:sample}

In this paper, we focus on four massive, bright, dynamically-active galaxy clusters: Abell\,3667, Abell\,2319, Abell\,520, and Abell\,2146. All these clusters show prominent high-brightness-contrast edges in their X-ray surface brightness. Abell\,3667 and Abell\,2319 are known as merging clusters with high-brightness-contrast cold fronts, whereas Abell\,520 and Abell\,2146 are recognized as major merging clusters with apparent bow shock fronts. Therefore, these four clusters are suitable for investigating the capability of our forward modeling approach, and exploring the mechanisms of shock-heating and the processes to generate cold fronts, based on the observed 3D thermodynamic structures of the ICM across the interface. Here, we provide a brief summary of previous studies on each galaxy cluster.

\begin{table*}[t]
\begin{center}
\caption{
Summary of \Chandra\ X-ray observations for our sample.
}\label{tab:list}
\begin{tabular}{lcccc}
\hline\hline	
Cluster		&	Redshift	&	Physical scale (kpc/$''$)	&	Expo. time (ksec)\tablenotemark{a}	&	ObsID\tablenotemark{b}\\\hline
A3667		&	0.0557	&	1.082				&	498.2						&	889, 5751, 5752, 5753, 6292, 6295, 6296, 7686	\\
A2319		&	0.0557	&	1.082				&	89.6							&	3231, 15187								\\
A520			&	0.2020	&	3.325				&	528.2						&	528, 4215, 7703, 9424, 9425, 9426, 9430			\\
A2146		&	0.2340	&	3.722				&	419.5						&	10464, 10888, 12245, 12246, 12247,			\\
			&			&						&								&	13020, 13021, 13023, 13120, 13138				\\
\hline
\end{tabular}
\end{center}
\tablenotetext{a}{Total net exposure time of \Chandra\ observations after masking flare time intervals.}
\tablenotetext{b}{\Chandra\ observation identification (ObsID) numbers.}
\end{table*}

\subsection{Abell\,3667}
\label{sec:A3667intro}

Abell\,3667 (hereafter A3667) is known to exhibit the highest-brightness-contrast cold front in the southeastern direction from the center, and this cluster is one of the first cold fronts discovered \citep{Vikhlinin01}. Since A3667 is one of nearby galaxy clusters ($z = 0.0557, 1'' = 1.082$\,kpc), Deep \Chandra\ observations provided us with a unique opportunity to explore the cold front as a probe for understanding the microphysics of the ICM \citep{Ichinohe17}. A3667 is also known to host a prominent radio relic in the northwestern direction \citep{Rottgering97}, which is known as the brightest radio relic. On the opposite side of this radio relic with respect to the cluster center, another radio relic was found \citep{Rottgering97}. Such two-sided radio relics are now recognized as evidence of past/ongoing merger activities in galaxy clusters. Thus, A3667 is widely considered as a merging cluster, and its cold front is considered to originate from an infalling subcluster.

\subsection{Abell\,2319}
\label{sec:A2319intro}

Abell\,2319 (hereafter A2319) is one of nearby galaxy clusters ($z = 0.0557, 1'' = 1.082$\,kpc), and known to exhibit a high-brightness-contrast cold front. \Chandra\ observations revealed the overall ICM temperature in this cluster is high ($> 8$\,keV) and the presence of a cold front $\sim 300$\,kpc away from the core toward the southeastern direction \citep{OHara04, Govoni04}. Along with the presence of a giant radio halo \citep{Harris78, Feretti97, Govoni01, Farnsworth13}, A2319 is recognized as a merging cluster based on its disturbed X-ray surface brightness \citep{Markevitch96, Million09, Sugawara09, Yan14, Storm15}. \cite{Ichinohe21} probed the microphysical properties of the ICM with this cold front. This cold front is often considered to be generated by gas sloshing, i.e., a sloshing cold front, \citep[e.g.,][]{Ichinohe21}.
A comparison of the thermodynamic properties of the ICM across the cold front between A3667 and A2319 is important not only for distinguishing stripping and sloshing cold fronts but also for finding the similarities between them.

\subsection{Abell\,520}
\label{sec:A520intro}

Abell\,520 (hereafter A520, $z = 0.2020, 1'' = 3.325$\,kpc) is known as one of three galaxy clusters showing a prominent bow shock induced by an ongoing major merger \citep[e.g.,][]{Markevitch05, Wang16b, Wang18b}. In addition, prominent substructures have been discovered in the X-ray surface brightness \citep{Govoni04}. The most prominent substructure is in the form of a dense, compact, cool gas that is considered as a remnant of a cool core originally in an infalling subcluster. A tail-like distribution of low-temperature gas associated with the most prominent substructure is found in the global temperature map of A520 \citep{Wang16b}. Therefore, A520 is an ideal laboratory to study major merger activities and shock fronts as a probe for understanding merger-induced high-energy phenomena.

\subsection{Abell\,2146}
\label{sec:A2146intro}

Abell\,2146 (hereafter A2146, $z = 0.2340, 1'' = 3.722$\,kpc) is also known as one of three galaxy clusters showing a prominent bow shock induced by an ongoing major merger (the last one is the Bullet cluster). There has been a wide range of studies in this context \citep{Russell10b, Russell12b, White15, King16, Coleman17, Russell22, Rojas23}. Radio observations of this cluster revealed the presence of diffuse synchrotron emission \citep{Russell11, Hlavacek-Larrondo18, Hoang19}. The microphysical properties of the ICM was studied using the shock in this cluster \citep{Chadayammuri22, Chadayammuri22b, Richard-Laferriere23}. Similar to A520, A2146 hosts disturbed substructures with low-temperature gas below the bow shock front \citep{Russell22}. The bow shock front in A2146 is complementary to that in A520, making it suitable experimental setups for exploring high-energy phenomena associated with shocks.

\section{Observations and data reduction}
\label{sec:obs}

We use the archival X-ray datasets of our sample taken with the Advanced CCD Imaging Spectrometer \citep[ACIS;][]{Garmire03, Grant24} on board the \Chandra\ X-ray Observatory. This paper employs a list of Chandra datasets, obtained by the Chandra X-ray Observatory, contained in~\dataset[DOI: 10.25574/cdc.265]{https://doi.org/10.25574/cdc.265}. All datasets analyzed are summarized in Table~\ref{tab:list}\footnote{In this paper, we do not use datasets taken from deep 2\,Msec \Chandra\ observations of A2146 conducted in 2018 and 2019. \cite{Russell22} investigated systematic uncertainty arising from the stability of the focal-plane temperature of the ACIS for these 2\,Msec datasets. They found that the ICM temperature and abundance are consistent with those derived from the datasets of earlier observations, but the ICM electron density became $\sim 2$\,\% higher. Since our forward modeling appraoch simultaneously estimates both the ICM density and temperature, this level of the systematic uncertainty impacts on our analysis. Therefore, we use only the first $\sim 420$\,ksec datasets in this paper.}. We use the versions of 4.15 and 4.10.2 for Chandra Interactive Analysis of Observations \citep[CIAO;][]{Fruscione06} and the calibration database (CALDB), respectively. After applying the \texttt{lc\_clean} task in CIAO to the data to exclude the duration of the flare, we obtain the net exposure times for each cluster summarized in Table~\ref{tab:list}. We adopt the blank-sky data included in the CALDB to estimate the background contribution to each galaxy cluster. We extract the X-ray spectrum of the ICM from the region of interest in each dataset with the \texttt{specexctract} task in CIAO and combined them after making individual spectrum, response, and ancillary response files for the spectral fitting. We use \texttt{XSPEC} version 12.13.0c \citep{Arnaud96} and the atomic database for plasma emission modeling version 3.0.9 in the X-ray spectral analysis, assuming that the ICM is in collisional ionization equilibration. Thus, we use the \texttt{APEC} model in \texttt{XSPEC} \citep{Smith01, Foster12}. The abundance table of \cite{Anders89} is used.  Here, the abundance of a given element is defined as $Z_{i} = (n_{i, {\rm obs}}/n_{\rm H, obs}) / (n_{i, \odot} / n_{\rm H, \odot})$, where $n_{i}$ and $n_{\rm H}$ are the number densities of the $i$th element and hydrogen, respectively. We use the iron abundance to represent the ICM metal abundance, such that the abundance of other elements is tied to the iron abundance as $Z_i = Z_{\rm Fe}$ \citep{Ueda21}. The Galactic absorption (i.e., \NH) for each galaxy cluster is estimated using \cite{HI4PI16} and fixed at the estimated value in the X-ray analysis as follows: \NH\ $= 4.25 \times 10^{20}$\,cm$^{-2}$ for A3667, \NH\ $= 1.36 \times 10^{21}$\,cm$^{-2}$ for A2319, \NH\ $= 5.51 \times 10^{20}$\,cm$^{-2}$ for A520, and \NH\ $= 2.62 \times 10^{20}$\,cm$^{-2}$ for A2146.

\section{Analyses and Results}
\label{sec:ana}

In this Section, we first present the \Chandra\ X-ray images of our sample, and define sectors analyzed using our forward modeling approach. Next, we describe our extended forward modeling approach, which was originally developed in \cite{Umetsu22} (see their Section~6.3). Then, we apply this extended approach to the datasets to determine the position of the interface, and infer the 3D thermodynamic profiles of the ICM across the interface simultaneously and self-consistently. We also carry out the X-ray spectral analysis using \texttt{XSPEC} to measure the projected temperature profile of the ICM, allowing us to validate our forward modeling approach.

\begin{figure*}[ht]
 \begin{center}
  \includegraphics[width=7.5cm]{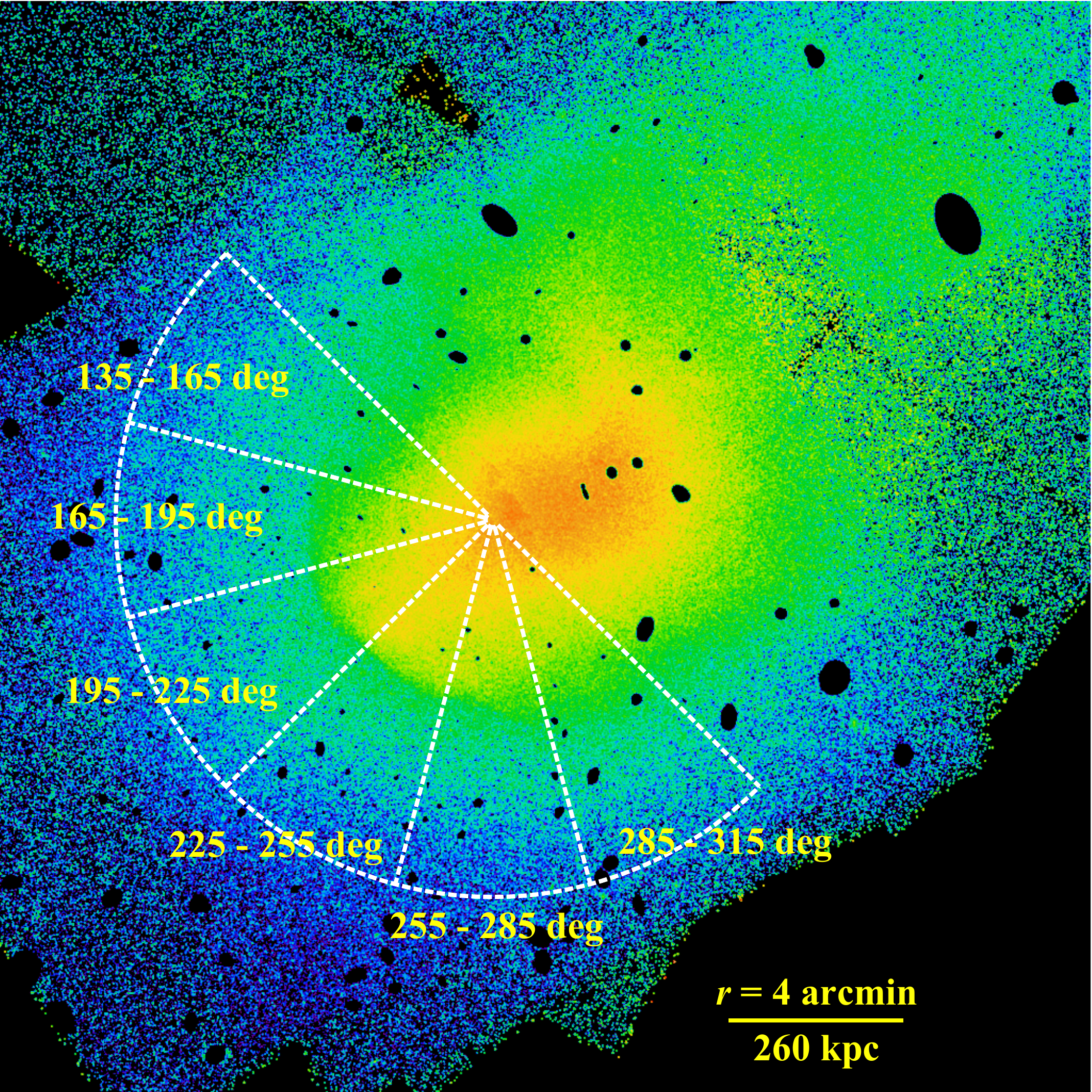}
  \includegraphics[width=7.5cm]{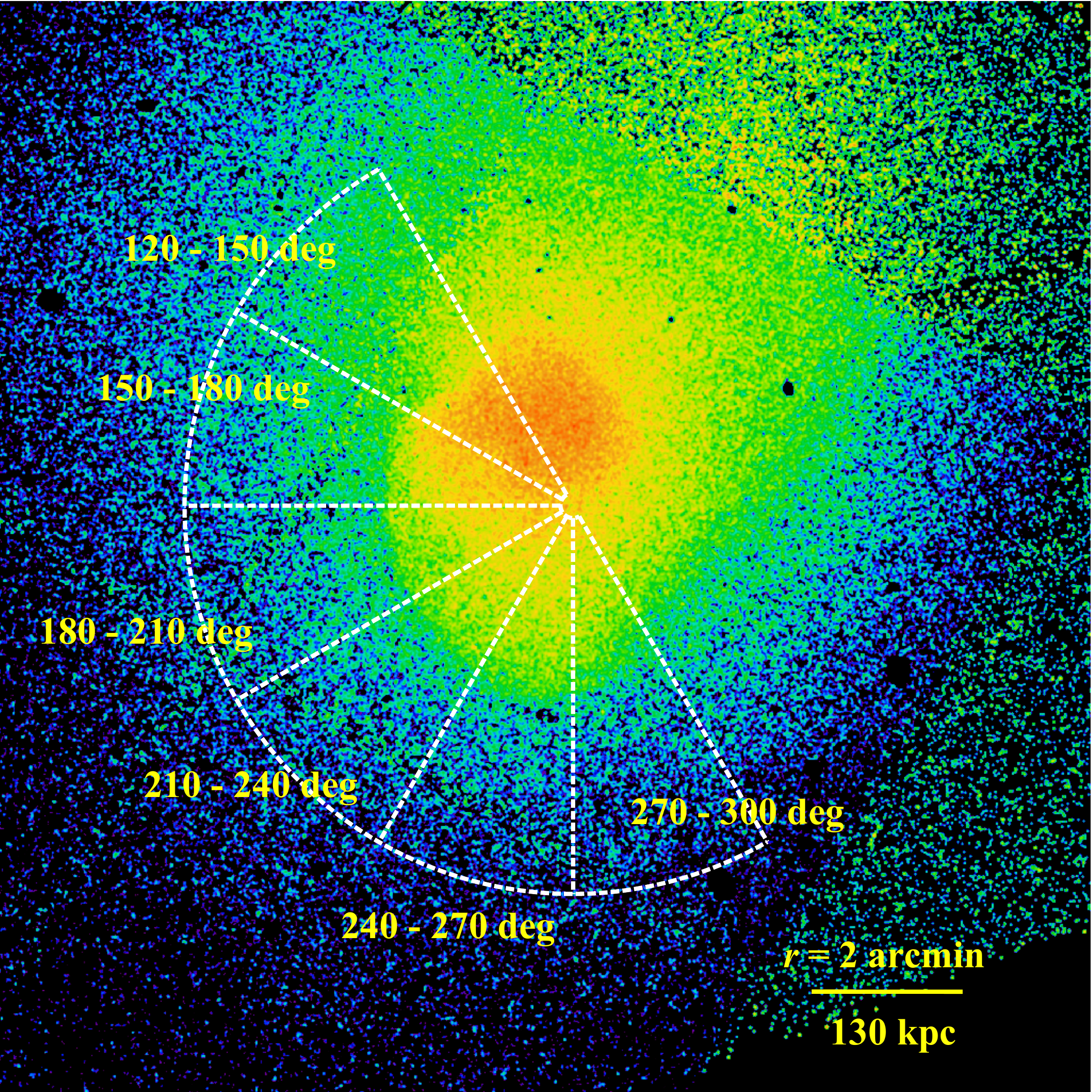}
  \includegraphics[width=7.5cm]{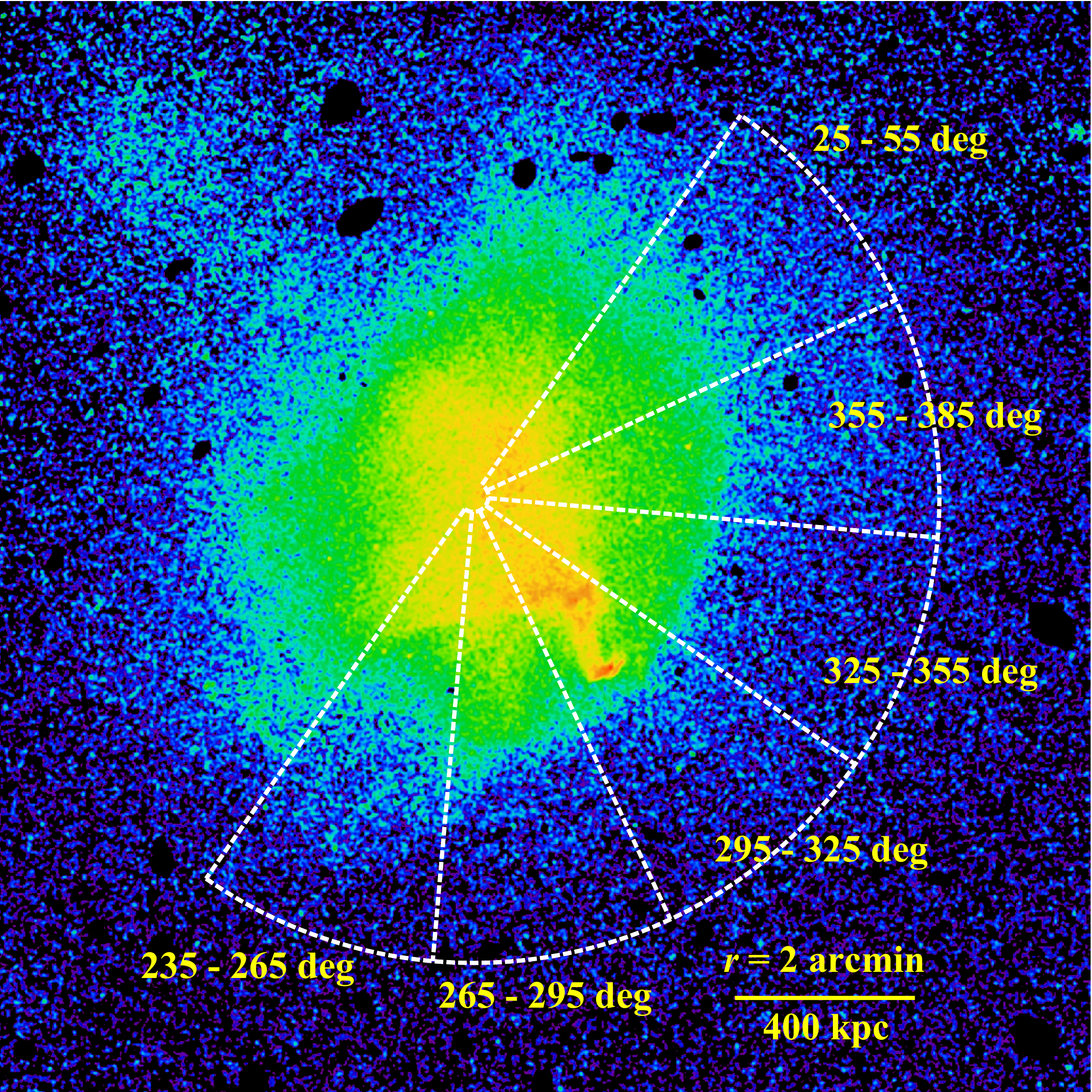}
  \includegraphics[width=7.5cm]{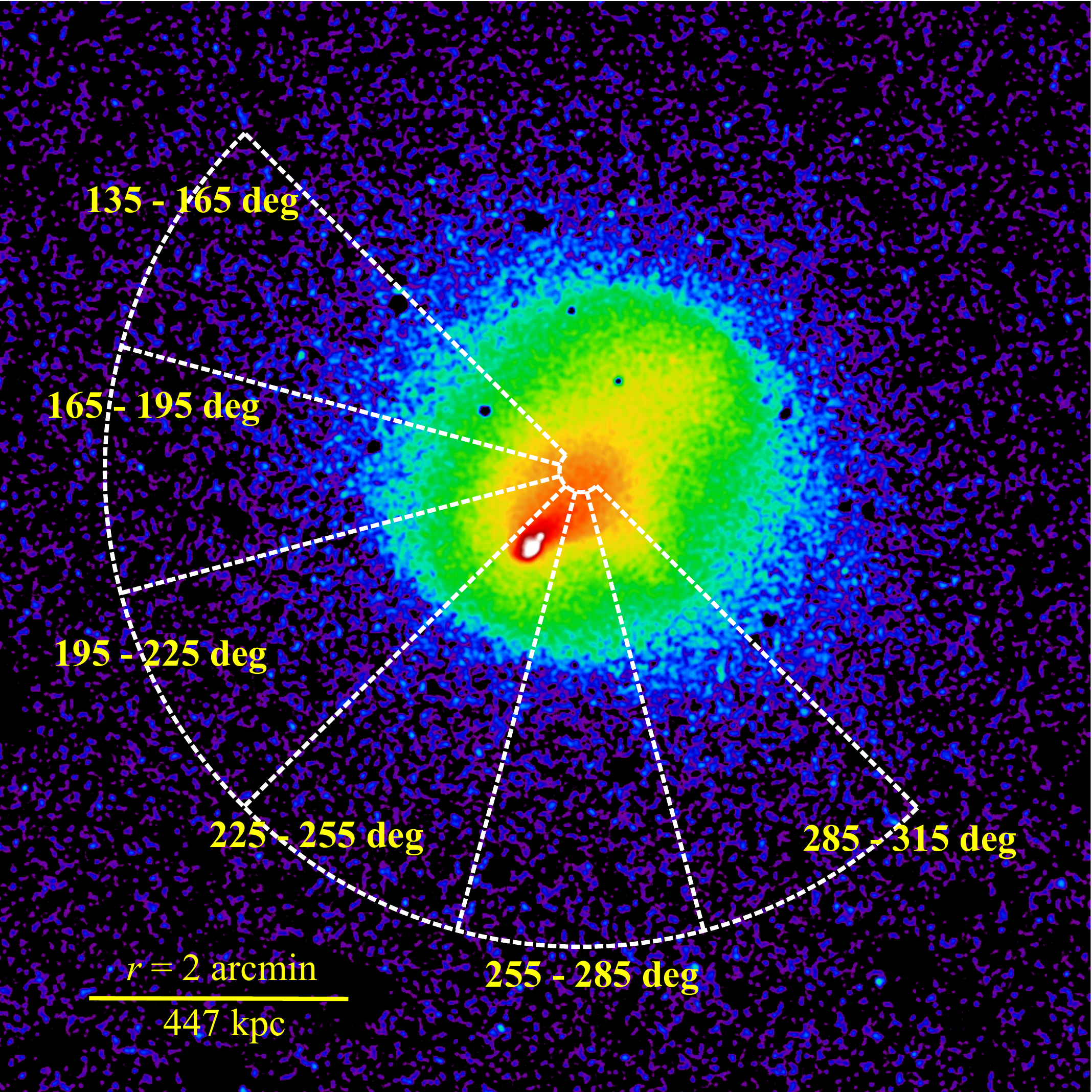}
 \end{center}
\caption{
Exposure corrected, background subtracted \Chandra\ X-ray surface brightness in the $0.5 - 7.0$\,keV band of A3667 (top left), A2319 (top right), A520 (bottom left), and A2146 (bottom right). The point sources identified are masked with black ellipses. These images are smoothed by a Gaussian kernel with $3.5''$\,FWHM. The overlaid, dashed white sectors denote the directions toward the edges to extract the X-ray surface brightness profiles. The yellow bars in each panel show the corresponding spatial scale.
(Top left): A3667. The size of each sector is $\theta\in [10\arcsec, 520\arcsec]$.
(Top right): A2319. The size of each sector is $\theta\in [10\arcsec, 310\arcsec]$.
(Bottom left): A520. The size of each sector is $\theta\in [10\arcsec, 310\arcsec]$.
(Bottom right): A2146. The size of each sector is $\theta\in [10\arcsec, 220\arcsec]$.
}
\label{fig:img}
\end{figure*}

\subsection{X-ray images of our sample}
\label{sec:img}

To extract the radial profiles of the X-ray surface brightness of our sample, we define the center of the sectors, assuming that the interface is axisymmetric. Following the literature \citep[e.g.,][]{Ichinohe17, Ichinohe21}, to make the radial directions perpendicular to the fronts, we define the centers of the curvature for our sample as summarized in Table~\ref{tab:pos}. Figure~\ref{fig:img} shows the \Chandra\ X-ray images of our sample. The sectors where we extract and analyze the X-ray surface brightness profiles are also plotted in Figure~\ref{fig:img}.

\begin{table}[ht]
\begin{center}
\caption{
Sky positions of the centers of the curvature and size of the sectors for our sample.
}\label{tab:pos}
\begin{tabular}{lccc}
\hline\hline	
Cluster		&	R.A.			&	Decl.			&	Size					     		\\	\hline
A3667		&	20:12:45.4493	&	-56:50:56.748	&	$\theta\in [10\arcsec, 520\arcsec]$	\\
A2319		&	19:21:10.1179	&	+43:55:46.724	&	$\theta\in [10\arcsec, 310\arcsec]$	\\
A520		&	4:54:11.2429	&	+2:55:31.663	&	$\theta\in [10\arcsec, 310\arcsec]$	\\
A2146		&	15:56:10.8882	&	+66:21:24.117	&	$\theta\in [10\arcsec, 220\arcsec]$	\\
\hline
\end{tabular}
\end{center}
\end{table}

\begin{figure*}[ht]
 \begin{center}
  \includegraphics[width=8.5cm]{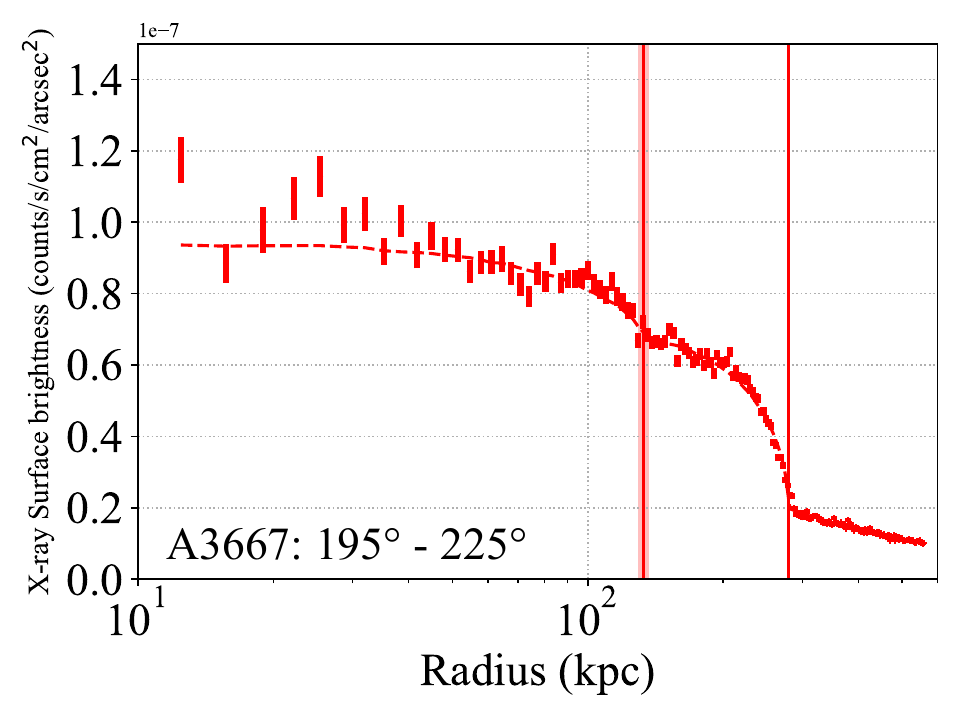}
  \includegraphics[width=8.5cm]{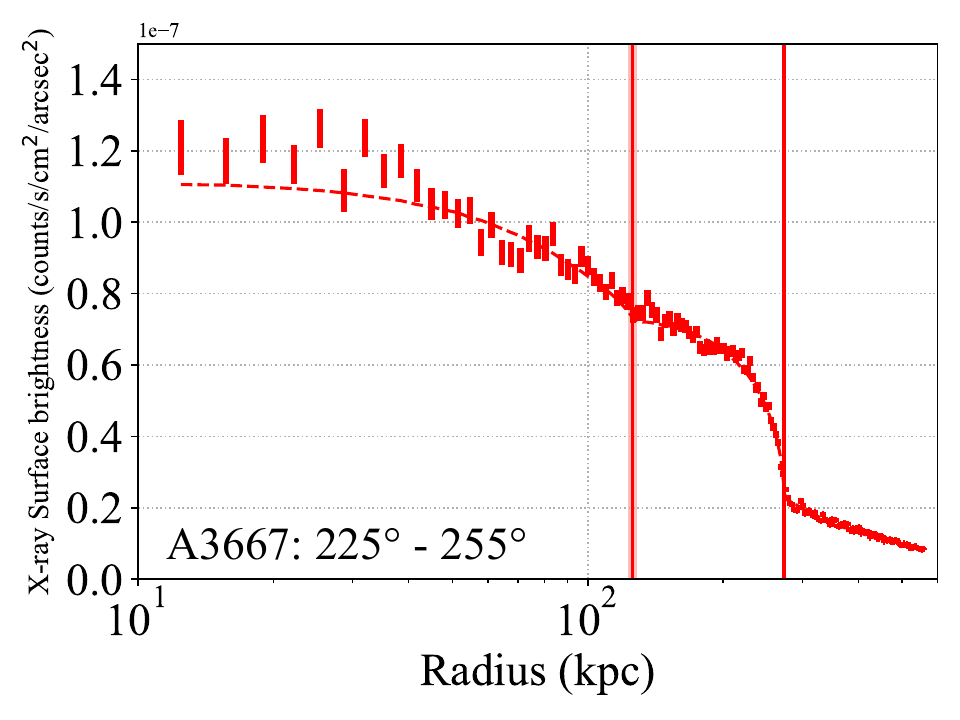}
  \includegraphics[width=8.5cm]{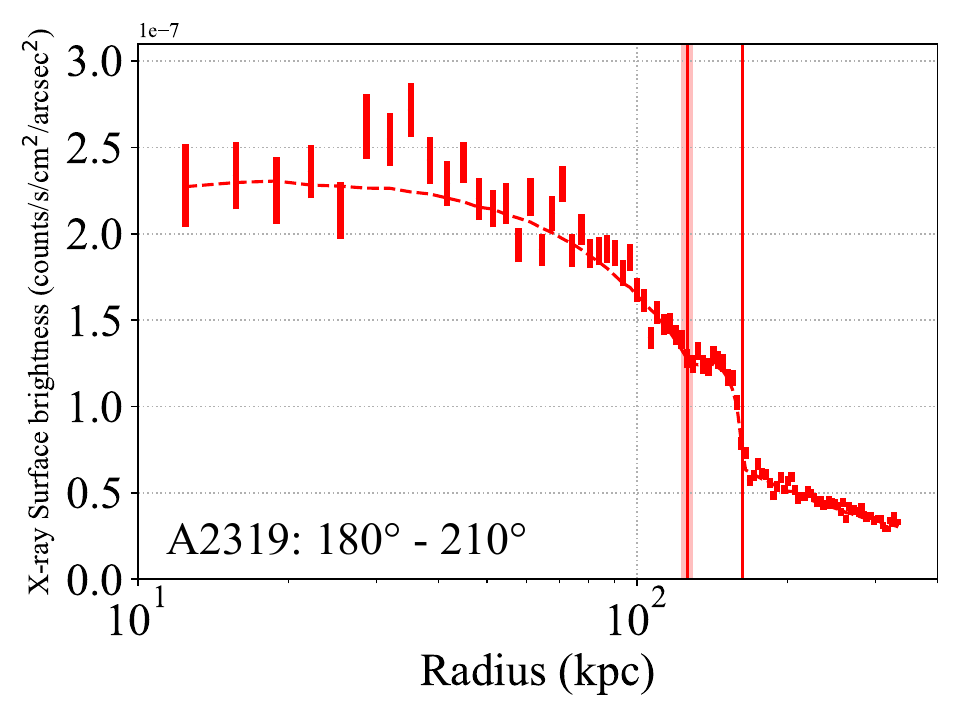}
  \includegraphics[width=8.5cm]{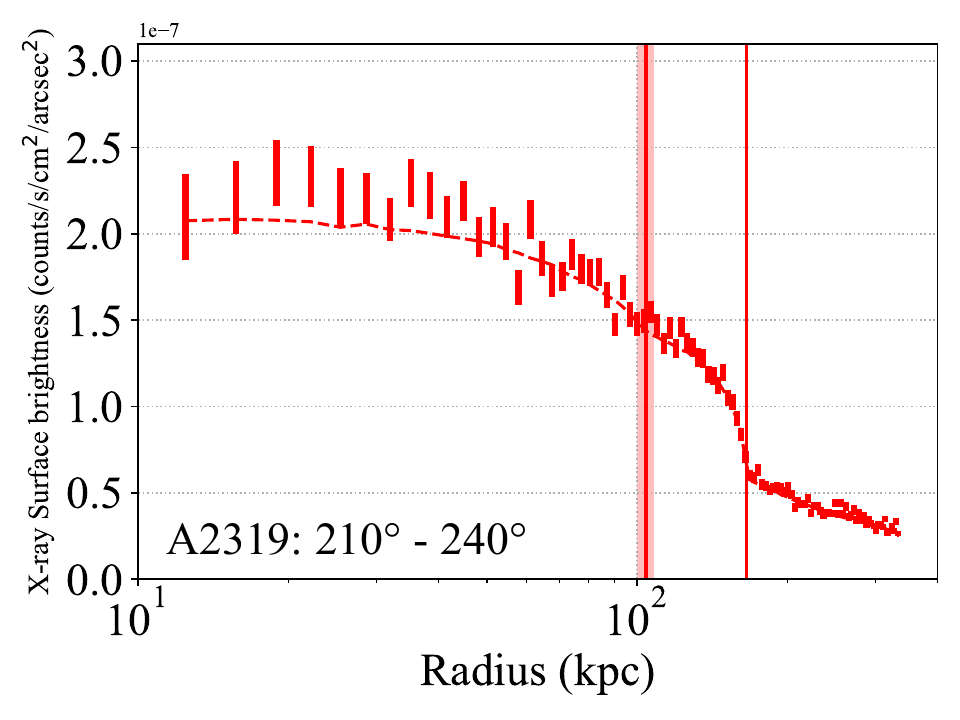}
 \end{center}
\caption{
Azimuthally averaged X-ray surface brightness profiles in the $0.5 - 7.0$\,keV band extracted from two representative sectors in A3667 (top) and A2319 (bottom), respectively. The red vertical bars show the $1\sigma$ confidence range of the azimuthally averaged X-ray surface brightness in radial bins. The red dashed line shows the best-fit profile obtained from the forward modeling analysis. The two red vertical lines and corresponding shaded areas show the positions of $r_{12}$ and $r_{23}$ and their $1\sigma$ confidence ranges, respectively.
(Top left): the $195^{\circ} - 225^{\circ}$ sector in A3667. 
(Top right): the $225^{\circ} - 255^{\circ}$ sector in A3667.
(Bottom left): the $180^{\circ} - 210^{\circ}$ sector in A2319. 
(Bottom right): the $210^{\circ} - 240^{\circ}$ sector in A2319. 
}
\label{fig:Sx_CF}
\end{figure*}

\begin{figure*}[ht]
 \begin{center}
  \includegraphics[width=8.5cm]{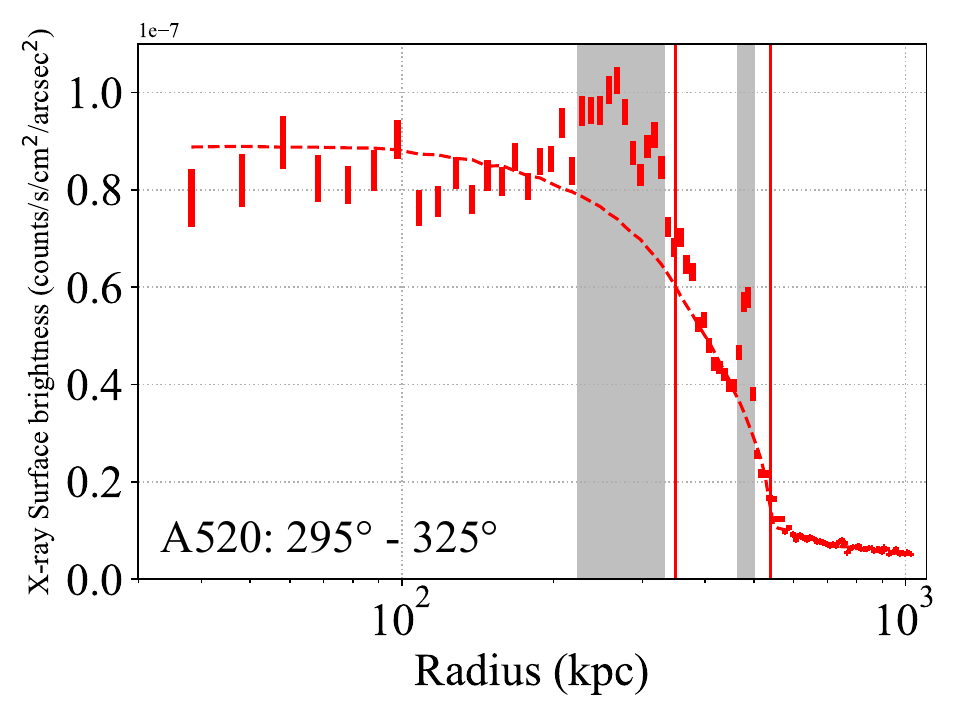}
  \includegraphics[width=8.5cm]{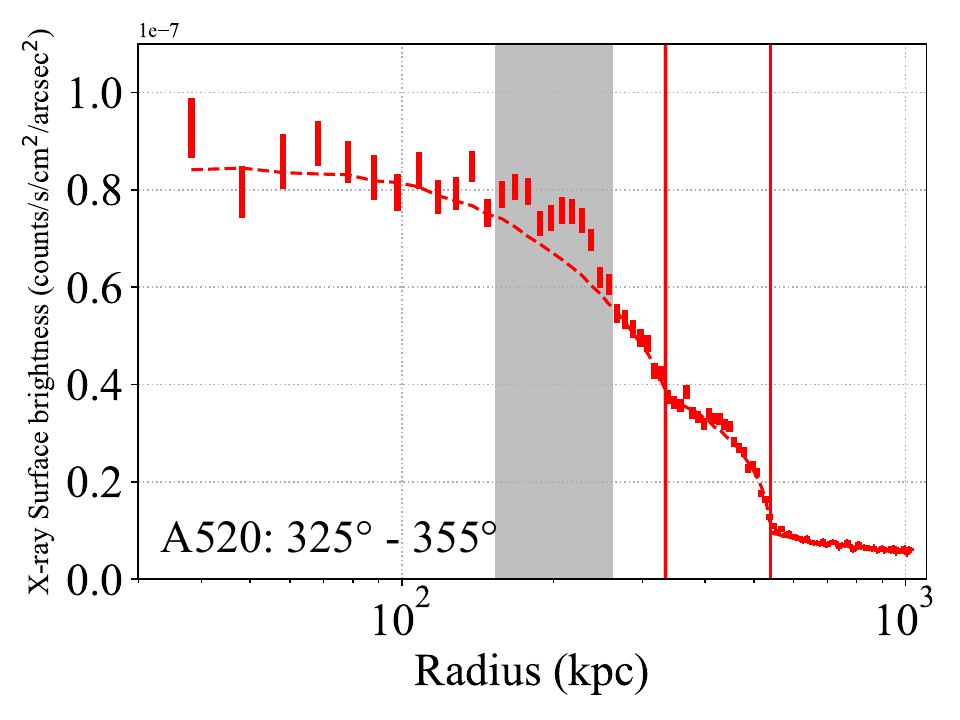}
  \includegraphics[width=8.5cm]{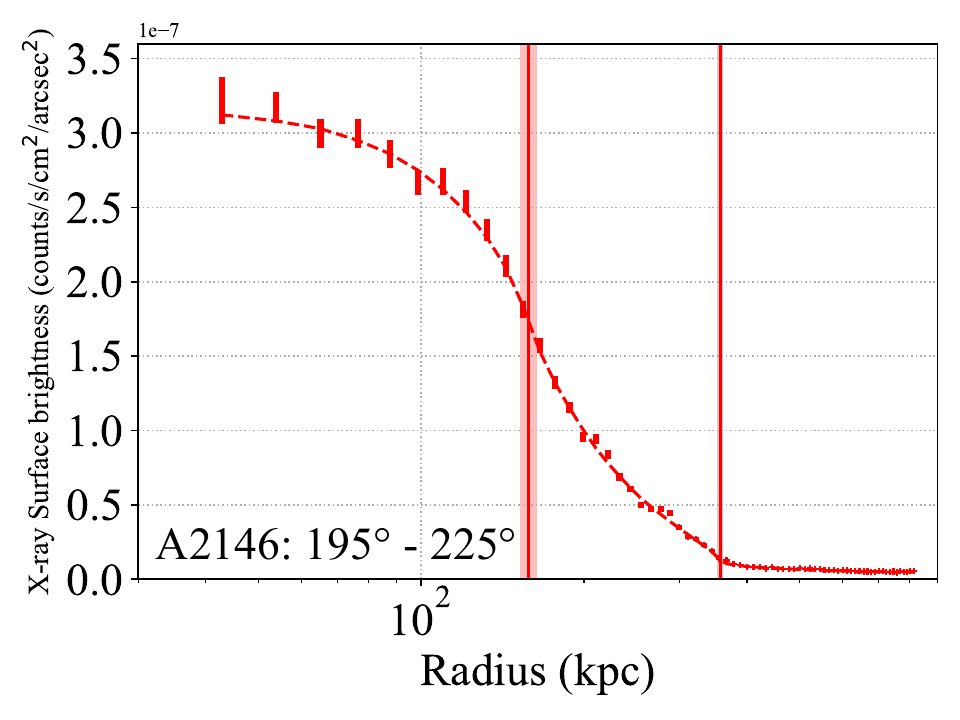}
  \includegraphics[width=8.5cm]{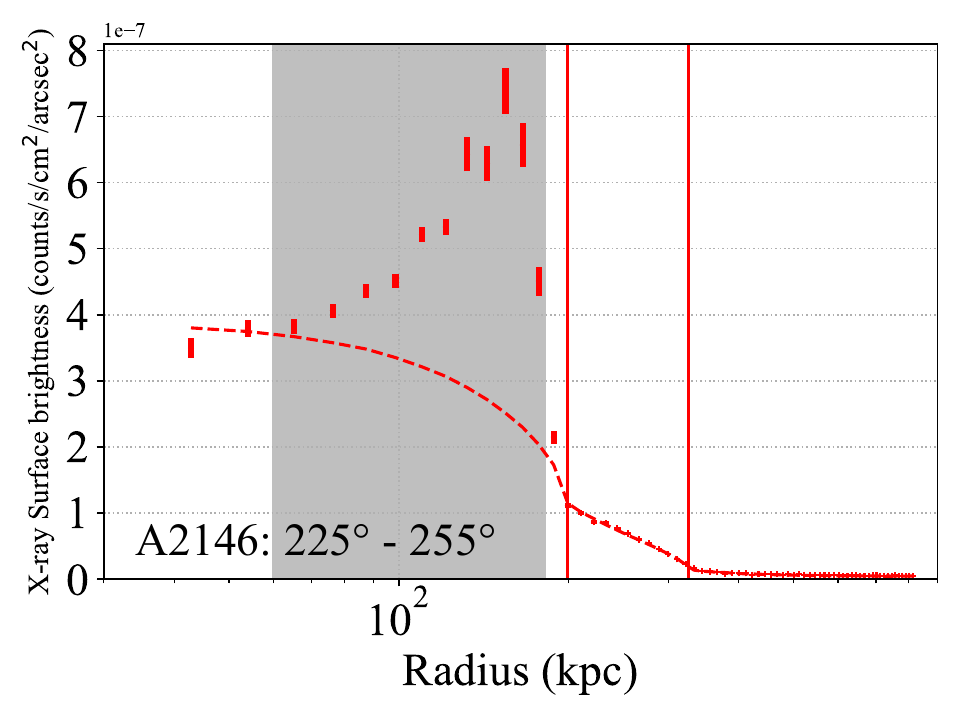}
 \end{center}
\caption{
Same as Figure~\ref{fig:Sx_CF} but for two representative sectors in A520 (top) and A2146 (bottom), respectively. The gray vertical shaded area shows a masked region that is strongly contaminated by the substructures. We removed the masked region from the forward modeling analysis.
(Top left): the $295^{\circ} - 325^{\circ}$ sector in A520. 
(Top right): the $325^{\circ} - 355^{\circ}$ sector in A520.
(Bottom left): the $195^{\circ} - 225^{\circ}$ sector in A2146. 
(Bottom right): the $225^{\circ} - 255^{\circ}$ sector in A2146. 
}
\label{fig:Sx_SF}
\end{figure*}

Figure~\ref{fig:Sx_CF} shows two representative examples of the azimuthally averaged X-ray surface brightness profiles in the $0.5 - 7.0$\,keV band for A3667 and A2319, while Figure~\ref{fig:Sx_SF} shows those for A520 and A2146, respectively. Sharp edges in the X-ray surface brightness profiles are observed. In addition, bump-like substructures are found at below the fronts in all clusters in our sample. The sectors almost perpendicular to the fronts appear not to host such bump-like substructures. These bump-like substructures are reported in the literature: e.g., \cite{Ichinohe17} for A3667 and \cite{Wang18b} for A520. As we will discuss later, a density jump and a slope change for the density profile are found at these points, respectively. Therefore, we will take into account these bump-like substructures in the forward modeling analysis (see Section~\ref{sec:model}).


\subsection{Parametric forward fitting}
\label{sec:model}

We perform forward modeling analysis of the \Chandra\ observations for our sample. In \cite{Umetsu22}, we developed a forward-modeling algorithm to simultaneously fit the X-ray surface brightness profiles binned in multiple energy bands to infer the 3D gas density and temperature profiles in a parametric form, without assuming hydrostatic equilibrium. In that algorithm, we assumed spherical symmetry. In this paper, we extend their approach to handle more generalized profiles of the ICM to adjust to our sample.

We model the 3D gas density profile $n_\mathrm{e}(r)$ as a combination of a $\beta$ model and two power-law profiles for representing an inner region ($r \le r_{12}$), an intermediate region ($r_{12} < r \le r_{23}$), and an outer region ($r_{23} < r$), respectively. In this context, $r_{12}$ and $r_{23}$ correspond to the positions of a bump-like substructure and the interface, respectively. Thus, our 3D gas density profile can be written as:
\begin{equation}
\label{eq:ne}
	n_\mathrm{e}(r)	 = 
	\begin{cases}
		n_\mathrm{e0} \left[1 + (r/r_\mathrm{c})^2 \right]^{-3\beta/2}	& : r \le r_{12} \\
		n_{\mathrm{e}, r_{12}} ~j_{12} ~(r/r_{12})^{-2s_{12}}			& : r_{12} < r \le r_{23} \\
		n_{\mathrm{e}, r_{23}} ~j_{23} ~(r/r_{23})^{-2s_{23}} 			& : r_{23} < r,
	\end{cases}
\end{equation}
where $n_\mathrm{e0}$ is the central electron number density, $\beta$ is the slope parameter, and $r_\mathrm{c}$ is the core radius of the $\beta$ model, $r_{12}$ is the 3D position of the inner density edge, $n_{{\rm e}, r_{12}}$ is the electron number density at $r_{12}$, $j_{12}$ is the amplitude of the inner density jump, $s_{12}$ is the slope parameter for the intermediate power-low model, $r_{23}$ is the 3D position of the outer density edge, $n_{{\rm e}, r_{23}}$ is the electron number density at $r_{23}$, $j_{23}$ is the amplitude of the outer density jump, and $s_{23}$ is the slope parameter for the outer power-law model.

We also model the 3D temperature profile $T(r)$. Following Equation~2 of \cite{McDonald19}, we introduce a modified version of the temperature model from \cite{Vikhlinin06}:
\begin{equation}
\label{eq:kT}
	T(r) = 
	\begin{cases}
		T_{0} \frac{(r / r_{\rm in})^{\alpha} + T_{\rm min} / T_{0}}{(r / r_{\rm in})^{\alpha} +1} \frac{1}{1 + (r / r_{\rm out})^{\beta}}	& : r \le r_{23} \\
		T_{\rm out}	& : r_{23} < r,
	\end{cases}
\end{equation}
where $T_{0}$ is the scaled temperature, $T_{\rm min}$ is the central temperature, $r_{\rm in}$ is the inner scaled radius, $r_{\rm out}$ is the outer scaled radius, $\alpha$ is the inner slope parameter, $\beta$ is the outer slope parameter, and $T_{\rm out}$ is the isothermal temperature in the outer region. This profile has seven free parameters, whereas the more generalized profile from \cite{Vikhlinin06} has nine free parameters. Note that \cite{McDonald19} fixed at $\beta = 2$ to reduce the number of free parameters. We present an example of the behaviors of our temperature model in Appendix~\ref{sec:kT}. Our model can reproduce specific temperature profiles, such as a profile with a maximum in an inner region and another with a maximum at the endpoint of the profile. For the ICM temperature in the outer region that represents the ambient component of the ICM in the primary galaxy cluster, we assume an isothermal profile (i.e., $T_{\rm out}$) to reduce computational time. 

We employ the spectroscopic-like temperature $T_\mathrm{2D}$ of \citet{Mazzotta04}, based on the inferred 3D temperature profile, to approximate spectroscopic projected temperatures extracted from \Chandra\ X-ray observations:
\begin{equation}
T_\mathrm{2D} = \frac{\int\! w T_\mathrm{3D}  dV}{\int\! w dV}, \label{eq:w}
\end{equation}
with $w = n_\mathrm{e}^2(r) T_\mathrm{3D}^{-3/4}(r)$. This $T_\mathrm{2D}$ is used as projected temperatures in each radial bin to fit the observed X-ray surface brightness profiles. We use the \texttt{pyatomdb} python package \citep{Foster20} to evaluate the cooling function $\Lambda_{\rm X}(T_{\rm 2D}, Z)$ in each energy band for a given value of the spectroscopic-like temperature $T_{\rm 2D}(r_{\perp})$, where $r_{\perp}$ is a projected radius on the plane of the sky. We assume the ICM abundance of 0.3\,solar to calculate the cooling function. Thus, the modeled X-ray surface brightness can be obtained by integrating the density profile along the line-of-sight (LOS) multiplied by the cooling function.

We have extracted the radial profiles of X-ray surface brightness in $N_\mathrm{spec}=10$ energy bands between neighbouring energies of 0.5, 0.75, 1.0, 1.25, 1.5, 2.0, 3.0, 4.0, 5.0, 6.0, and 7.0\,keV. In each energy band, the X-ray surface brightness profiles of A3667, A2319, A520, and A2146 are sampled in 170, 100, 100, and 70 linearly spaced radial bins (i.e., $3''$ per bin), respectively, centered on their curvature center (see Table~\ref{tab:pos}), respectively. Following \citet{Sanders18}, we have chosen these bands so as to capture most of the spectral information without overly increasing the computational time. In this work, we use the standard error based on the estimated variance to characterize the uncertainty in the mean X-ray surface brightness in each bin, following \cite{Umetsu22}.

We generate exposure maps of each energy band using the \texttt{fluximage} task in CIAO. We set an effective energy for each exposure map to the midpoint of the corresponding energy band. Thus, we extract the radial profiles of the exposure maps using the same approach as the X-ray surface brightness.

The background contribution in each energy band is determined from the blank-sky data included in \texttt{CALDB}. We estimate the count rate of the blank-sky data in the spectral range of $9-12$\,keV dominated by the particle background \citep{Hickox06}. Using the ratio between the count rates observed in our sample and the background ones in the $9 - 12$\,keV band, we rescale the background contribution in each energy band to match the observations of our sample, respectively, accounting for the difference in exposure times. We then construct the azimuthally averaged radial profile of the background map in each energy band. Similarly, we create azimuthally averaged radial profiles of exposure maps in the 10 energy bands.

We simultaneously fit the observed X-ray surface brightness profiles in the 10 energy bands with our model using affine-invariant MCMC sampling \citep{Goodman10} implemented by the \textsc{emcee} python package \citep{Foreman-Mackey13}. Following \cite{Umetsu22}, the log-likelihood function for the data is defined by (up to a normalization constant)
\begin{equation}
-2 \ln \mathcal{L} = \sum_{i,j} \frac{\left[d_{ij} - \left( w_{ij} \tau_i \widehat{S}_{\mathrm{X},ij} + {\cal N}_i \times \mathrm{BGD}_{ij} \right)  \right]^{2}}{\sigma_{ij}^{2}},
\end{equation}
where $i$ and $j$ run over all energy bands and all radial bins, respectively, $d_{ij}$ is the binned X-ray brightness measured in units of counts per pixel, $\sigma_{ij}$ is the statistical uncertainty of the measurement in each bin, $\tau_i$ represents the Galactic transmission in the $i$th energy band calculated by \texttt{XSPEC} using the photoionization cross sections of \citet{Verner96}, $\widehat{S}_{\mathrm{X},ij}$ is the model prediction in each bin for the X-ray surface brightness in units of counts\,cm$^{-2}$\,sec$^{-1}$,  $w_{ij}$ is the conversion factor proportional to the product of the effective area and the net exposure time in each bin in units of cm$^{2}$\,sec\,pixel$^{-1}$, $\mathrm{BGD}_{ij}$ denotes the background contribution in each bin given in units of counts per pixel, and ${\cal N}_i$ is a dimensionless calibration factor of the background in the $i$th energy band.

\subsection{Results of the forward modeling analysis}
\label{sec:results}

We have performed the forward modeling analysis to infer the 3D ICM electron number density and temperature profiles across the interface. Since the ICM pressure ($p$) and entropy ($K$) are computed as $p(r) = n_{\rm e} (r) \times T(r)$ and $K(r) = T(r) \times n_{\rm e}(r)^{-2/3}$, respectively, we also estimate the 3D profiles of the ICM pressure and entropy.

Here, we summarize the results of each cluster obtained from the forward modeling analysis.

\subsubsection{A3667}
\label{sec:A3667}

The cold front in A3667 is suitable for validating our forward modeling analysis by comparison with previous measurements.

\begin{figure*}[ht]
 \begin{center}
  \includegraphics[width=8.5cm]{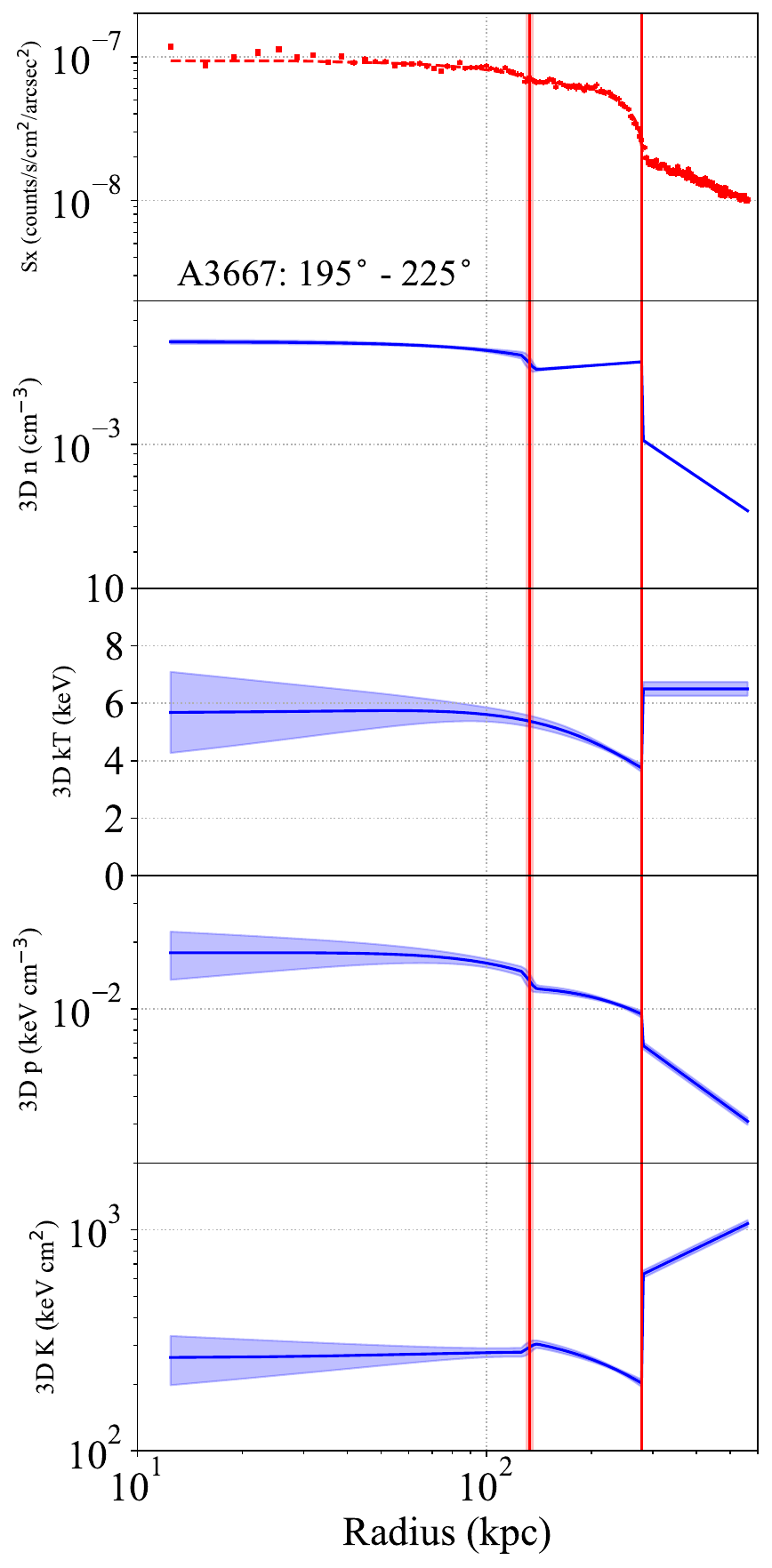}
  \includegraphics[width=8.5cm]{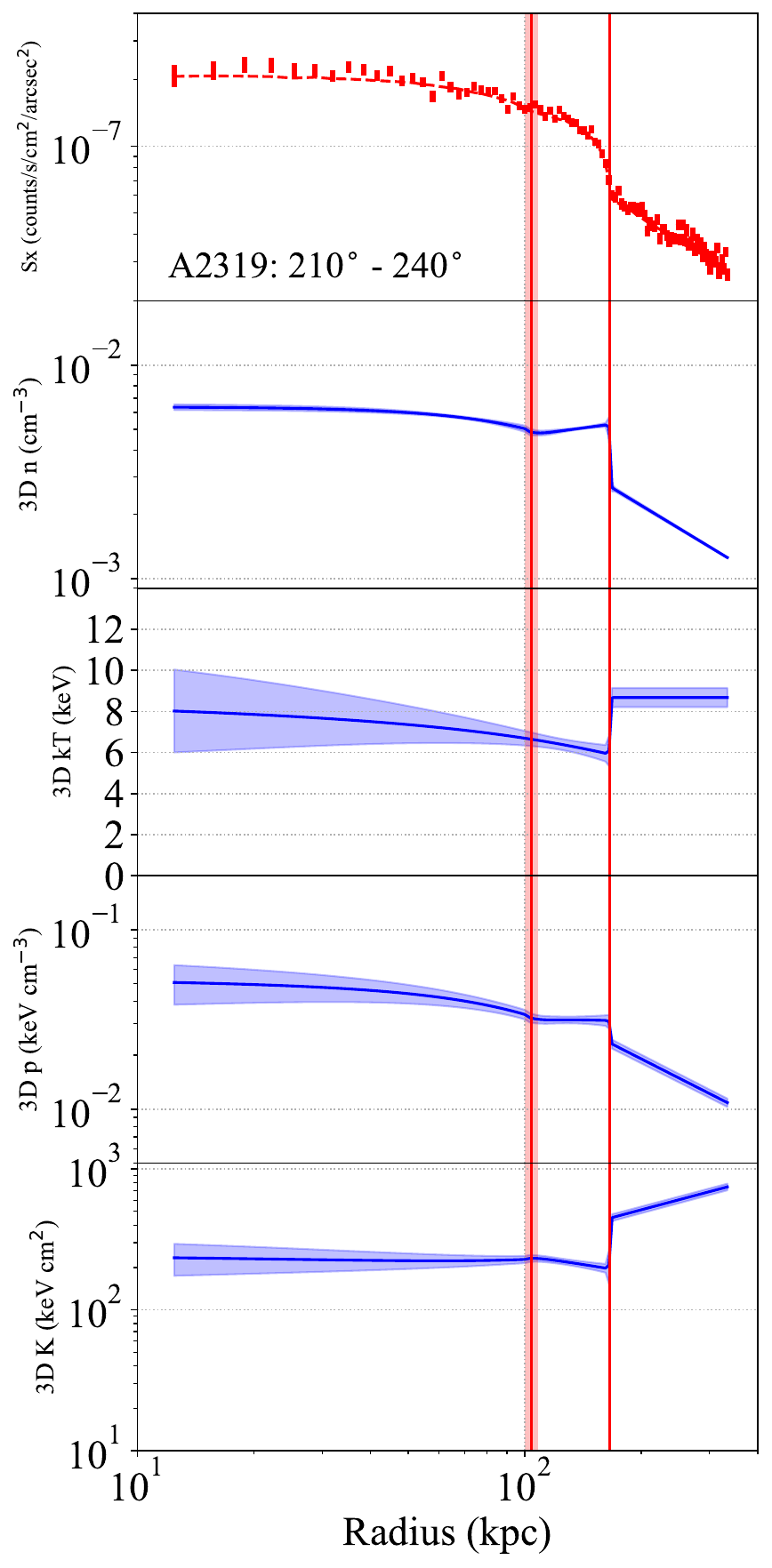}
 \end{center}
\caption{
Representative examples of the radial profiles of the X-ray surface brightness and 3D thermodynamic profiles of the ICM across the cold fronts in A3667 (left: $195^{\circ}$ to $225^{\circ}$) and A2319 (right: $210^{\circ}$ to $240^{\circ}$), respectively. From top to bottom, the panels show the X-ray surface brightness, the 3D ICM electron number density, temperature, pressure, and entropy profiles. The red dashed line in the top panel shows the best-fit profile of the X-ray surface brightness profile. The blue line and the bule shaded area correspond to the best-fit profiles and their $1\sigma$ confidence ranges for the thermodynamic properties, respectively. The two red vertical lines and the red shaded areas correspond to the positions of $r_{12}$ and $r_{23}$ and their $1\sigma$ confidence ranges, respectively.
}
\label{fig:CF_profile}
\end{figure*}

The left panel of Figure~\ref{fig:CF_profile} shows the radial profiles of one of the representative sectors in A3667 ($195^{\circ}$ to $225^{\circ}$): the observed and best-fit X-ray surface brightness profiles, as well as the inferred 3D profiles of the thermodynamic properties, i.e., the ICM electron number density, temperature, pressure, and entropy. Clear jumps in both the ICM density and temperature are detected. The contrasts of these density and temperature jumps are consistent with the previous measurements reported by \cite{Ichinohe17} that they performed spectral deprojection analyses across the edge to measure the density and temperature jumps. We found that the ICM temperature at the edge position is lowest, and the ICM temperature in the inner region ($r \le r_{12}$) is consistent with an isothermal profile. As indicated by the observed density and temperature jumps, the entropy profile shows a clear jump at the edge position. All the radial profiles are shown in Appendix~\ref{sec:all_radial} (Figure~\ref{fig:A3667_Appendix}).

\cite{Vikhlinin01} measured the ICM pressures inside and outside the cold front in A3667, respectively, and detected a pressure jump across the cold front. If a pressure gradient exists at the interface, it is difficult to maintain sharp edges in the X-ray surface brightness. Therefore, the presence of such sharp edges along with pressure jumps indicate that non-thermal support significantly contributes to maintaining the sharpness of cold fronts. In the case of the cold front in A3667, ram pressure is considered to provide non-thermal support \citep{Vikhlinin01}. Thus, \cite{Vikhlinin01} estimated the velocity of gas flow in the plane of the sky for the first time. \cite{Datta14} and \cite{Ichinohe17} applied this hypothesis to constrain the velocity of the cool gas in A3667. 

\begin{figure*}[ht]
 \begin{center}
  \includegraphics[width=8.5cm]{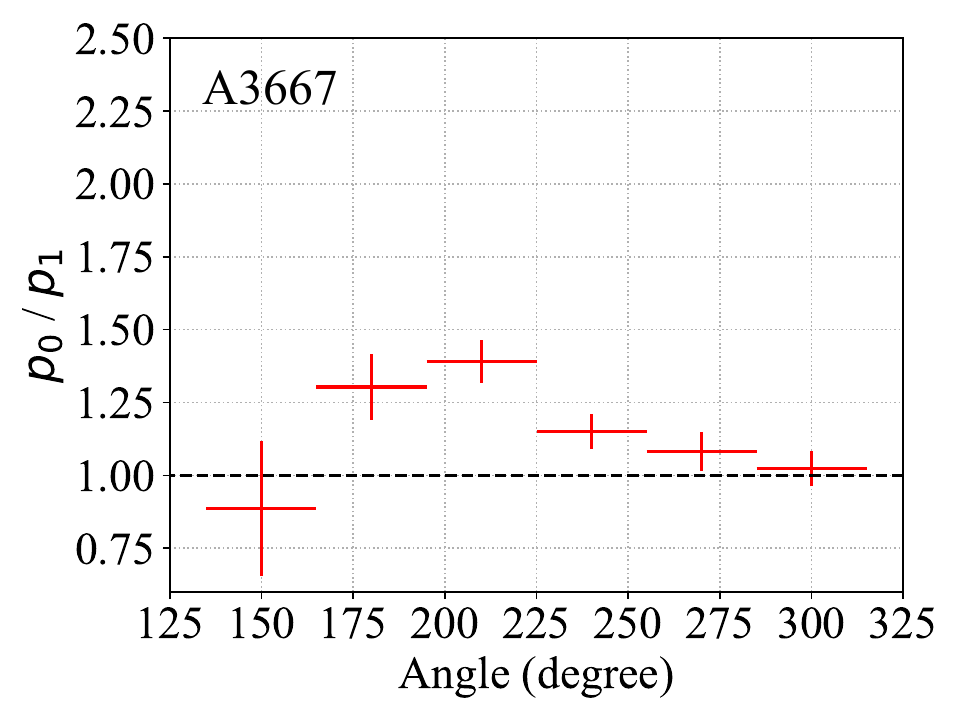}
  \includegraphics[width=8.5cm]{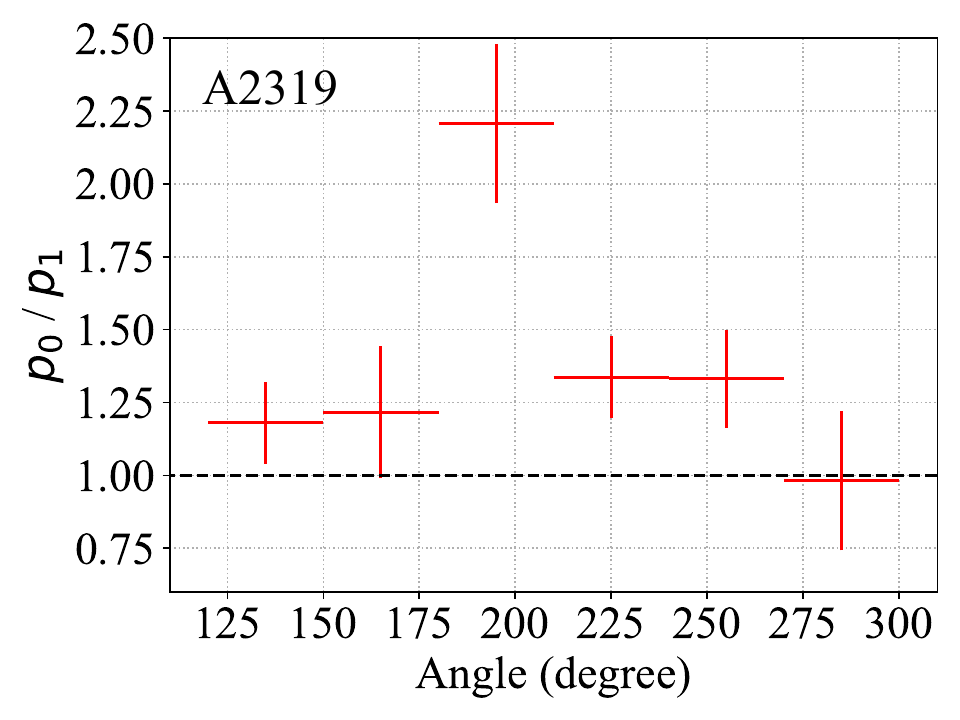}
 \end{center}
\caption{
Azimuthal distributions of the ICM thermal pressure ratio at the cold fronts in A3667 (left) and A2319 (right), respectively. The indices, $0$ and $1$, denote the values just below and above the cold fronts, respectively.
}
\label{fig:P}
\end{figure*}

We measure the pressure jump in the 3D profile at the interface of the cold front. The left panel of Figure~\ref{fig:P} shows the azimuthal distribution of the ICM thermal pressure ratio at the cold front in A3667. The values of $p_{0}$ and $p_{1}$ are extracted from the 3D ICM pressure profile at just below and above the interface of the cold front, respectively. The $195^{\circ} - 225^{\circ}$ sector exhibits the maximum value of $p_{0}/p_{1}$, indicating the highest thermal pressure gradient at the interface in this sector. On the other hand, the ratios in the sectors where are almost perpendicular to the cold front are consistent with unity, indicating that no ICM thermal pressure gradient exists at the interface.

Following \cite{Vikhlinin01}, we constrain the velocity of the gas flow using the observed pressure ratio. Following Equations~(122.1) and (122.2) in \cite{Landau59} \citep[see also][]{Vikhlinin01}, the relation between $p_{0} / p_1$ and the Mach number is expressed as
\begin{equation}
\label{eq:p0/p1}
\frac{p_0}{p_1} = 
	\begin{cases}
	\Bigl( 1 + \frac{\gamma -1}{2} \mathcal{M}^2 \Bigr)^{\gamma / (\gamma - 1)}		& : \mathcal{M} \le 1 \\
	\Bigl( \frac{\gamma + 1}{2} \Bigr)^{(\gamma + 1) / (\gamma - 1)} \mathcal{M}^2 \Bigl[ \gamma - \frac{\gamma - 1}{2 \mathcal{M}^2} \Bigr]^{-1 / (\gamma -1)}	& : \mathcal{M} > 1,
	\end{cases}
\end{equation}
where $\mathcal{M}$ is the Mach number of the free stream and $\gamma = 5/3$ is the adiabatic index of the monatomic gas. Since $\mathcal{M} = 1.0$ provides $p_0/p_1 = 2.05$, the observed values of $p_0/p_1$ indicate a subsonic motion of the gas flow. Assuming that $p_0/p_1 = 1.39 \pm 0.07$ in the sector ($195^{\circ}$ to $225^{\circ}$) represents the gas flow, the expected Mach number is $\mathcal{M} = 0.65^{+0.05}_{-0.07}$. \cite{Ichinohe17} reported that the Mach number of the free stream is $0.70 \pm 0.06$, which is consistent with that measured by \cite{Datta14}. The sound speed $c_{s1}$ in the free stream is calculated using $c_{s1} = \sqrt{\gamma kT_1 / \mu m_{\rm p}}$, where $kT_1$ is the ICM temperature of the free stream, $\mu = 0.6$ is the mean particle weight with respect to the proton mass $m_{\rm p}$. Using $kT_1 = 6.5 \pm 0.2$\,keV, the velocity of the gas flow is estimated at $850 ^{+70}_{-90}$\,km\,s$^{-1}$. This inferred velocity is in good agreement with the previous measurements by \cite{Datta14} and \cite{Ichinohe17}. Therefore, we demonstrate the validity and capability of our forward modeling analysis with the cold front in A3667.

We also found a density discontinuity at the position of $r_{12}$, and a change of the slope for the density profile between the inner ($r \le r_{12}$) and the intermediate region ($r_{12} < r \le r_{23}$). The obtained density slope for the intermediate region indicates an increase of the ICM density toward the cold front. This result is consistent with the previous measurements by \cite{Ichinohe17}.


\subsubsection{A2319}
\label{sec:A2319}

The right panel of Figure~\ref{fig:CF_profile} shows the radial profiles of one of the representative sectors in A2319 ($210^{\circ}$ to $240^{\circ}$). Clear density and temperature jumps are found. The observed 3D thermodynamic profiles of the ICM are consistent with those expected from cold fronts. The contrasts of the ICM density and temperature jumps at the cold front are measured at $n_{\rm e0} / n_{\rm e1} = 1.96 \pm 0.06$ and $T_{0} / T_{1} = 0.70 \pm 0.09$, respectively. The ICM temperature in the inner regions appear isothermal, similar to A3667. All the radial profiles are shown in Appendix~\ref{sec:all_radial} (Figure~\ref{fig:A2319_Appendix}).

The azimuthal distribution of the ICM thermal pressure ratio at the cold front in A2319 is shown in the right panel of Figure~\ref{fig:P}. The trend of the azimuthal distribution is similar to that in A3667, except for the $180^{\circ} - 210^{\circ}$ sector, where exhibits $p_{0} / p_{1} = 2.21 \pm 0.27$. The pressure ratio in the $210^{\circ} - 240^{\circ}$ sector is measured at $p_{0} / p_{1} = 1.34 \pm 0.14$, corresponding to $\mathcal{M} = 0.61^{+0.10}_{-0.14}$ using Equation~\ref{eq:p0/p1}. Hence, if ram pressure provides non-thermal support to maintaining the sharpness of the cold front, the velocity of gas flow is inferred to be $920^{+160}_{-210}$\,km\,s$^{-1}$. We will discuss more details about the cold front in A2319 in Section~\ref{sec:CF_vs_SF}.

Similar to A3667, we also found a density jump and a change of the slope of the density profile at the position of $r_{12}$.

\subsubsection{A520}
\label{sec:A520}

\begin{figure*}[ht]
 \begin{center}
  \includegraphics[width=8.5cm]{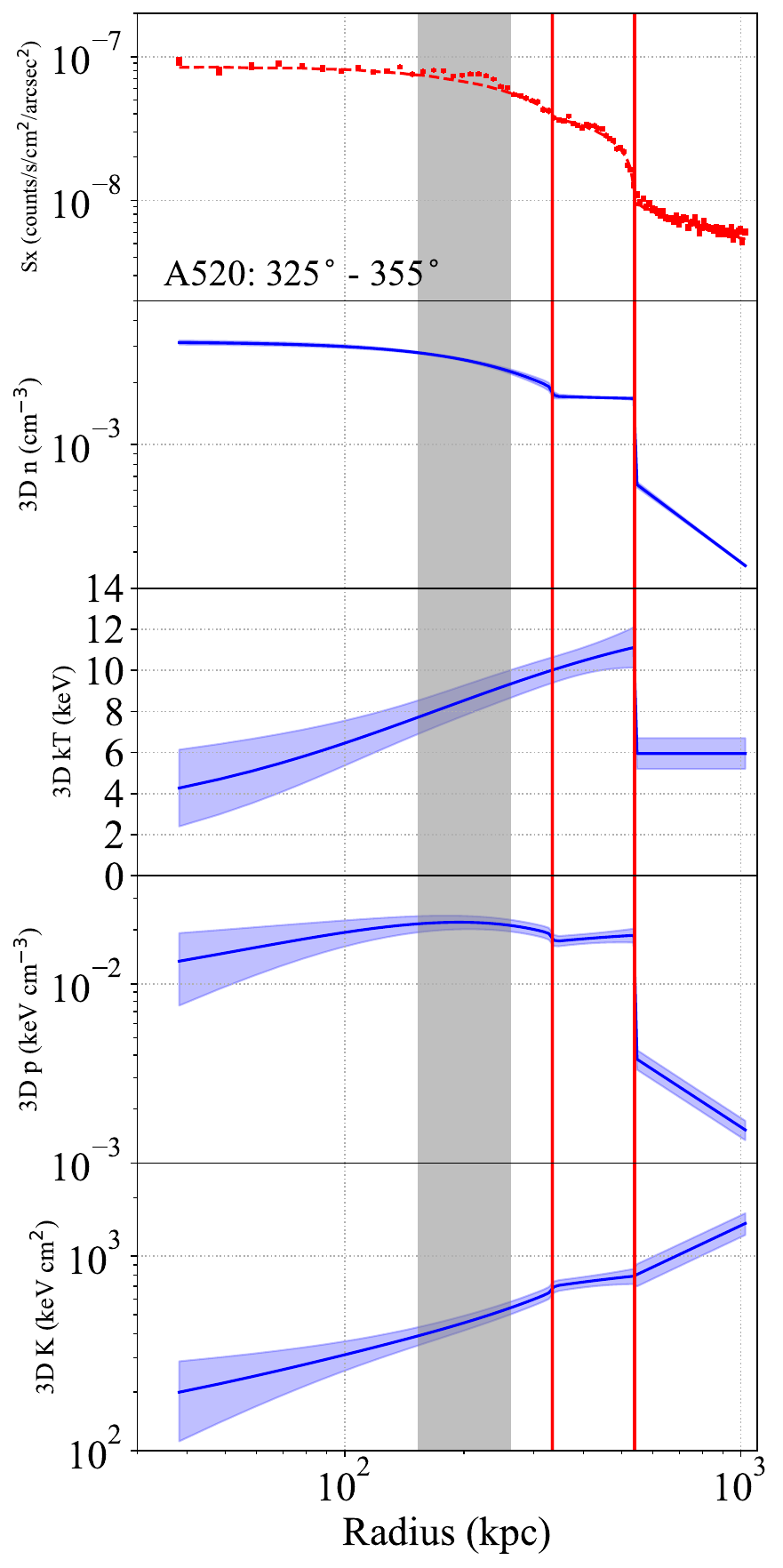}
  \includegraphics[width=8.5cm]{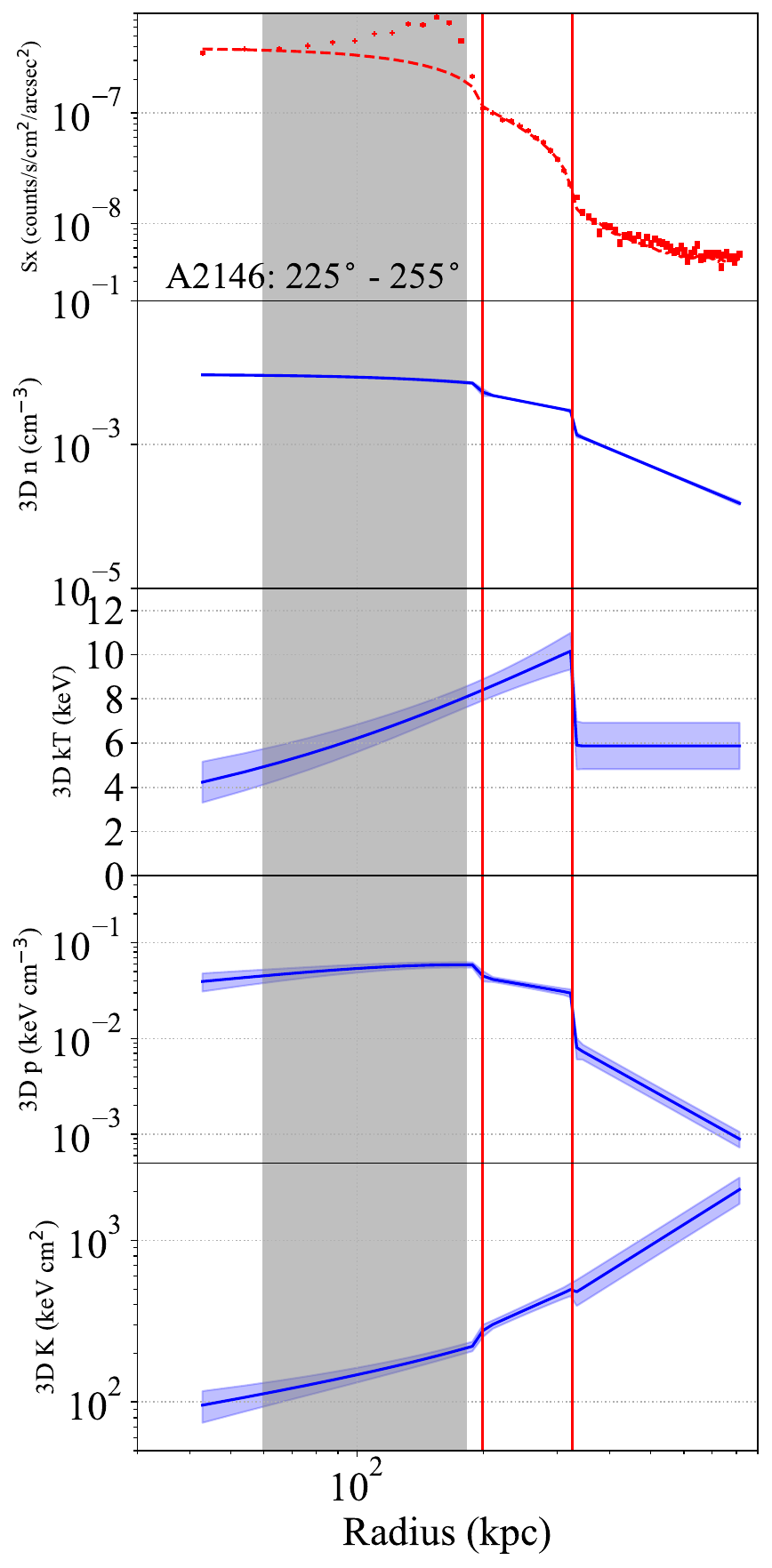}
 \end{center}
\caption{
Same as Figure~\ref{fig:CF_profile}, but for the representative sectors in A520 (left: $325^{\circ}$ to $355^{\circ}$) and A2146 (right: $225^{\circ}$ to $255^{\circ}$), respectively. The gray vertical shaded area shows a masked region where is strongly contaminated by the substructures. We removed the masked region from the forward modeling analysis.
}
\label{fig:SF_profile}
\end{figure*}

The left panel of Figure~\ref{fig:SF_profile} shows the radial profiles of one of the representative sectors in A520 ($325^{\circ}$ to $355^{\circ}$). Clear density, temperature, and pressure jumps are detected in their profiles, respectively. The obtained 3D thermodynamic structures of the ICM are consistent with those expected from shock fronts. In contrast to A3667 and A2319, the ICM temperature in the innermost region is lowest. The ICM temperature increases toward the shock fronts, namely, the highest temperature of the ICM is found at the position of the shock front. No significant entropy decrease across the interface is observed. Note that some sectors exhibits a hint of an entropy decrease across the interface. All the radial profiles are shown in Appendix~\ref{sec:all_radial} (Figure~\ref{fig:A520_Appendix}). We will discuss the behavior of the 3D temperature profiles in the post-shock region in Section~\ref{sec:3D_kT}.

A520 is known as one of the three galaxy clusters exhibiting a prominent bow shock to date. The previous measurements by \cite{Wang16b} and \cite{Wang18b} revealed the temperature map of the ICM in A520, and estimated the Mach number of the shock, based on the density jump at the edge position of the shock front. Additionally, \cite{Wang18b} conducted the spectral deprojection analysis of the ICM across the shock fronts.

Based on the results from the forward modeling analysis, we estimate the Mach number of the shock. The Mach number, $\mathcal{M}$, of the shock relative to the upstream flow is derived from the density jump using the Rankine-Hugoniot jump conditions \citep{Landau59}:
\begin{equation}
\label{eq:Mach}
\mathcal{M} = \Bigl ( \frac{2x}{(\gamma + 1) - (\gamma - 1)x} \Bigr)^{1/2},
\end{equation}
where $x$ is the density jump between the post-shock and pre-shock regions and $\gamma$ is the adiabatic index for monoatomic gas (i.e., $\gamma = 5/3$). 

\begin{figure*}[ht]
 \begin{center}
  \includegraphics[width=8.5cm]{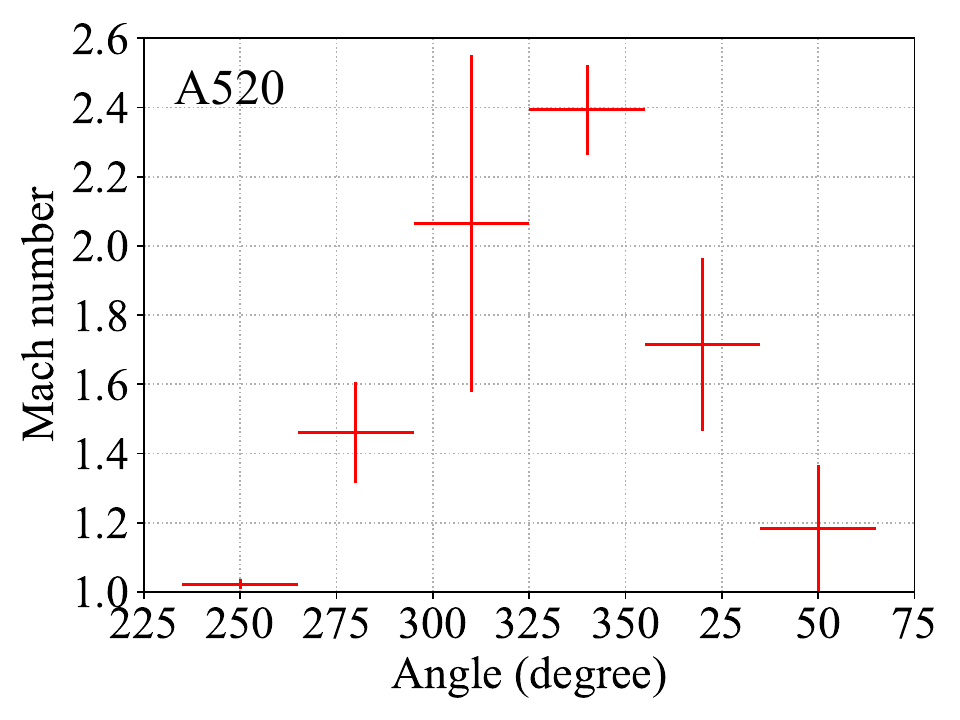}
  \includegraphics[width=8.5cm]{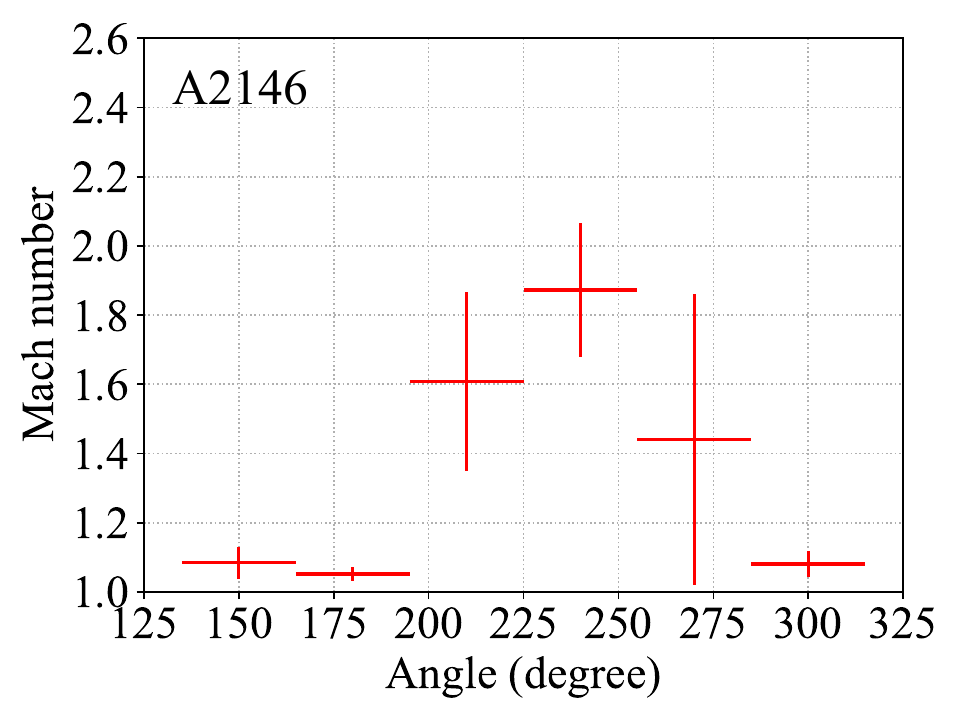}
 \end{center}
\caption{
Same as Figure~\ref{fig:P}, but for azimuthal distributions of the Mach number associated with shock in A520 (left) and A2146 (right), respectively.
}
\label{fig:Mach}
\end{figure*}

The azimuthal distribution of the Mach number associated with the shock in A520 is shown in the left panel of Figure~\ref{fig:Mach}. The Mach number in the $325^{\circ} - 355^{\circ}$ sector is estimated at $\mathcal{M} = 2.39 \pm 0.13$, which is consistent with that measured by \cite{Wang18b} ($\mathcal{M} = 2.4^{+0.4}_{-0.2}$). The Mach number in the $295^{\circ} - 325^{\circ}$ sector that hosts the apparent substructures just below the shock front has a large uncertainty, compared to the other sectors. This is due to the difficulty in determining the edge position of the shock front. This sector is heavily contaminated by a low-temperature ICM associated with the substructures (see the top right panel of Figure~\ref{fig:vs_A520}), as presented in  \cite{Wang16b} and \cite{Wang18b}.

Similar to the cold fronts in A3667 and A2319, we also found a density jump and a change of the slope of the density profile at the position of $r_{12}$.


\subsubsection{A2146}
\label{sec:A2146}

The right panel of Figure~\ref{fig:SF_profile} shows the radial profiles of one of the representative sectors in A2146 ($225^{\circ}$ to $255^{\circ}$). Similar to A520, we found clear density, temperature, and pressure jumps. The observed 3D thermodynamic structures of the ICM are consistent with those expected from shock fronts. The ICM temperature is highest at the position of the shock front and decreases toward the inner region (i.e., the post-shock region). The contrasts of the 3D ICM density and temperature jumps at the shock front are smaller than those observed in A520, respectively. Although the ICM entropy seems to be statistically consistent across the interface, there is a slight indication of an entropy decrease at the interface. All the radial profiles are shown in Appendix~\ref{sec:all_radial} (Figure~\ref{fig:A2146_Appendix}). We will discuss the behavior of the 3D temperature profiles  in the post-shock region in Section~\ref{sec:3D_kT}.

Based on the observed density jumps at the shock front and using Equation~\ref{eq:Mach}, we estimate the Mach number of the shock in each sector. The right panel of Figure~\ref{fig:Mach} shows the azimuthal distribution of the Mach number in each sector in A2146. The observed Mach number is consistent with previous measurements \citep[][]{Russell10b, Russell12b, Russell22}. As expected from the contrast of the 3D thermodynamic structures at the shock front, the observed Mach number in A2146 is systematically lower than that in A520. The largest Mach number is measured at $\mathcal{M} = 1.87 \pm 0.19$ in the $225^{\circ} - 255^{\circ}$ sector.

Similar to the other clusters, we also found a density jump and a change of the slope of the density profile at the position of $r_{12}$ in A2146.

\subsection{X-ray spectral analyses}
\label{sec:xspec}

Here, we extract the X-ray spectra of the ICM from each sector of our sample, and analyze them to measure projected temperature profiles of the ICM using \texttt{XSPEC}. Then, we compare these profiles with the projected temperature profiles derived from the forward modeling analysis using Equation~\ref{eq:w}. In the X-ray spectral analysis, we fix the column density of the Galactic absorption, ICM abundance, and cluster redshift. The ICM abundance is fixed at 0.3\,solar. Additionally, for A520 and A2146, we extract the X-ray spectra of the ICM from the masked regions where we did not use the data points in these regions for the forward modeling analysis.

We found that the projected ICM temperature profiles obtained from the forward modeling analysis are in good agreement with those measured by the spectral analysis with \texttt{XSPEC}. All the comparison plots are shown in Appendix~\ref{sec:comp_kT_p}. We emphasize that the spectral analysis also reveals a clear temperature jump at the edge positions of the cold fronts and shock fronts determined by the forward modeling analysis, whereas the contrast of the ICM temperature jump across the shock fronts is smaller than that observed in the 3D profiles. This is due to contamination by low-temperature gas along the LOS. Moreover, the isothermal profiles assumed for outside the cold fronts or shock fronts appear to agree with the observed temperature profiles by the spectral analysis.

The results of the spectral analysis indicate that the ICM temperatures in the radial bins that host the apparent substructures just below the shock fronts in A520 and A2146 are significantly lower than those in the neighboring radial bins. These features are consistent with previous measurements for A520 \citep{Wang18b} and A2146 \citep{Russell22}, respectively. These results indicate that such substructures originate from a cool core of an infalling subcluster and such lower-temperature gas is being stripped and disrupted by ongoing mergers.

\section{Discussion}
\label{sec:discussion}

\subsection{3D thermodynamic structures across the interfaces}
\label{sec:CF_vs_SF}

To infer the 3D thermodynamic structures of the ICM across the interface of cold fronts and shock fronts, we have extended the forward modeling approach originally developed in \cite{Umetsu22} to uniformly apply to the X-ray surface brightness profiles of our sample taken with the \Chandra\ X-ray Observatory. We have validated that our forward modeling analysis can reproduce well the expected characteristics of the ICM thermodynamic profiles across cold fronts and shock fronts. Thus, our forward modeling analysis enables us to uniformly classify edges in the X-ray surface brightness as either cold fronts or shock fronts.

In previous studies, the X-ray spectral analysis, particularly, the spectral deprojection analysis with the \texttt{projct} routine implemented in \texttt{XSPEC} was typically conducted to infer the 3D thermodynamic structures of the ICM across the interface. However, it has been reported that artificial behaviors are often seen in the obtained deprojected profiles of the ICM thermodynamic properties \citep[e.g.,][]{Russell08}. Although the analysis of the projected profiles of the ICM thermodynamic properties is relatively straightforward, the projection effect from different temperature gases along the LOS makes significant contamination in measuring these properties \citep[e.g.,][]{Molnar15}. In addition, additional modeling is needed to obtain the radial profiles of the ICM properties. Moreover, in general, the ICM density and temperature decrease toward the outer regions of galaxy clusters, and a size of radial bins is required to be large to collect enough photon counts for the X-ray spectral analysis. As a result, it is difficult to distinguish the real pressure jump generated by shocks from the artificial gradients produced by the coarse discretization of radial bins.

The forward modeling approach developed in this paper allows us to infer the 3D thermodynamic structures of the ICM in a parametric form from the inner region (below the interface) to the outer region (above the interface), based on Equations~\ref{eq:ne} and \ref{eq:kT}. Thus, we are able to measure the density and temperature jumps at the interface, and the 3D profiles of the thermodynamic properties simultaneously and self-consistently. The ICM thermodynamic properties in the inner region is important not only for distinguishing between cold fronts and shock fronts but also for exploring the interior 3D thermal structures of the primary galaxy clusters.

The obtained 3D entropy profiles in A3667 and A2319 indicate a constant distribution in the inner regions with $K = 200 - 300$\,keV\,cm$^{2}$, which is significantly larger than that measured in cool cores \citep[$K \ll 100$\,keV\,cm$^{2}$; e.g.,][]{Hudson10}. These ICM entropy features are consistent with those of merging clusters. In addition, the ICM entropy starts decreasing at $r_{12}$ toward the interface of the cold front, and shows a clear jump at the interface. The sectors with no apparent interface (i.e., $135^{\circ}$ to $165^{\circ}$ and $285^{\circ}$ to $315^{\circ}$) in A3667 show no such entropy profile as well as the isothermal profile of the ICM temperature (see the top left and bottom right panels of Figure~\ref{fig:A3667_Appendix}). 

In A520 and A2146, the 3D ICM temperature profiles reveal lower-temperature gas ($\sim 4 - 5$\,keV) in the innermost regions ($r \lesssim 50$\,kpc), with a gradual increase toward the interface of the shock front. This lower-temperature gas may originate from a cool core of the primary cluster. Since the infalling subcluster is now passing through the core region of the primary cluster, the cool core of the primary cluster may not be disrupted yet or heated well completely. This implies that the very central region of a cool core might be still surviving. In addition to this hypothesis, the merger perhaps might have non-zero impact parameter along the LOS. Numerical simulations will provide further insights into the survival of cool cores during major mergers with different merger conditions \citep[e.g.,][]{Hahn17, Valdarnini21}.

In this paper, we have assumed an isothermal temperature profile for the outer regions, i.e., regions above the interface (see Equation~\ref{eq:kT}), to reduce computational time. However, assuming such an isothermal temperature profile may be simplistic. In general, the ICM temperature decreases toward the outskirts. To investigate the systematic uncertainty due to the assumption of an isothermal temperature profile, we analyze the sectors exhibiting the highest Mach number in A520 and A2146, respectively, with a power-law temperature profile for the outer region (i.e., $T(r) = T_{\rm out} (r/r_{23})^{\gamma}: r_{23} < r$), instead of the isothermal temperature profile. We find no significant difference between the results derived from the isothermal and power-law temperature profiles. Further study is needed to extend a power-law temperature profile for our forward modeling approach and apply it to other interfaces.

\subsection{3D temperature profiles in the post-shock regions}
\label{sec:3D_kT}

The forward modeling analysis provides us with an opportunity to discuss the mechanisms of shock-heating in terms of not only the ICM temperature but also its 3D profile for the first time. All the observed 3D ICM temperature profiles associated with the shock fronts indicate an universal trend: (1) the peak of the ICM temperature is found at the position of the shock front, and (2) the ICM temperature gradually decreases toward inside the post-shock region, except for the $295^{\circ} - 325^{\circ}$ sector in A520 (top right panel of Figure~\ref{fig:A520_Appendix}) where is significantly affected by low-temperature gas of the substructures. In particular, the $325^{\circ} - 355^{\circ}$ sector with the largest Mach number ($\mathcal{M} \sim 2.4$) in A520 (left panel of Figure~\ref{fig:SF_profile}) indicates that (1) the highest ICM temperature with $11.1 \pm 1.0$\,keV is found at the position of the shock front, and (2) the ICM temperature appears isothermal with a temperature of $\sim 10$\,keV until $\sim 300$\,kpc away from the shock front in the post-shock region. 
The timescale of equilibration via Coulomb collisions between the ions and the electrons is expressed by
\begin{equation}
t_{\rm eq} \sim 2 \times 10^{8} \bigg(\frac{n_{\rm e}}{10^{-3}\,{\rm cm}^{-3}} \bigg)^{-1} \bigg(\frac{T_{\rm e}}{10^{8}\,{\rm K}} \bigg)^{3/2} \,{\rm yr},
\end{equation}
\citep[e.g.,][]{Wong09, Sarazin16, Wang18b}. Since the post-shock gas velocity is inferred to be $\sim 1150$\,km\,s$^{-1}$, the location of $\sim 300$\,kpc away from the shock front corresponds to the expected collision equilibration timescale of $t_{\rm eq} \sim 0.3$\,Gyr for the shock. Therefore, we found no significant gradient of the electron temperature within the expected collision equilibration timescale.

In general, two types of models for shock-heating mechanism are considered: instant-equilibration and adiabatic compression \citep{Markevitch07}. The instant-equilibration model predicts that the electrons rapidly equilibrate with the ions, through e.g., magnetic fields (i.e., collisionless heating), such that the electron temperature rapidly reaches the temperature determined by the Rankine-Hugoniot jump conditions immediately behind the shock front. On the other hand, the adiabatic compression model predicts that an adiabatic compression of the electron population at the shock and a subsequent slower equilibration with the ions on a timescale determined by Coulomb collisions. These two models have been studied in a variety of shock fronts in merging clusters \citep[e.g.,][]{Markevitch06, Russell12b, Sarazin16, Wang18b, Di_Mascolo19b, Russell22, Sarkar22, Sarkar24}.

The observed 3D profiles of the ICM temperature are in good agreement with those expected by the instant-equilibration model. The highest temperature at the shock front and the isothermal profile behind the shock front are expected to be seen in the instant-equilibration model. If the adiabatic compression model is the case, the highest temperature is expected to be found at the location where the electrons reach collision equilibration, and the temperature at the shock front is lower than at the equilibration point. These predicted profiles are not shown in the observed 3D temperature profiles of all sectors exhibiting the shock front. However, even though the instant-equilibration model seems to be better to explain the observed profiles, the observed peak temperature of the ICM is significantly lower than that expected by the instant-equilibration model. The expected temperature from a $\mathcal{M} = 2.4$ shock in the $325^{\circ} - 355^{\circ}$ sector in A520 is $15.7^{+1.6}_{-1.2}$\,keV, which is $\sim 2.3\,\sigma$ different from the observed temperature. This discrepancy may be due to that (1) the sensitivity of \Chandra\ is not sufficient to observe such hot gas ($\gg 10$\,keV) \citep[e.g.,][]{Wang18b}, and (2) even though the merger takes place in the plane of the sky, if the merger has a non-zero impact parameter along the LOS, low-temperature gas distributed along the LOS may not be taken into account fully under the assumption of spherical symmetry, such that contamination from such gas causes a bias to make the observed temperature lower \citep[e.g.,][]{Molnar15}. Note that even the case of the adiabatic compression model, the equilibration temperature of $\sim 16$\,keV is expected, but such temperature is not seen in the observed profile either. Therefore, we conclude that the trend of the observed 3D temperature profiles is consistent with the instant-equilibration model, whereas the adiabatic compression model is not completely ruled out. Numerical simulations of a binary cluster merger involving relatively high Mach number and high-sensitivity observations in the hard X-ray ($> 10$\,keV) band are needed to explore the shock-heating mechanisms. 

High-energy resolution spectroscopy with e.g., {\em XRISM} \citep{Tashiro18} and Athena \citep{Barcons17} is expected to reveal the thermodynamic properties of shock-heated gas more accurately. For instance, the X-ray microcalorimeter Resolve on board {\em XRISM} is expected to enable us to measure the ionization temperature of the electrons using the flux ratio between Fe XXV and Fe XXVI \citep[e.g.,][]{Molnar15}. Comparing the ionization temperature with the electron temperature measured by the continuum, we will be able to conduct a diagnosis of plasma conditions in the ICM \citep[e.g.,][]{Inoue16}.

\subsection{Azimuthal distributions of the ICM thermal pressure ratio at the cold fronts}
\label{sec:P_dist}

We have measured the azimuthal distribution of the ICM thermal pressure ratio across the cold fronts in A3667 and A2319, as shown in Figure~\ref{fig:P}. For A3667, the observed trend is consistent with previous measurements \citep[][]{Datta14, Ichinohe17}, indicating that the nature of subsonic gas motions. On the other hand, A2319 shows a significant pressure jump at the interface in the $180^{\circ} - 210^{\circ}$ sector. The pressure jump in this particular sector is measured at $p_{0} / p_{1} = 2.21 \pm 0.27$. The other sectors appear similar to the trend seen in A3667. If the sharpness of the cold front in A2319 is maintained by ram-pressure, the expected gas flow velocity in this particular sector is comparable to the transonic regime, given that $\mathcal{M} = 1.0$ provides $p_0/p_1 = 2.05$.

Cold fronts have been classified into two types: stripping and sloshing \citep{Markevitch07}. Stripping cold fronts are typically found at the front of an infalling subcluster. On the other hand, sloshing cold fronts are mostly found in the central regions of cool-core clusters, and considered to be generated by gas sloshing induced by stirring motions owing to the transport of angular momentum from an infalling subcluster \citep[e.g.,][]{Ascasibar06, ZuHone10}. Spiral patterns in the residual images of X-ray surface brightness are known as evidence of gas sloshing, and sloshing cold fronts are located at the edges of such spiral patterns \citep[e.g.,][]{Ascasibar06, ZuHone10, Roediger11, Roediger12, Keshet12, ZuHone13b, Ueda17, Naor20, Ueda20}. Thus, in contrast to stripping cold fronts, since sloshing gas originates in cool cores, contact discontinuities are generated between gases of different entropy originally at different places in the same cool-core cluster. Such different conditions may generate different thermodynamic and/or microphysical properties of the ICM across cold fronts.

To investigate the differences between stripping and sloshing cold fronts, we focus on the Perseus cluster as a reference for sloshing cold fronts. This is because the Perseus cluster is known to host a prominent sloshing cold front seen as a spiral-like feature in the residual image of the X-ray surface brightness \citep[][also see Figure~\ref{fig:Perseus}]{Churazov03, Zhuravleva14}. We apply the forward modeling approach to the observed X-ray surface brightness profile extracted along the direction toward the sloshing cold front. A detailed analysis is presented in Appendix~\ref{sec:Perseus}. We found a positive pressure jump at the interface, indicating a different behavior compared to the cold fronts in A3667 and A2319. Such positive pressure jump was found and discussed in \cite{Ichinohe19}. Thus, even though the sloshing cold front in the Perseus cluster exhibits a distinct trend of pressure jump from the cold fronts in A3667 and A2319, it is hard to look into possible differences between A3667 and A2319.

The higher pressure jump at the interface in the $180^{\circ} - 210^{\circ}$ sector of A2319 may be a hint of exploring the mechanisms to create cold fronts. If gas sloshing is the case, this sector is located at the head of the spiral pattern \citep{Ichinohe21}. The head of a sloshing spiral is expected to have the fastest tangential gas flow. In fact, transonic motions associated with gas sloshing are found in Abell\,907 \citep{Ueda19}. In this context, the higher pressure jump may be evidence of gas sloshing. However, the presence of clear X-ray substructures in the north-west direction may indicate a major merger in A2319 \citep[e.g.,][]{Markevitch96, Ghirardini18}. In fact, the observed flat entropy profiles in the inner region of A2319 support this hypothesis. If a major merger is the case, the spiral pattern can be interpreted as bulk motions of the core of A2319, in a similar fashion to those of A754 \citep[e.g.,][]{Markevitch03, Henry04, Macario11}. Further study is needed to explore the origin of the spiral pattern. On the other hand, although the cold front in A3667 is widely recognized as a stripping cold front, the hypothesis that this cold front results from gas sloshing either in the plane of the sky or the LOS can be considered. Since no apparent difference in the trend for the pressure ratio is found, it is difficult to distinguish stripping cold fronts from sloshing cold fronts, based on the results obtained from the forward modeling analysis. This hypothesis of gas sloshing is testable with {\em XRISM}.

\subsection{Position of the interface}
\label{sec:pos}

\begin{figure*}[ht]
 \begin{center}
  \includegraphics[width=8.0cm]{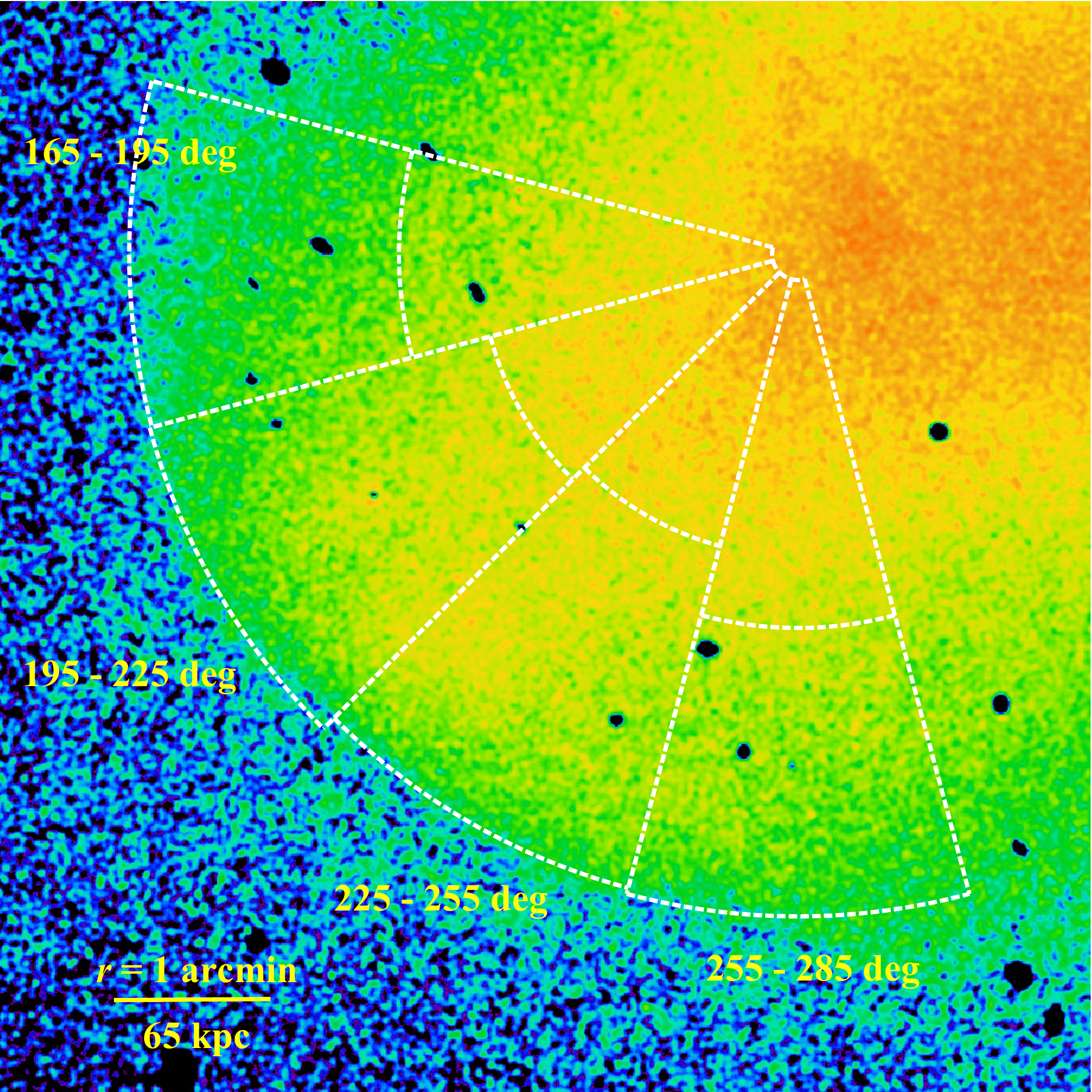}
  \includegraphics[width=8.0cm]{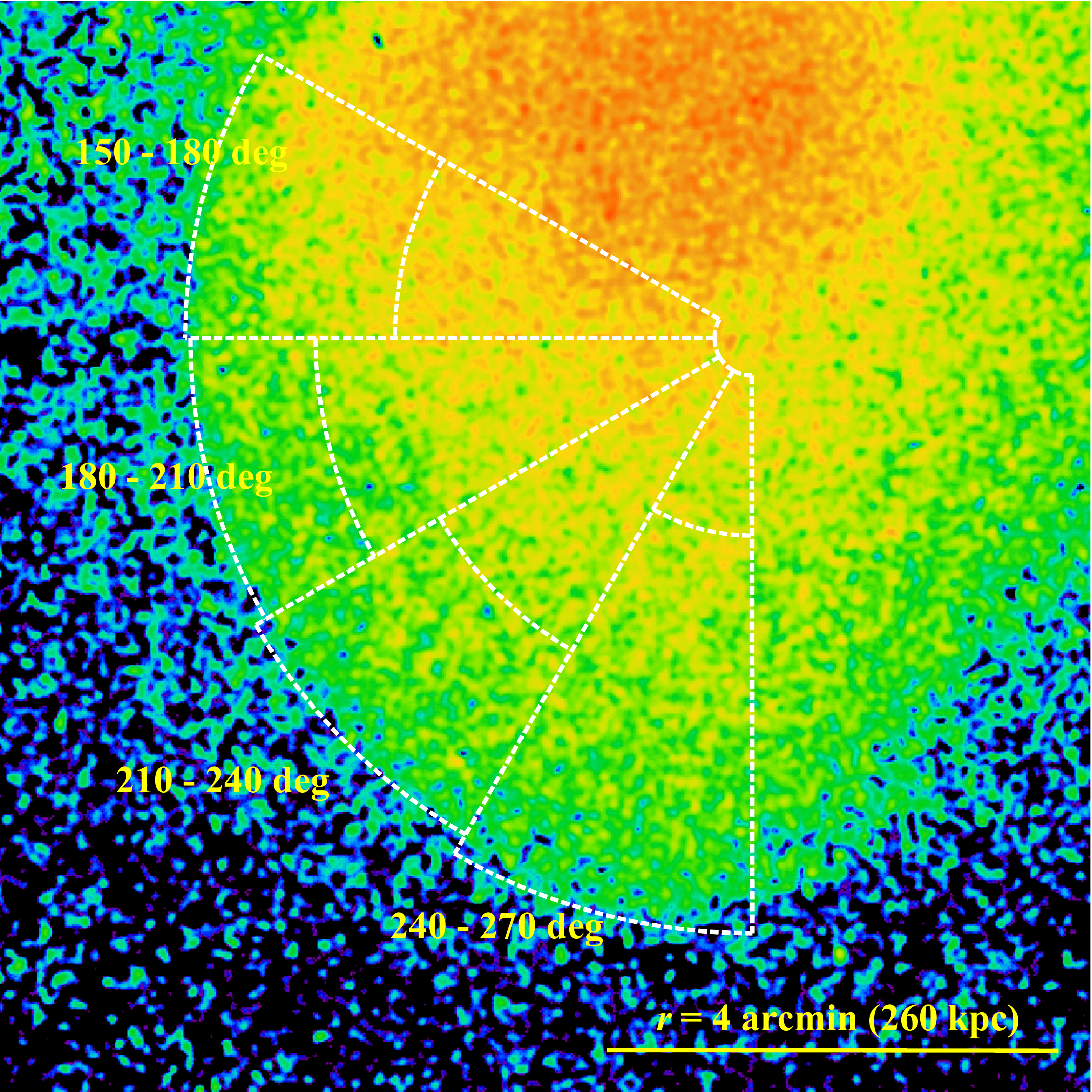}
  \includegraphics[width=8.0cm]{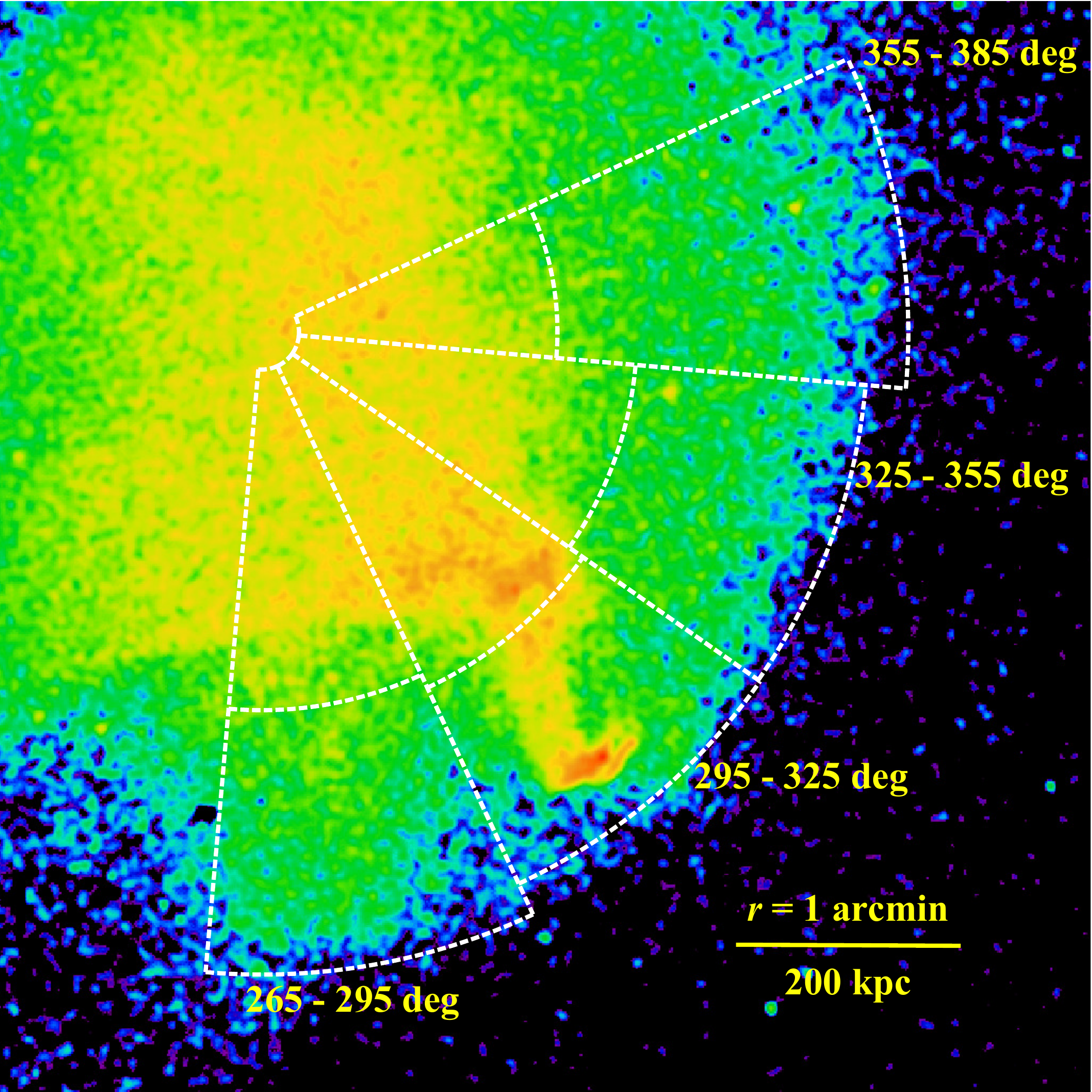}
  \includegraphics[width=8.0cm]{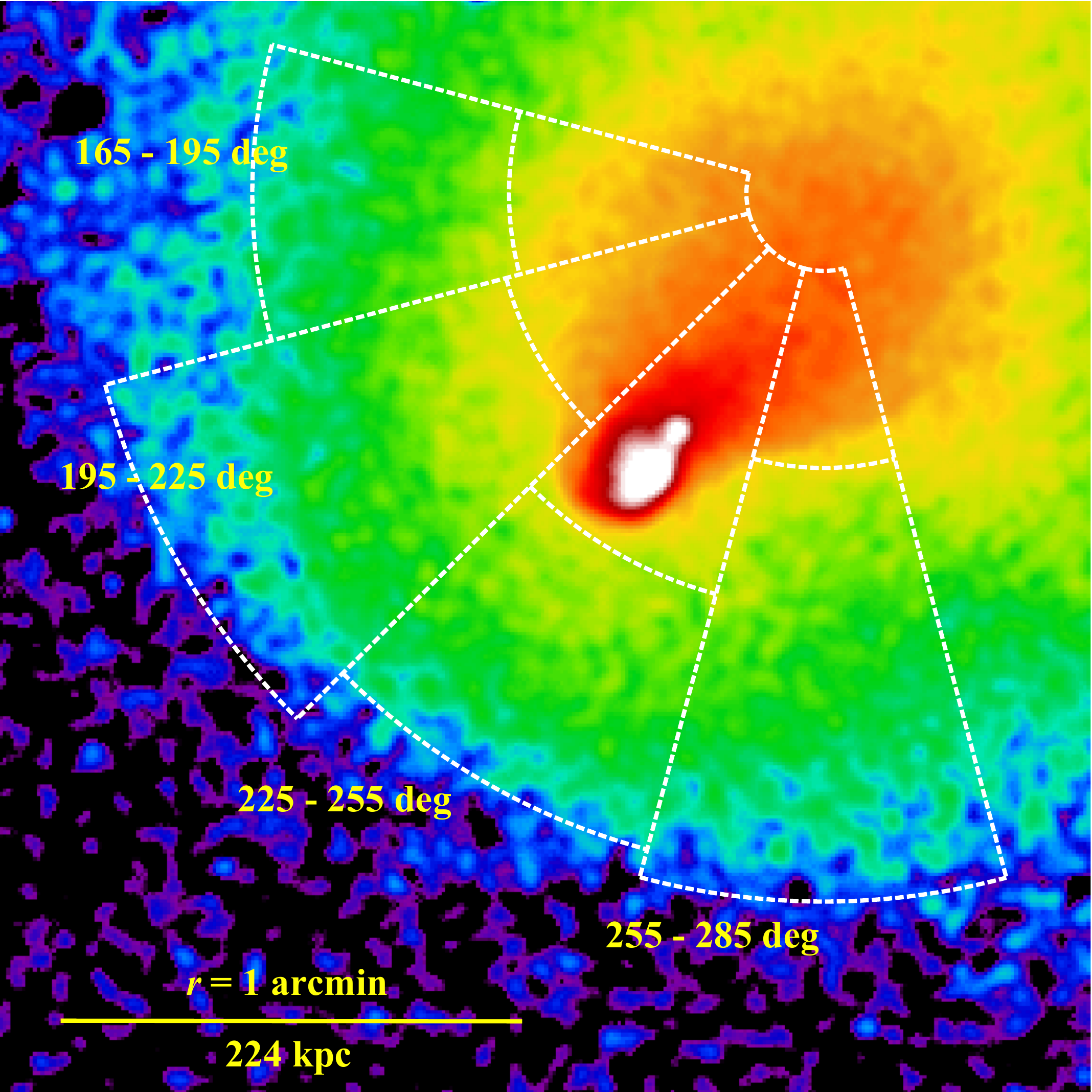}
 \end{center}
\caption{
Zoomed-in images around the cold fronts in A3667 (top left) and A2319 (top right), respectively, and around the shock fronts in A520 (bottom left) and A2146 (bottom right), respectively.
The overlaid, dashed white sectors denote the directions along each sector analyzed and the edge positions associated with the fronts. The middle, dashed white lines correspond to the positions of the bump-like substructures (i.e., $r_{12}$ in each sector), respectively. The yellow bar in each panel shows a spatial scale of the images. 
}
\label{fig:img_zoom}
\end{figure*}

Figure~\ref{fig:img_zoom} shows the azimuthal variations of the positions of the interface detected in each sector for our sample. For the cold fronts, the positions of the interface show spatial variations among the four sectors in A3667 and A2319, respectively. For A520, the positions of the interface located at the front of the shock front are statistically consistent with each other (i.e., $295^{\circ}$ to $325^{\circ}$ and $325^{\circ}$ to $355^{\circ}$). For A2146, the positions also show spatial variations among the four sectors. 

Such spatial variations of the interface for the cold fronts may be due to wiggles through the Kelvin-Helmholtz instability \citep[e.g.,][]{Ichinohe17, Ichinohe21}. For the shock front in A2146, the observed spatial variations of the positions of the shock front may also be seen in the previous measurements \citep{Russell22}. \cite{Russell22} shows the spatial variations in part of the sectors analyzed in this paper (i.e., part of $195^{\circ} - 285^{\circ}$). On the other hand, A520 shows no such variation at least the front of the shock front. This may be owing to stronger shock than A2146. However, it is possible that missalignment of the center of the curvature and/or the assumption of constant curvature approximation causes artificial spatial variations of the interface. Further dedicated study (e.g., more azimuthally-resolved study, and X-ray microcalorimeter spectroscopy with high-angular resolution) is needed to investigate the origin of the spatial variations of the interface.

\section{Summary and conclusions}
\label{sec:summary}

In this paper, we have conducted forward modeling analysis of the observed \Chandra\ X-ray surface brightness profiles across the cold fronts in A3667 and A2319 as well as the shock fronts in A520 and A2146. The goal of this study was to investigate the 3D thermodynamic structures of the ICM across these fronts and to examine the shock-heating mechanisms and the processes for creating cold fronts. To this end, we have measured the 3D thermodynamic profiles of the ICM across the fronts simultaneously and self-consistently. The main conclusions of this paper are summarized as follows:

\begin{enumerate}

\item We extended the previous forward modeling approach with more generalized ICM density and temperature profiles to uniformly apply to the X-ray surface brightness profiles of our sample, allowing us to infer the 3D thermodynamic structures of the ICM across the interface. We succeeded not only in determining the positions of the cold fronts and shock fronts but also in measuring the 3D thermodynamic structures simultaneously and self-consistently. 

\item The observed 3D thermodynamic profiles with the forward modeling analysis are consistent with the characteristic features of both the cold front and the shock front. The density and temperature jumps at the interface are clearly detected. The ICM thermal pressure jumps at the cold front in A3667 are consistent with the previous measurements. We also obtained the azimuthal distribution of the ICM thermal pressure ratio at the cold front in A2319. The Mach numbers of the shocks in A520 and A2146 are consistent with the previous works, respectively. 

\item The observed 3D temperature profiles in the post-shock region of all sectors exhibiting shocks in A520 and A2146 are in good agreement with those predicted by the instant-equilibration model. The ICM temperature is highest at the position of the shock front. In the case of the sector exhibiting $\mathcal{M} = 2.4$ in A520, the ICM temperature appears isothermal with a temperature of $\sim 10$\,keV until $\sim 300$\,kpc away from the shock front (i.e., inside the post-shock region) where the electrons reach to equilibration via Coulomb collisions between the ions and the electrons. However, the adiabatic compression model is not completely ruled out.

\item The observed azimuthal distribution of the ICM thermal pressure ratio in A2319 is in broadly agreement with that in A3667, except for one sector. This particular sector shows a high pressure jump, indicating the presence of transonic motions if the sharpness of the cold front is maintained by ram pressure. This result might imply different origins of the cold fronts, i.e., stripping and sloshing.

\end{enumerate}

\begin{acknowledgments}
We are grateful to the anonymous referee for the helpful and constructive comments.
We thank Maxim Markevitch for the fruitful discussions and constructive comments.
We also thank Keiichi Umetsu and Sandor Molnar for their helpful discussions.
The scientific results reported in this article are based on data obtained from the Chandra Data Archive\footnote{\url{https://doi.org/10.25574/cdc.265}}.
SU acknowledges the support from the National Science and Technology Council of Taiwan (NSTC 111-2112-M-001-026-MY3).

\end{acknowledgments}

%

\vspace{5mm}
\facilities{CXO}


\software{
astropy \citep{Astropy13, Astropy18},
CIAO \citep{Fruscione06},
XSPEC \citep{Arnaud96}
}



\appendix

\section{Behaviors of parametric temperature profiles}
\label{sec:kT}

\begin{figure}[ht]
 \begin{center}
  \includegraphics[width=8.5cm]{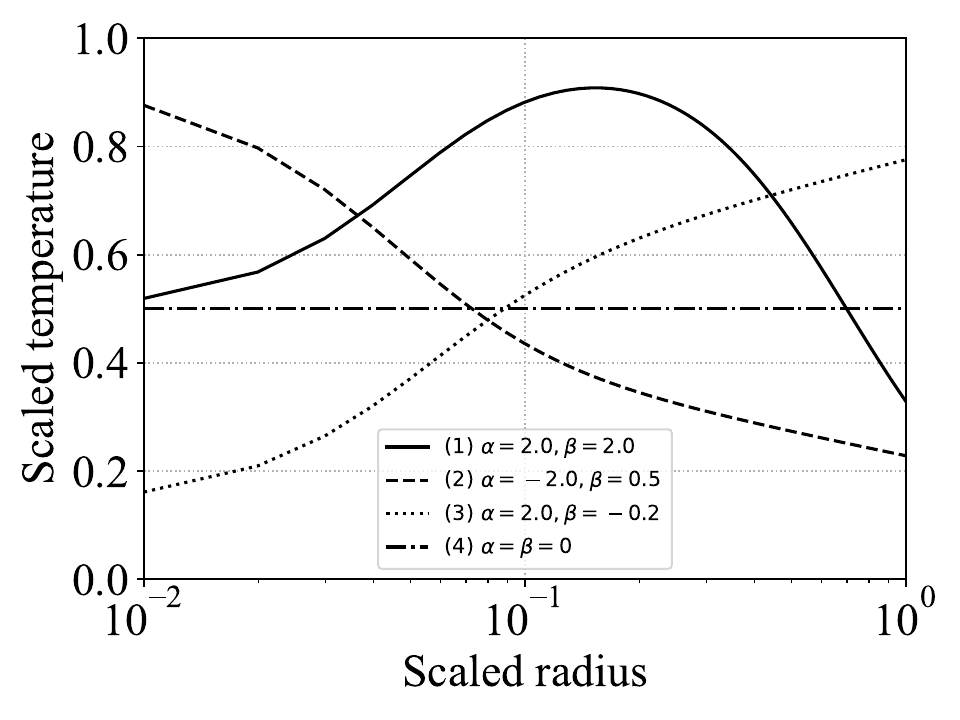}
 \end{center}
\caption{
Scaled temperature profiles provided by Equation~\ref{eq:kT}. Four profiles are shown with the slope parameters of (1) $\alpha = 2.0, \beta = 2.0$, (2) $\alpha = -2.0, \beta = 0.5$, (3) $\alpha = 2.0, \beta = -0.2$, and (4) $\alpha = \beta = 0$ (i.e., isothermal), respectively.
}
\label{fig:kTpro}
\end{figure}

Here, we present an example of the behaviors of the ICM temperature profile used in this paper. Figure~\ref{fig:kTpro} shows four examples of the scaled temperature profiles provided from Equation~\ref{eq:kT}. By adjusting the two slope parameters ($\alpha$ and $\beta$), our model can reproduce specific temperature profiles, such as a profile with a maximum in an inner region and another with a maximum at the endpoint of the profile.  

\section{Radial profiles}
\label{sec:all_radial}

\begin{figure*}[ht]
 \begin{center}
  \includegraphics[width=5cm]{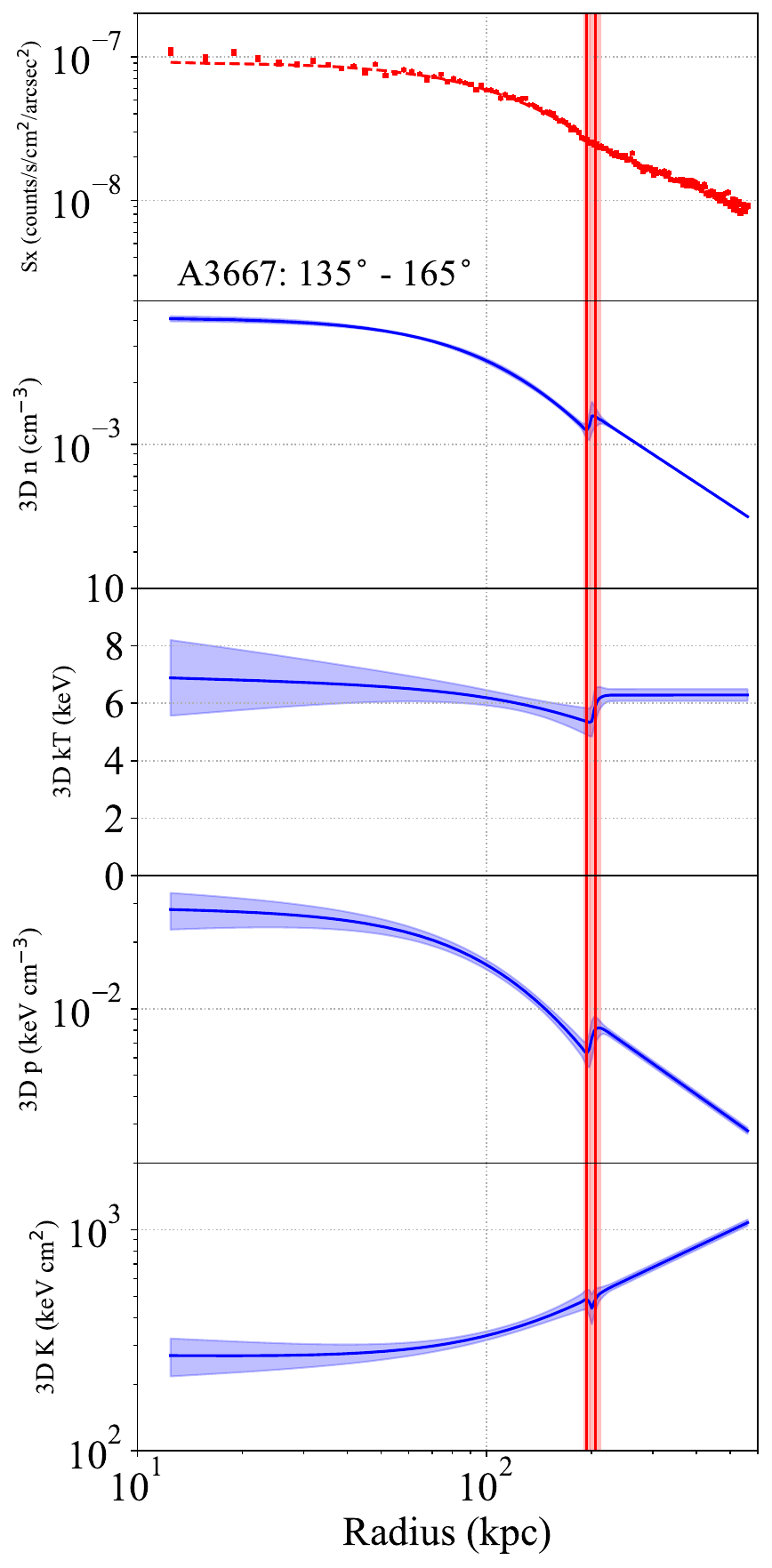}
  \includegraphics[width=5cm]{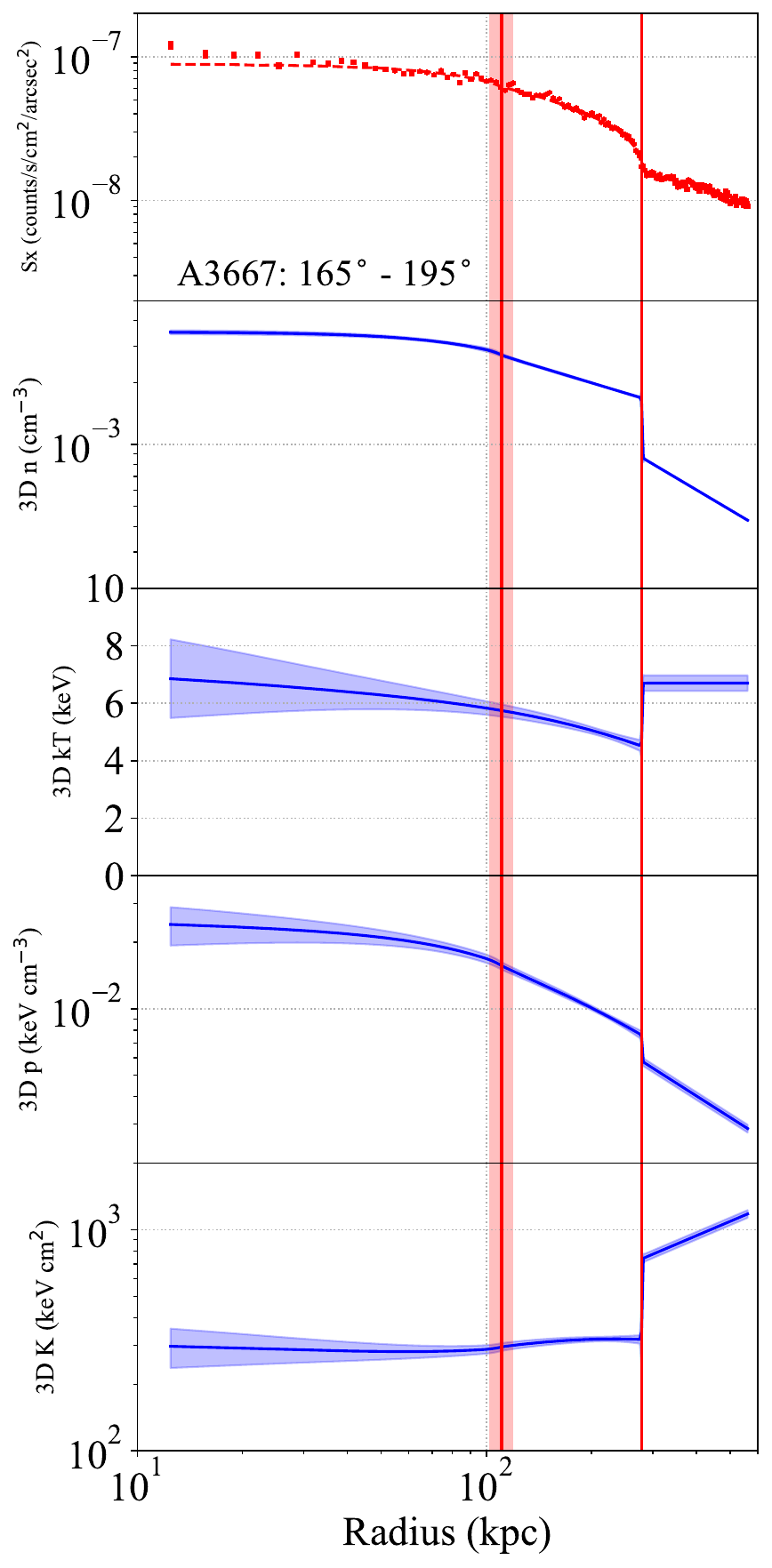}
  \includegraphics[width=5cm]{./A3667_195to225deg_all_profiles.pdf}
  \includegraphics[width=5cm]{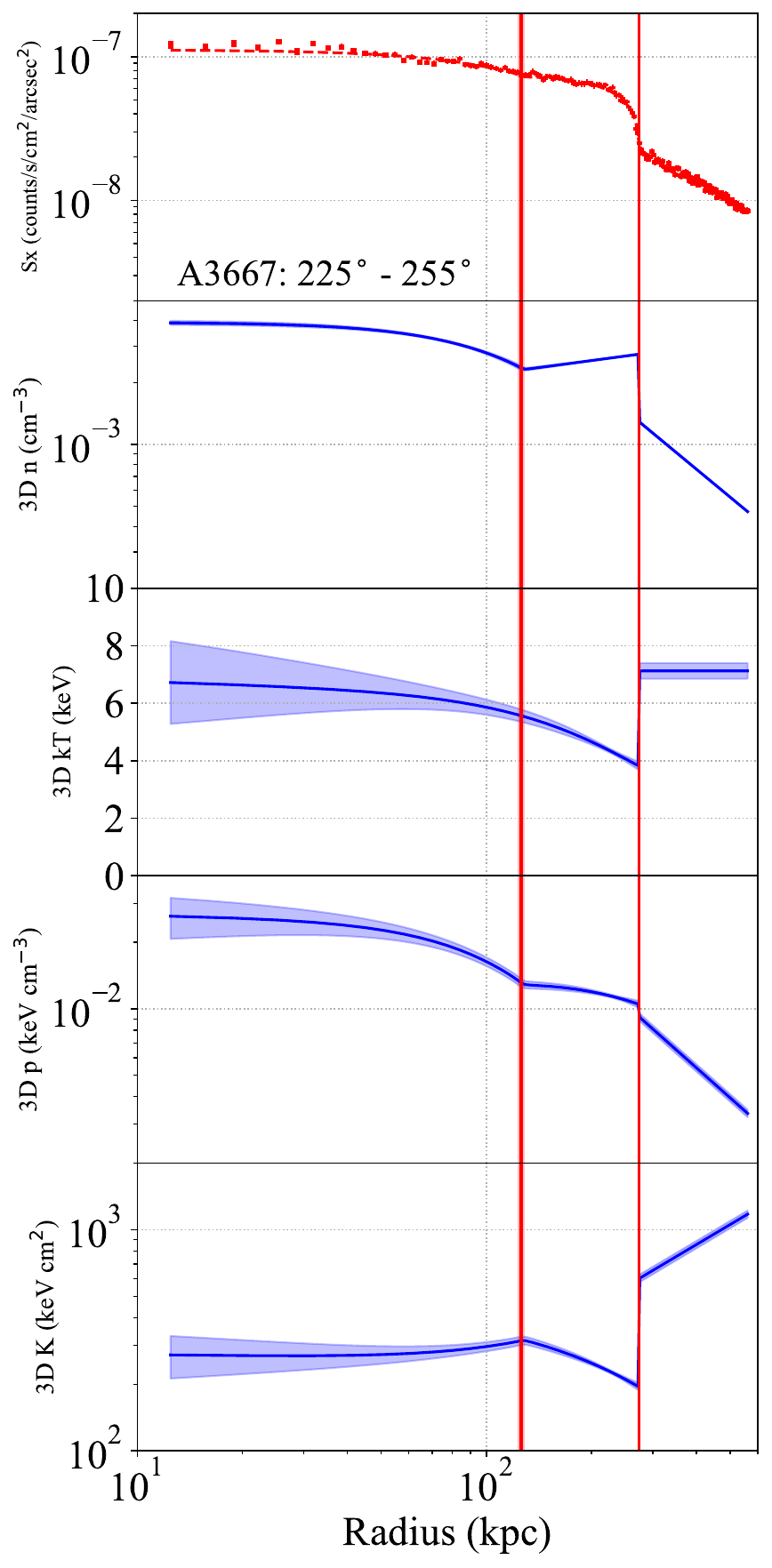}
  \includegraphics[width=5cm]{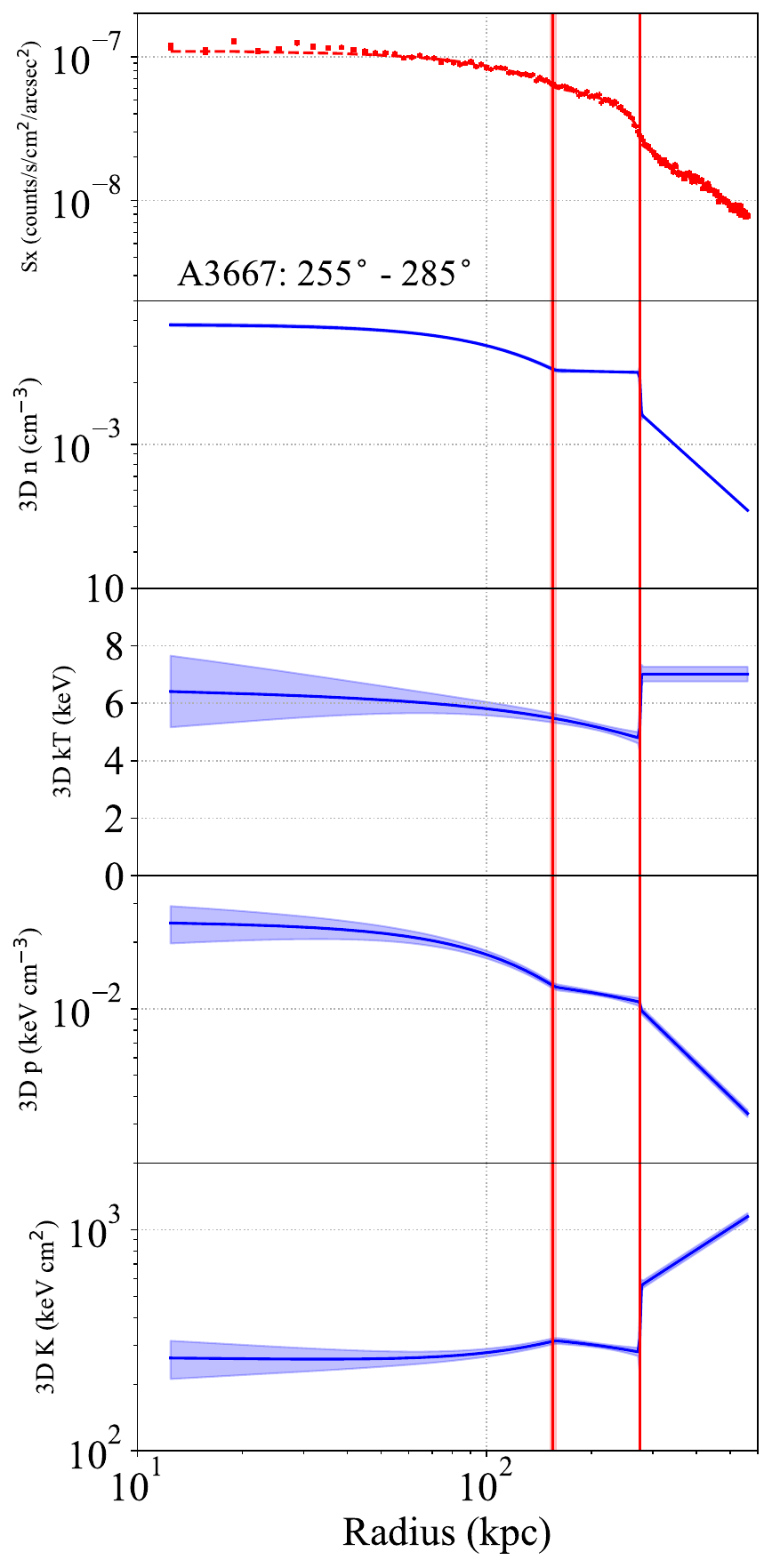}
  \includegraphics[width=5cm]{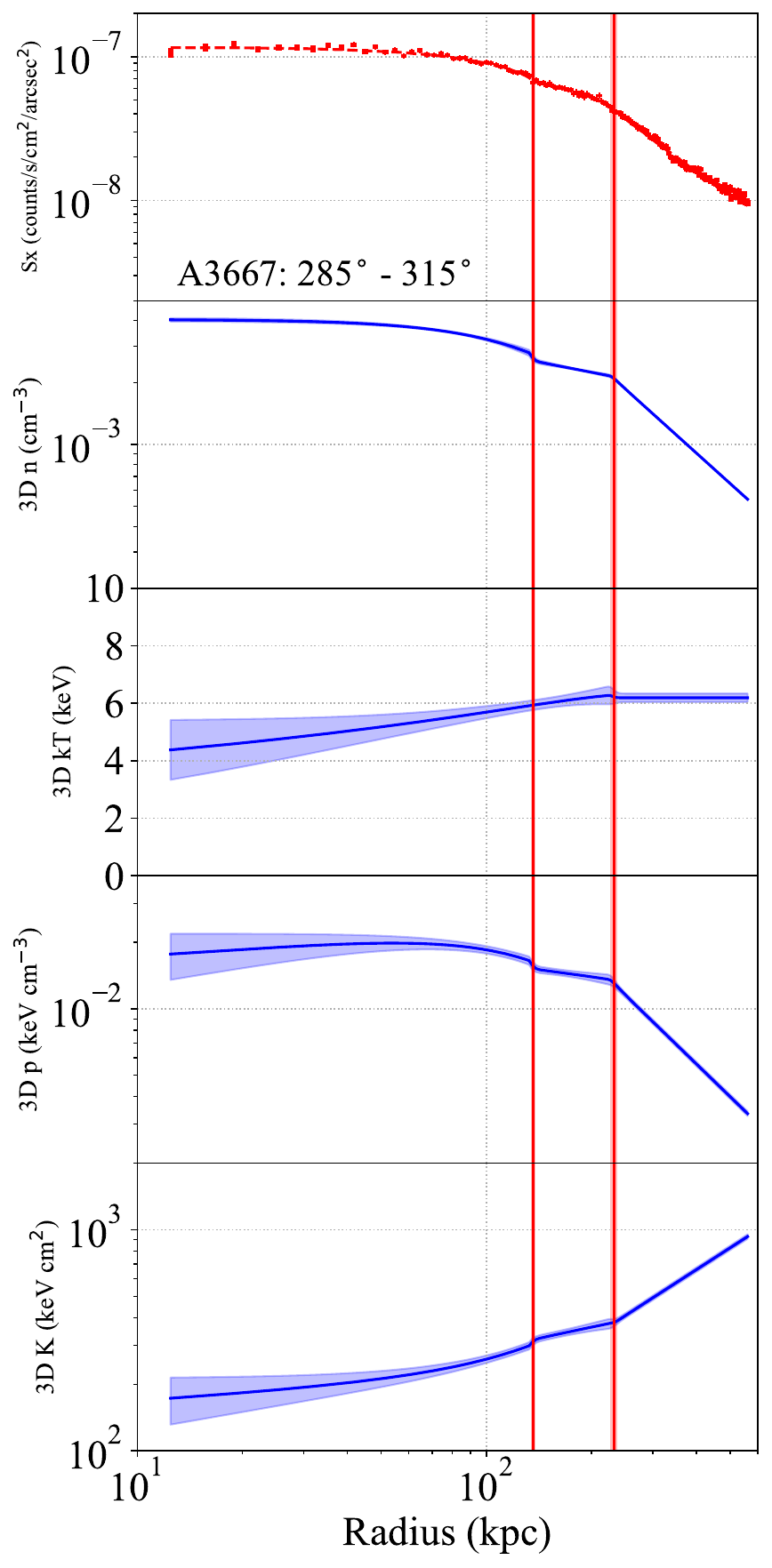}
 \end{center}
\caption{
Same as Figure~\ref{fig:CF_profile}, but for all sectors analyzed in A3667.
(Top left):  the $135^{\circ} - 165^{\circ}$ sector.
(Top middle): the $165^{\circ} - 195^{\circ}$ sector.
(Top right): the $195^{\circ} - 225^{\circ}$ sector.
(Bottom left): the $225^{\circ} - 255^{\circ}$ sector.
(Bottom middle): the $255^{\circ} - 285^{\circ}$ sector.
(Bottom right): the $285^{\circ} - 315^{\circ}$ sector.
The plot for the $195^{\circ} - 225^{\circ}$ sector is the same as the left panel of Figure~\ref{fig:CF_profile}.
}
\label{fig:A3667_Appendix}
\end{figure*}

\begin{figure*}[ht]
 \begin{center}
  \includegraphics[width=5cm]{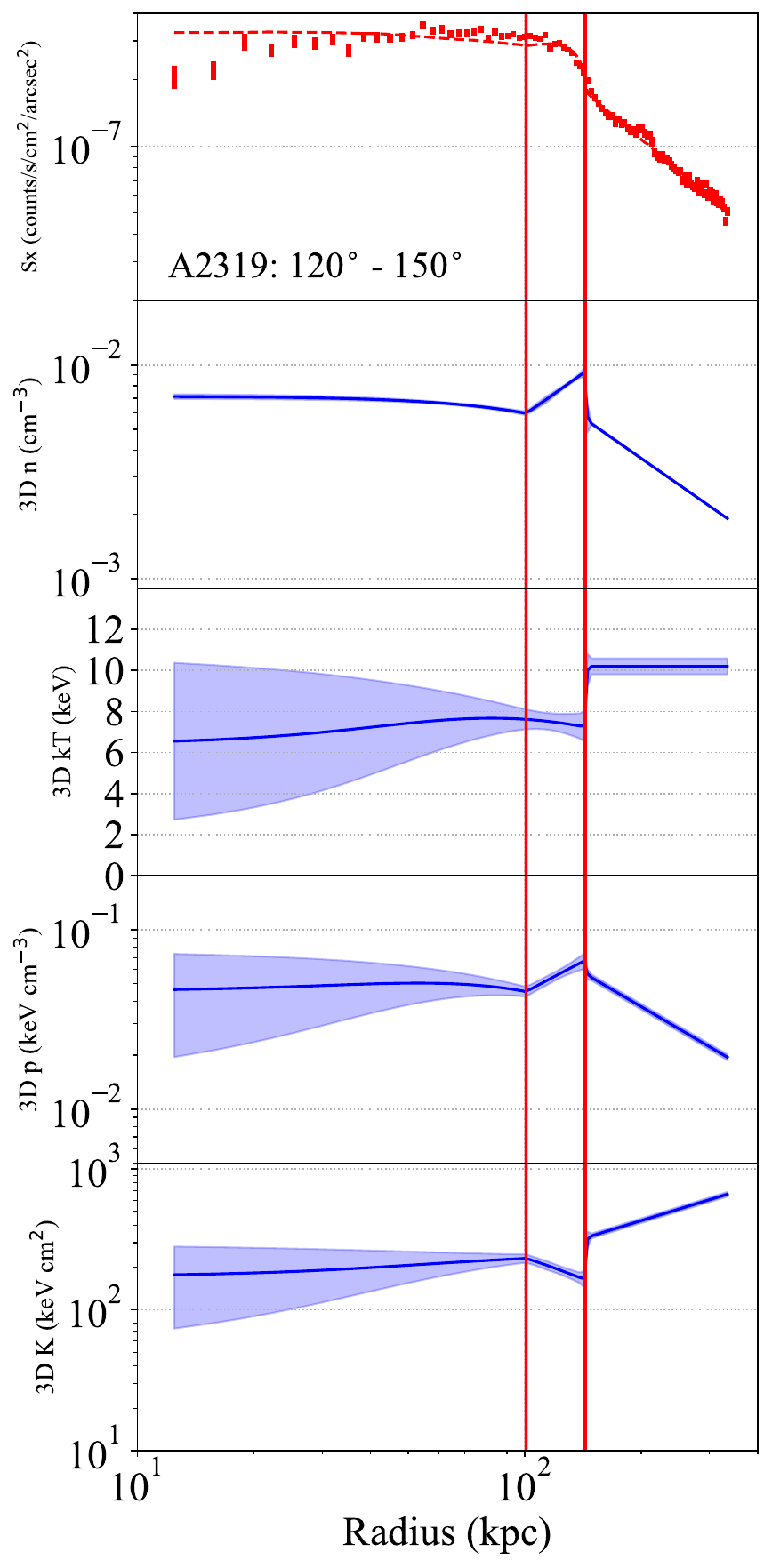}
  \includegraphics[width=5cm]{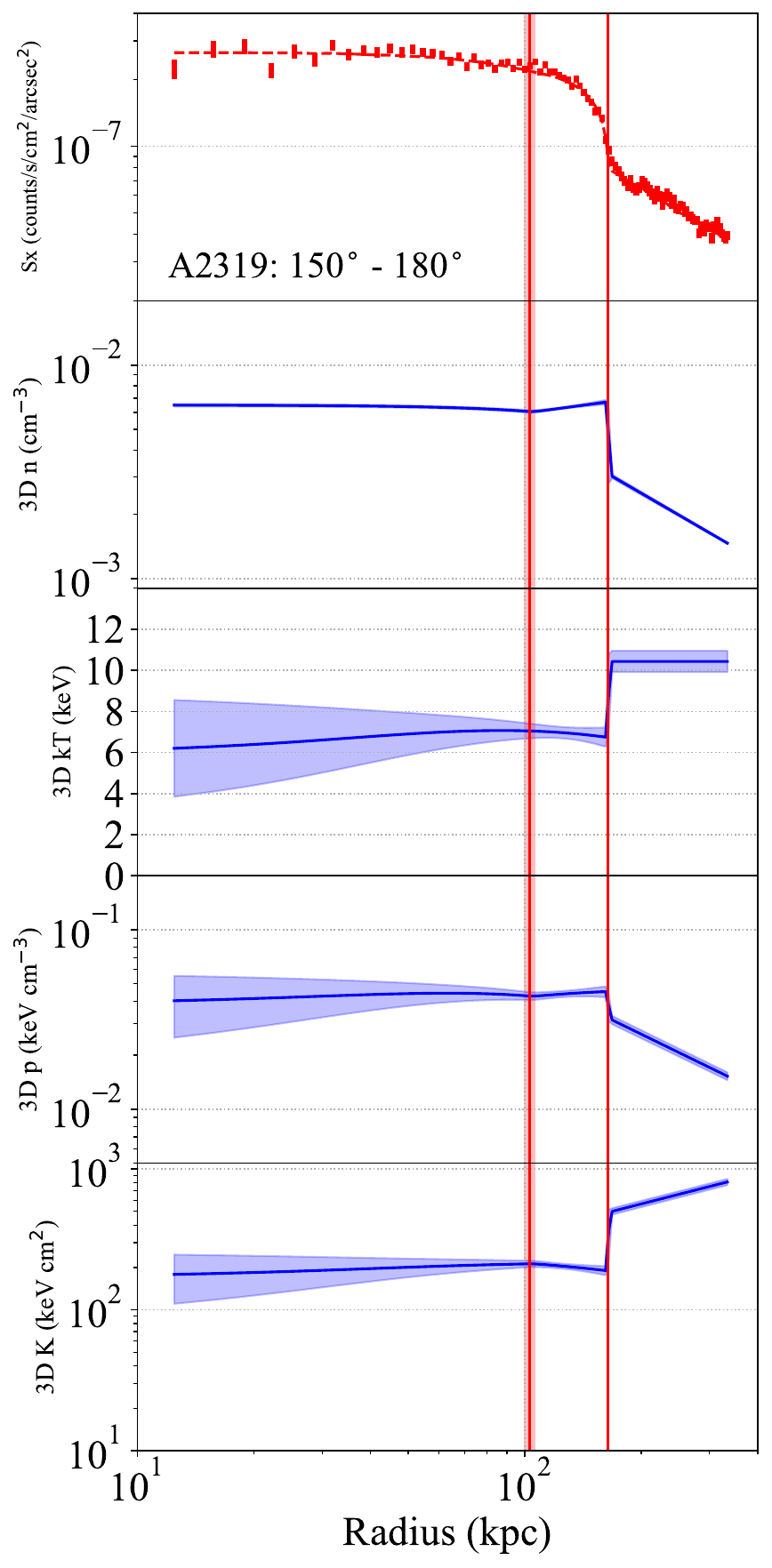}
  \includegraphics[width=5cm]{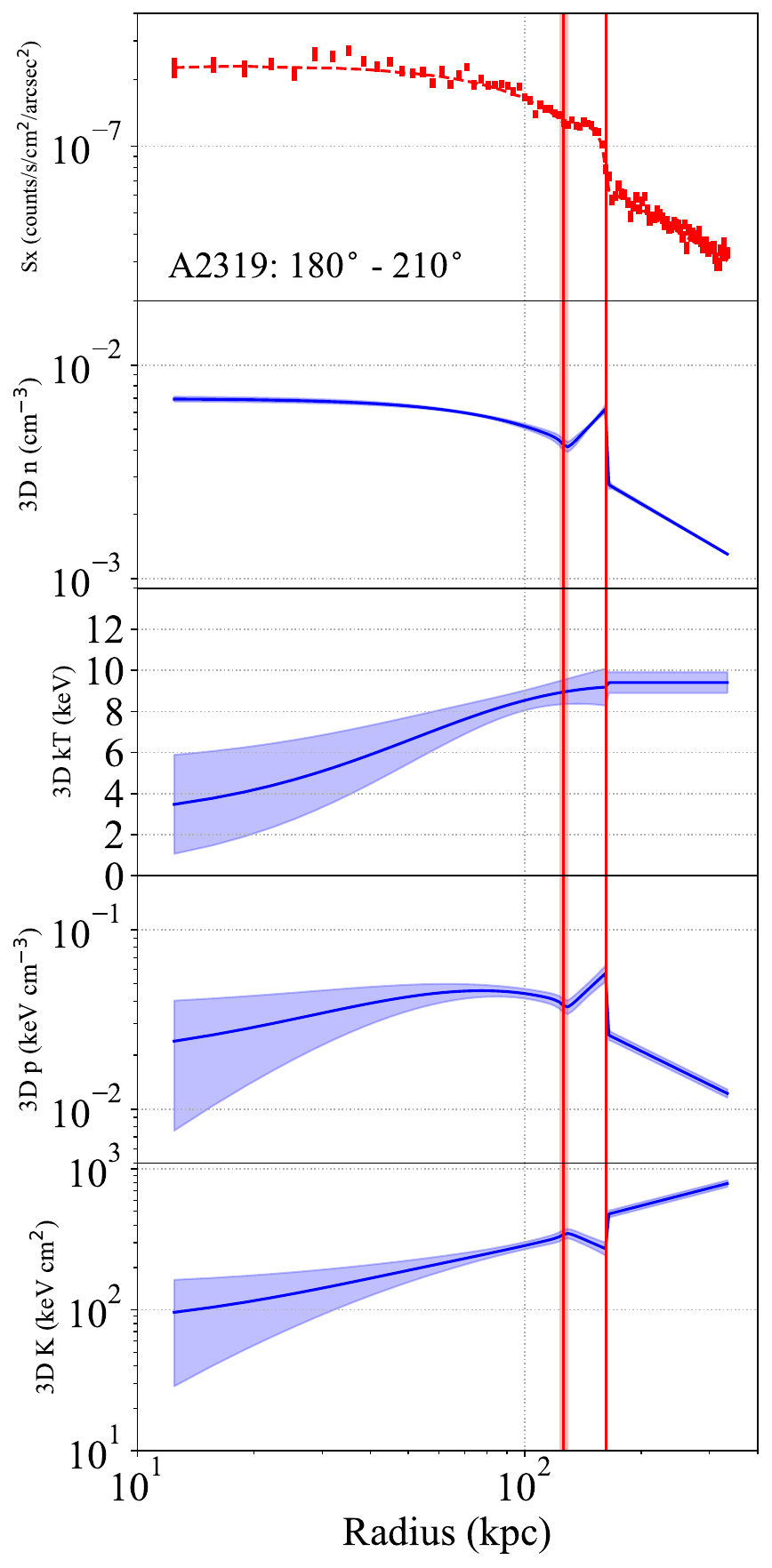}
  \includegraphics[width=5cm]{./A2319_210to240deg_all_profiles.pdf}
  \includegraphics[width=5cm]{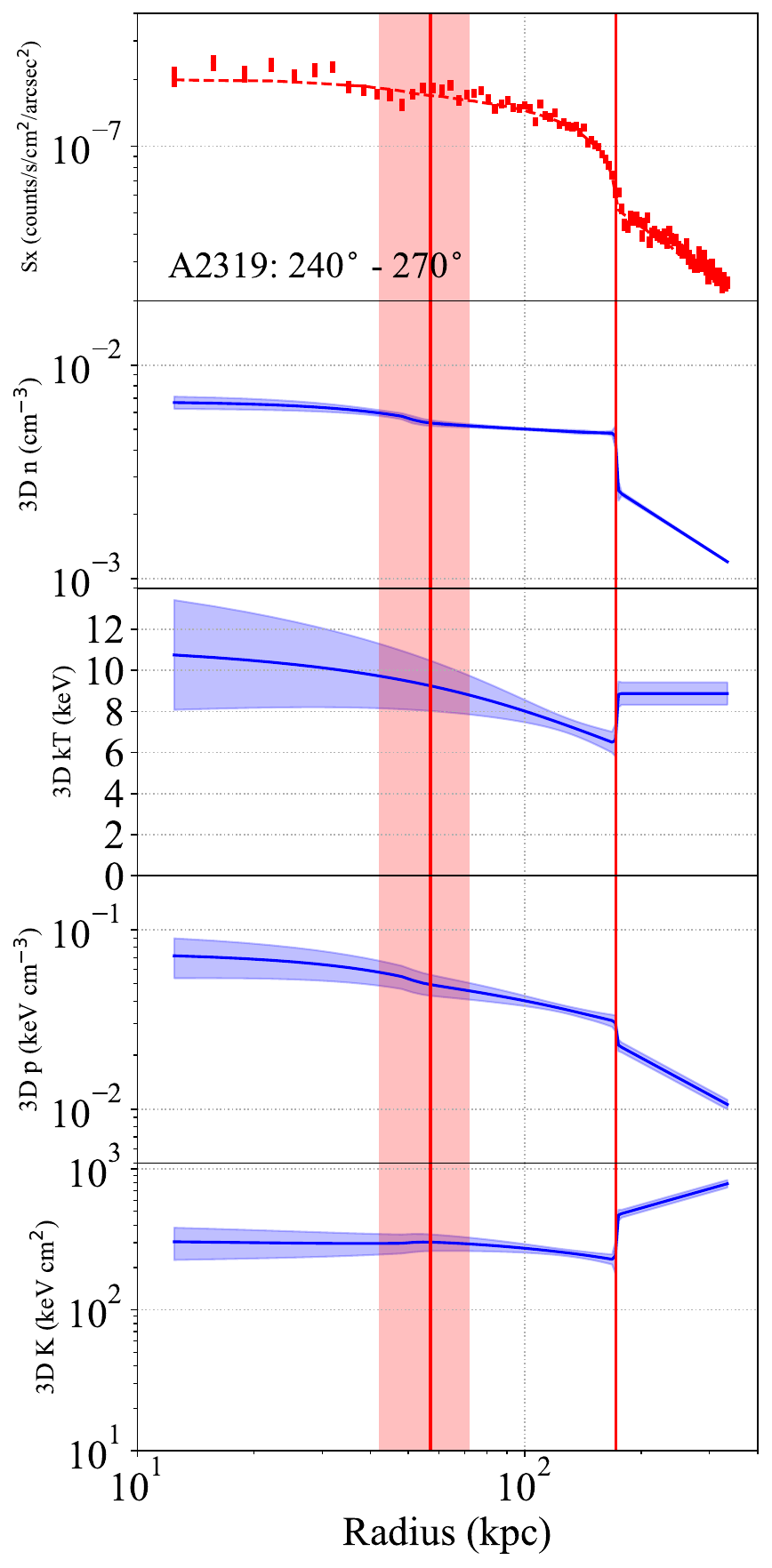}
  \includegraphics[width=5cm]{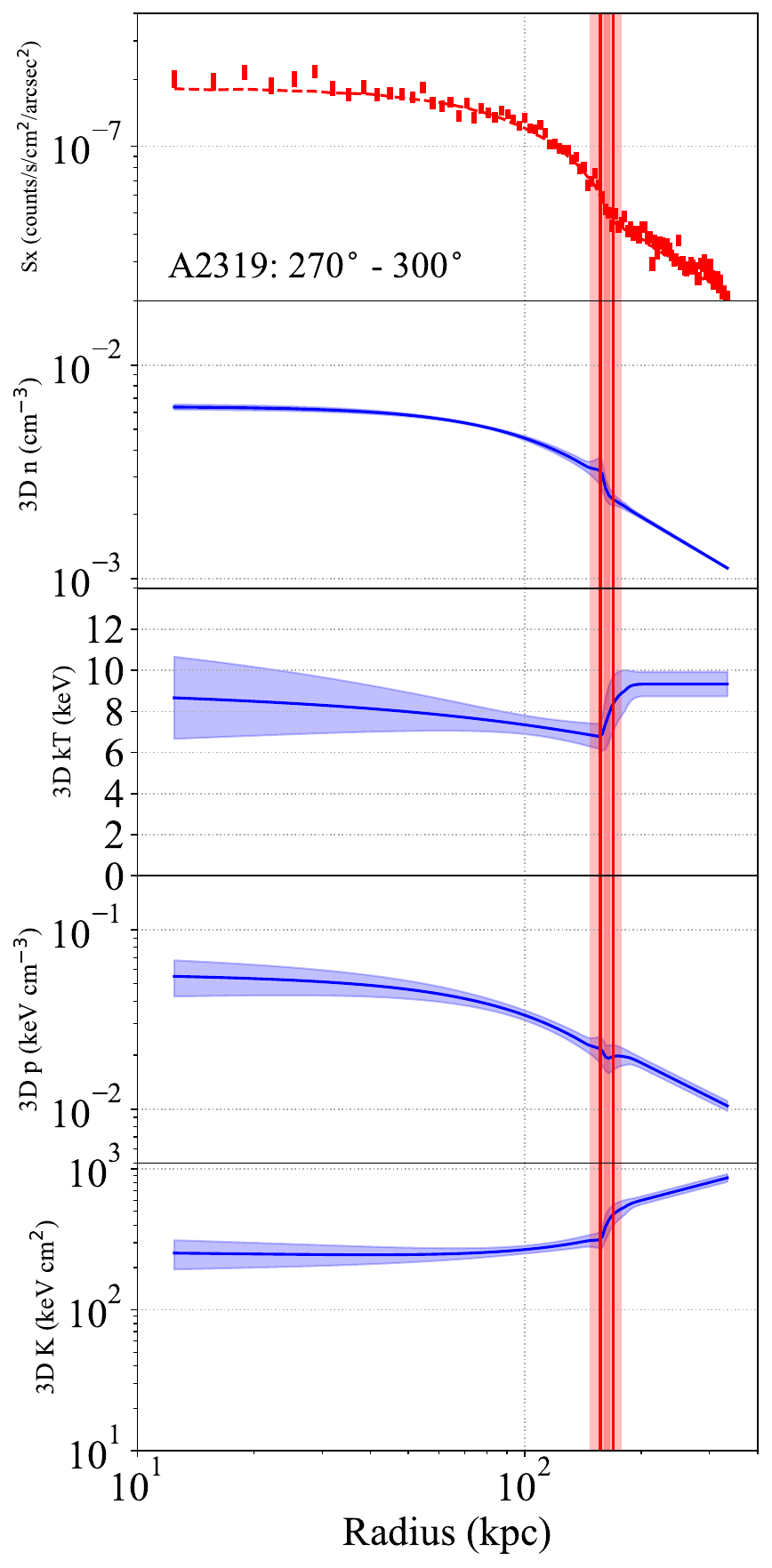}
 \end{center}
\caption{
Same as Figure~\ref{fig:A3667_Appendix}, but for A2319. 
(Top left):  $120^{\circ} - 150^{\circ}$ sector.
(Top middle): the $150^{\circ} - 180^{\circ}$ sector.
(Top right): the $180^{\circ} - 210^{\circ}$ sector.
(Bottom left): the $210^{\circ} - 240^{\circ}$ sector.
(Bottom middle): the $240^{\circ} - 270^{\circ}$ sector.
(Bottom right): the $270^{\circ} - 300^{\circ}$ sector.
The plots for the $180^{\circ} - 210^{\circ}$ sector is the same as the right panel of Figure~\ref{fig:CF_profile}.
}
\label{fig:A2319_Appendix}
\end{figure*}

\begin{figure*}[ht]
 \begin{center}
  \includegraphics[width=5cm]{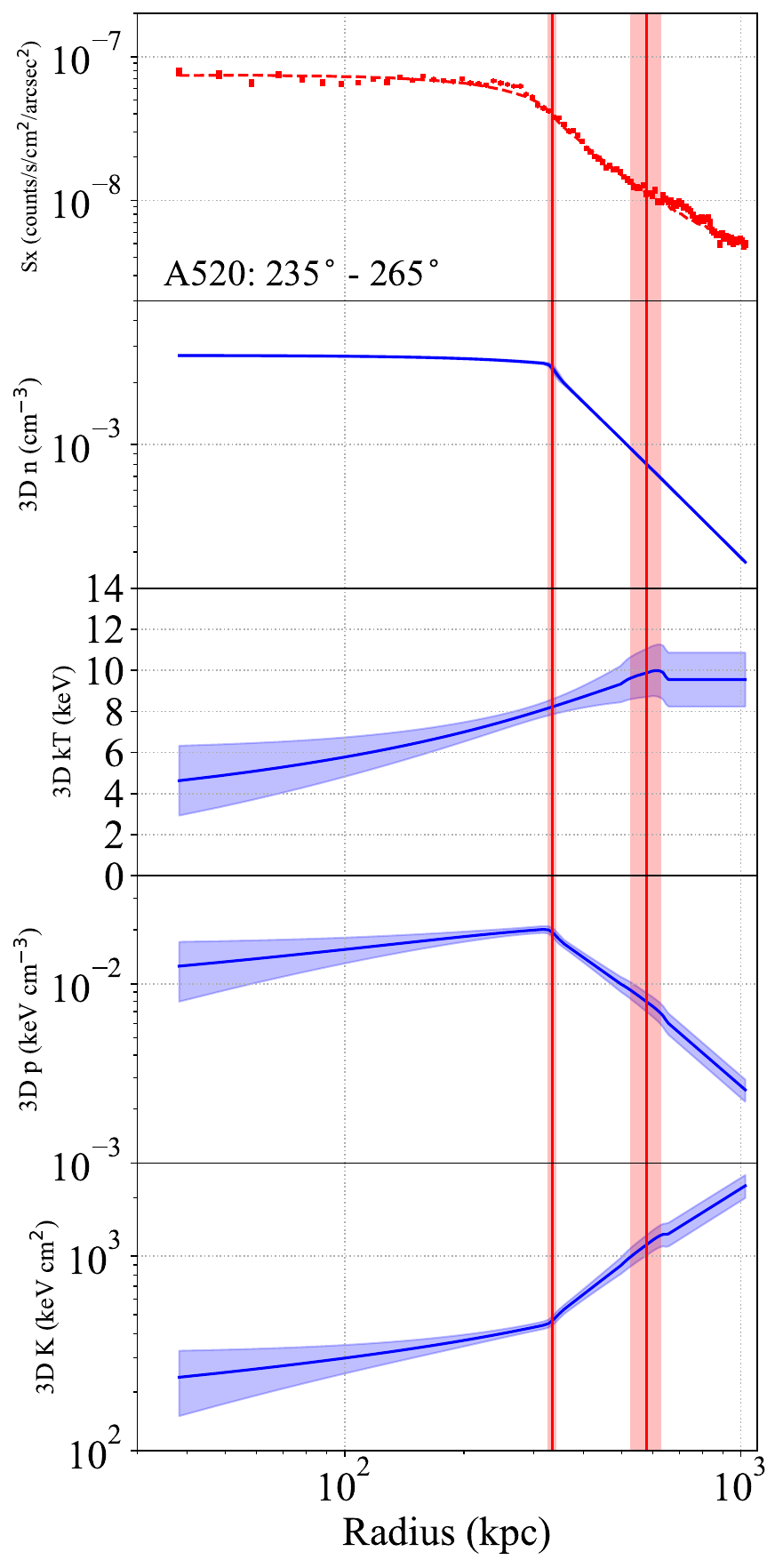}
  \includegraphics[width=5cm]{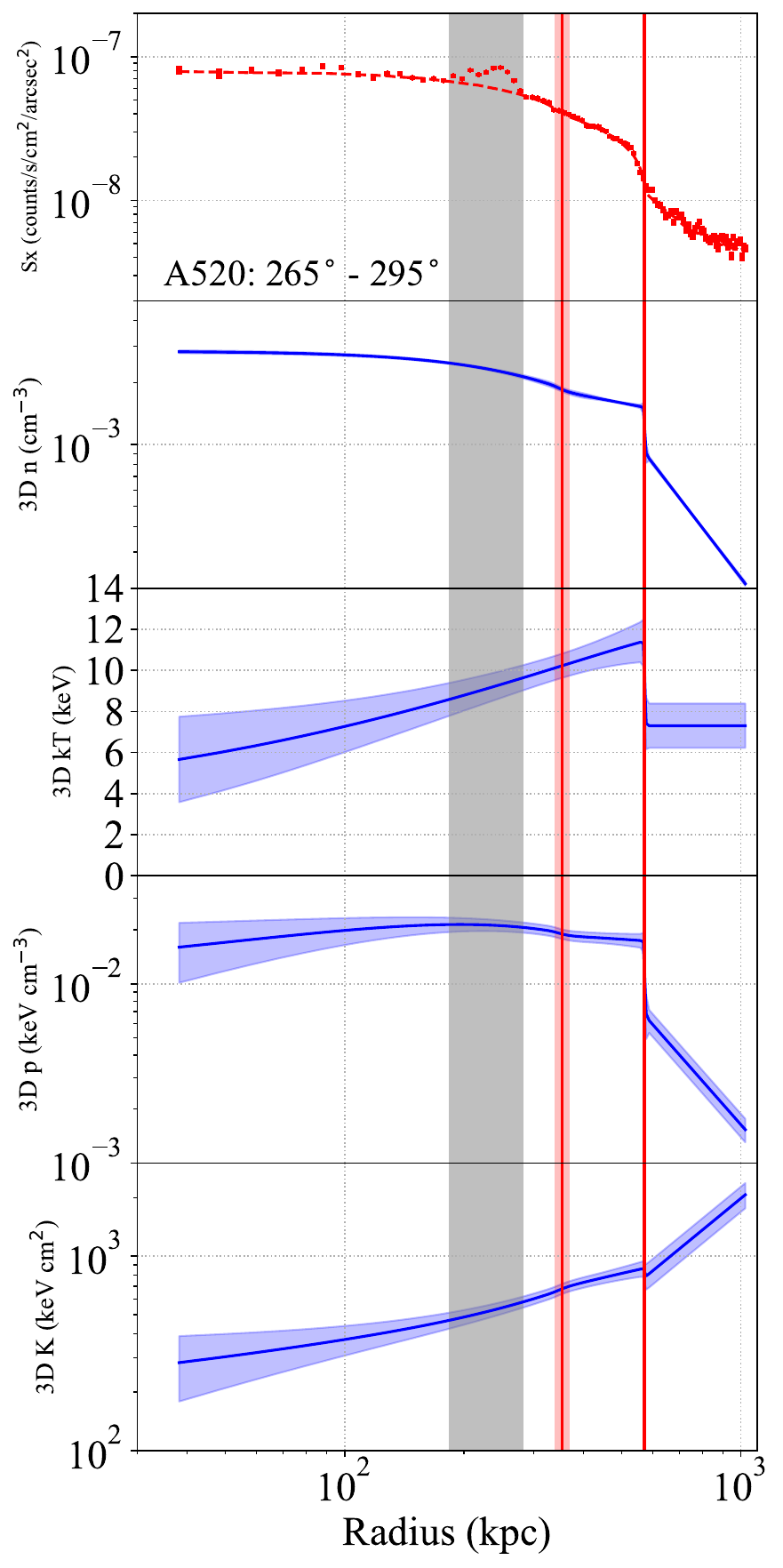}
  \includegraphics[width=5cm]{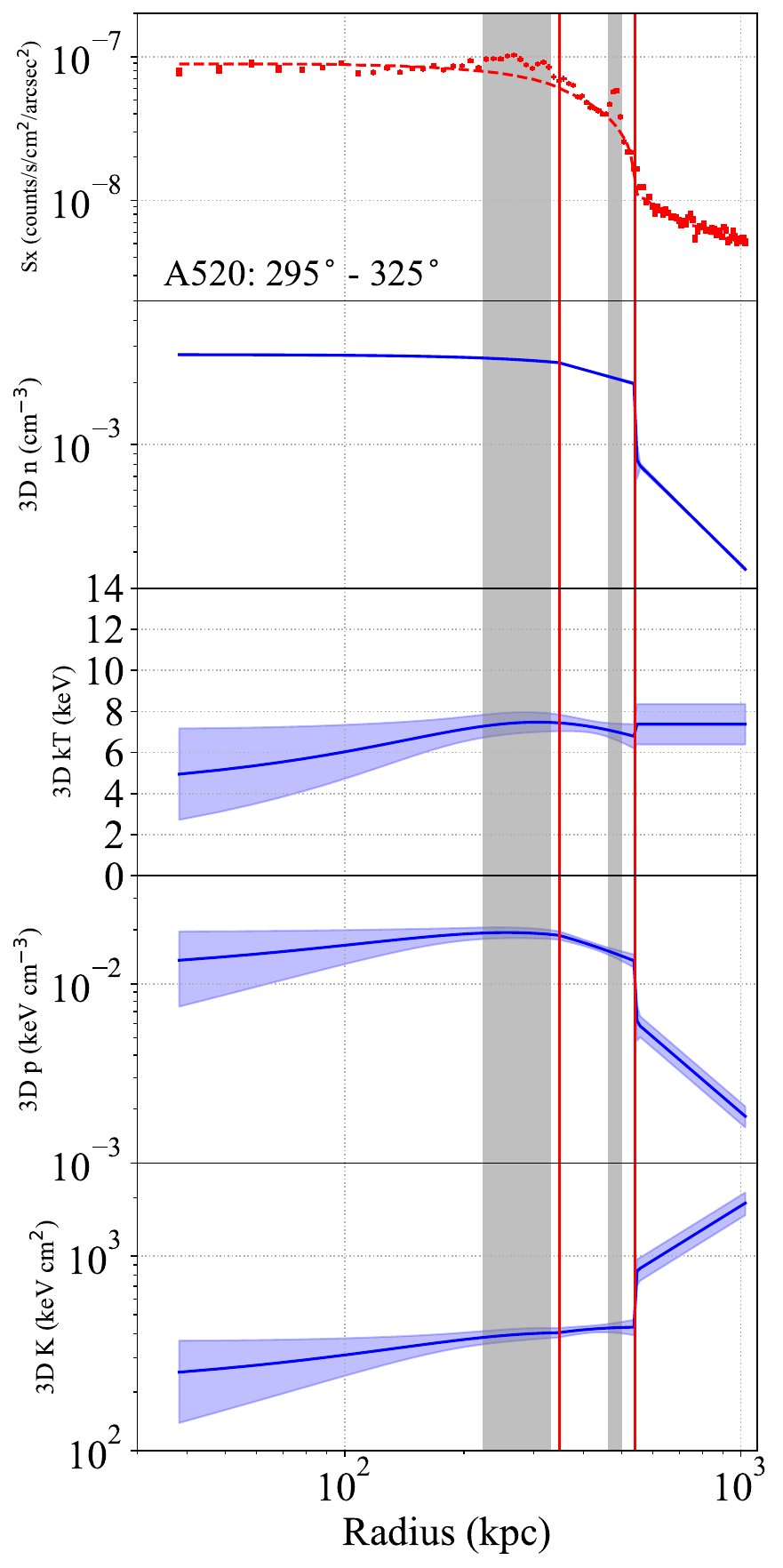}
  \includegraphics[width=5cm]{./A520_325to355deg_all_profiles.pdf}
  \includegraphics[width=5cm]{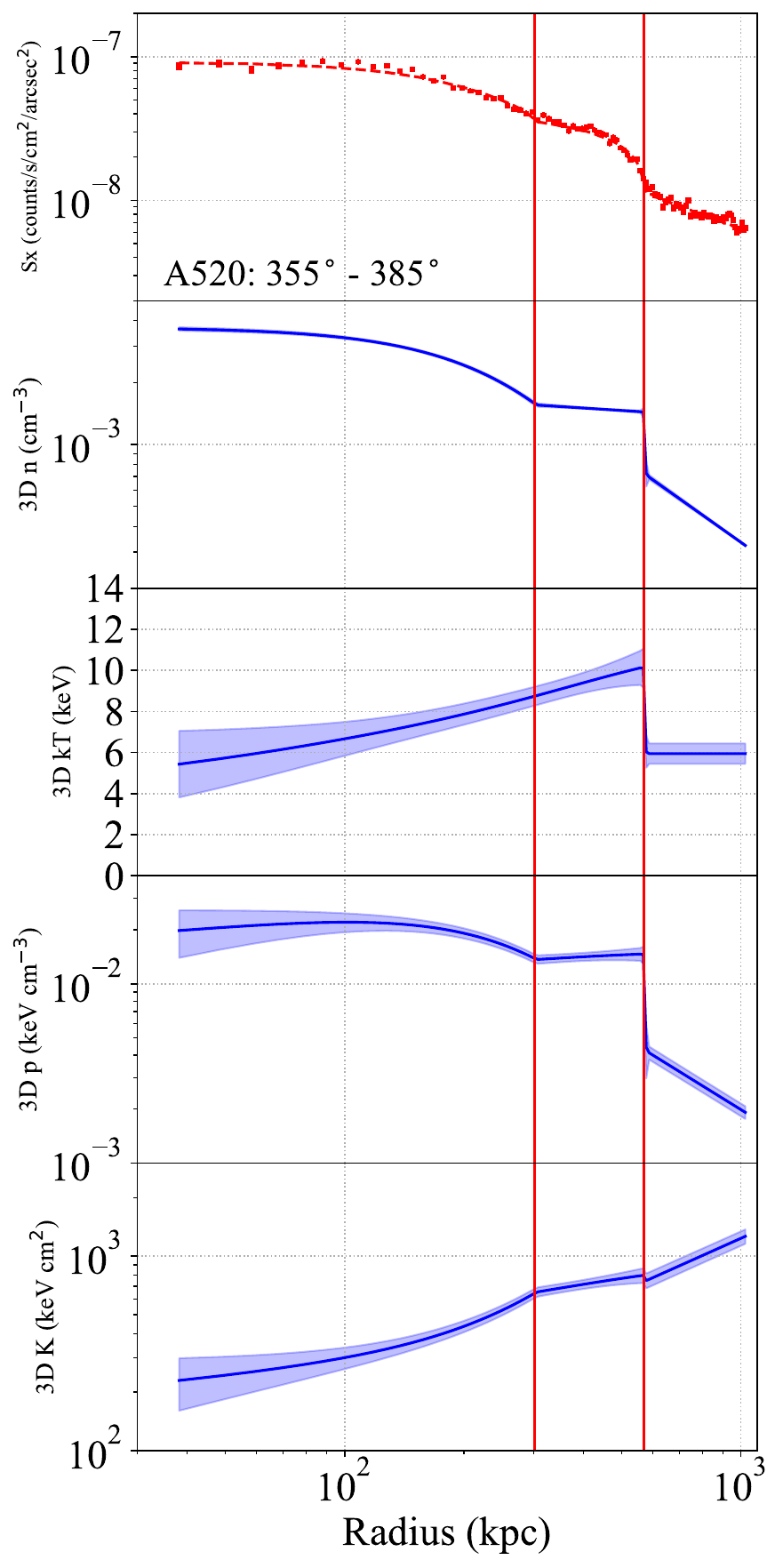}
  \includegraphics[width=5cm]{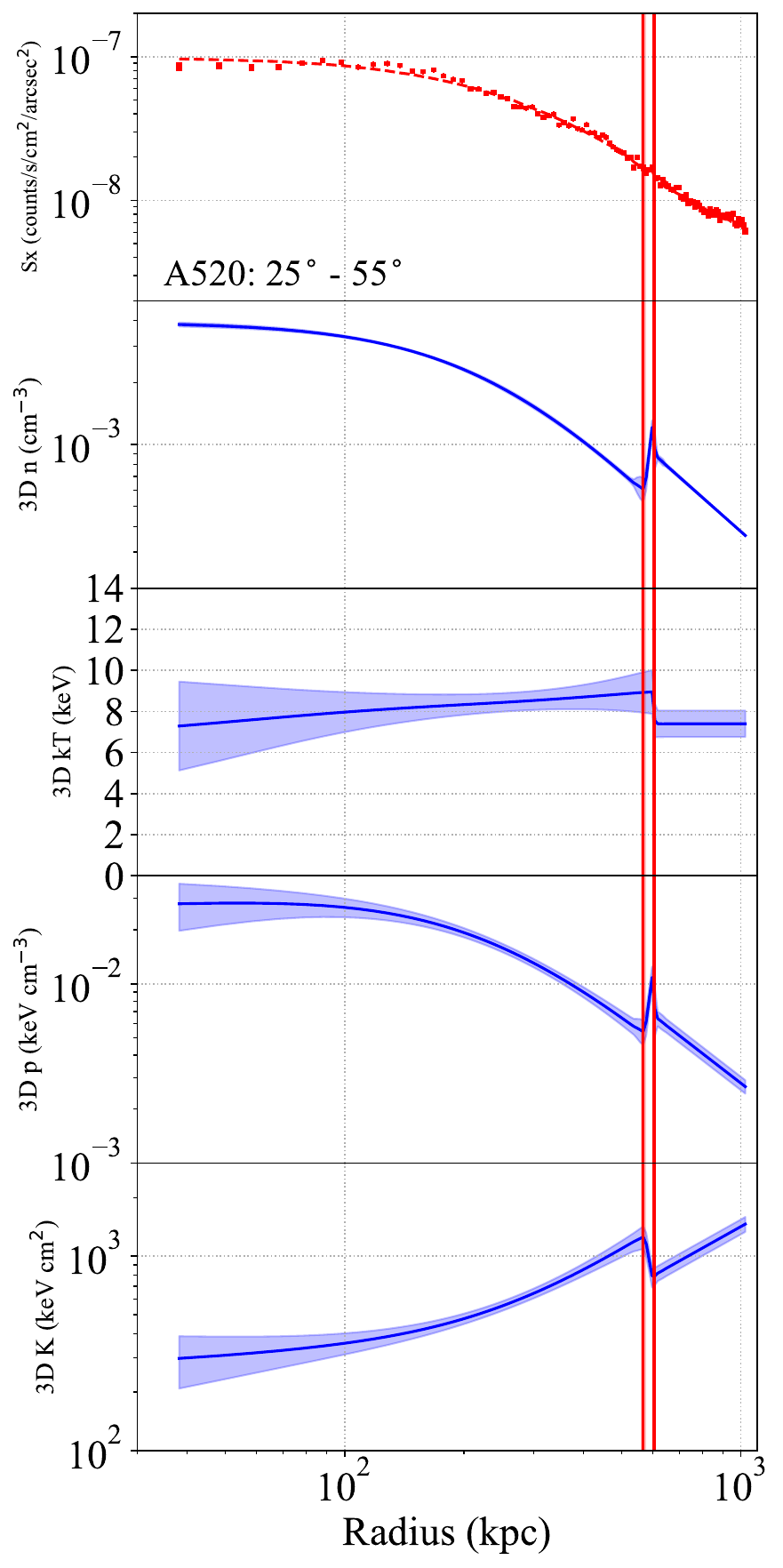}
 \end{center}
\caption{
Same as Figure~\ref{fig:SF_profile}, but for all sectors analyzed in A520.
(Top left):  the $235^{\circ} - 265^{\circ}$ sector.
(Top middle): the $265^{\circ} - 295^{\circ}$ sector.
(Top right): the $295^{\circ} - 325^{\circ}$ sector.
(Bottom left): the $325^{\circ} - 355^{\circ}$ sector.
(Bottom middle): the $25^{\circ} - 55^{\circ}$ sector.
(Bottom right): the $285^{\circ} - 315^{\circ}$ sector.
The plots for the $325^{\circ} - 355^{\circ}$ sector is the same as the left panel of Figure~\ref{fig:SF_profile}.
}
\label{fig:A520_Appendix}
\end{figure*}

\begin{figure*}[ht]
 \begin{center}
  \includegraphics[width=5cm]{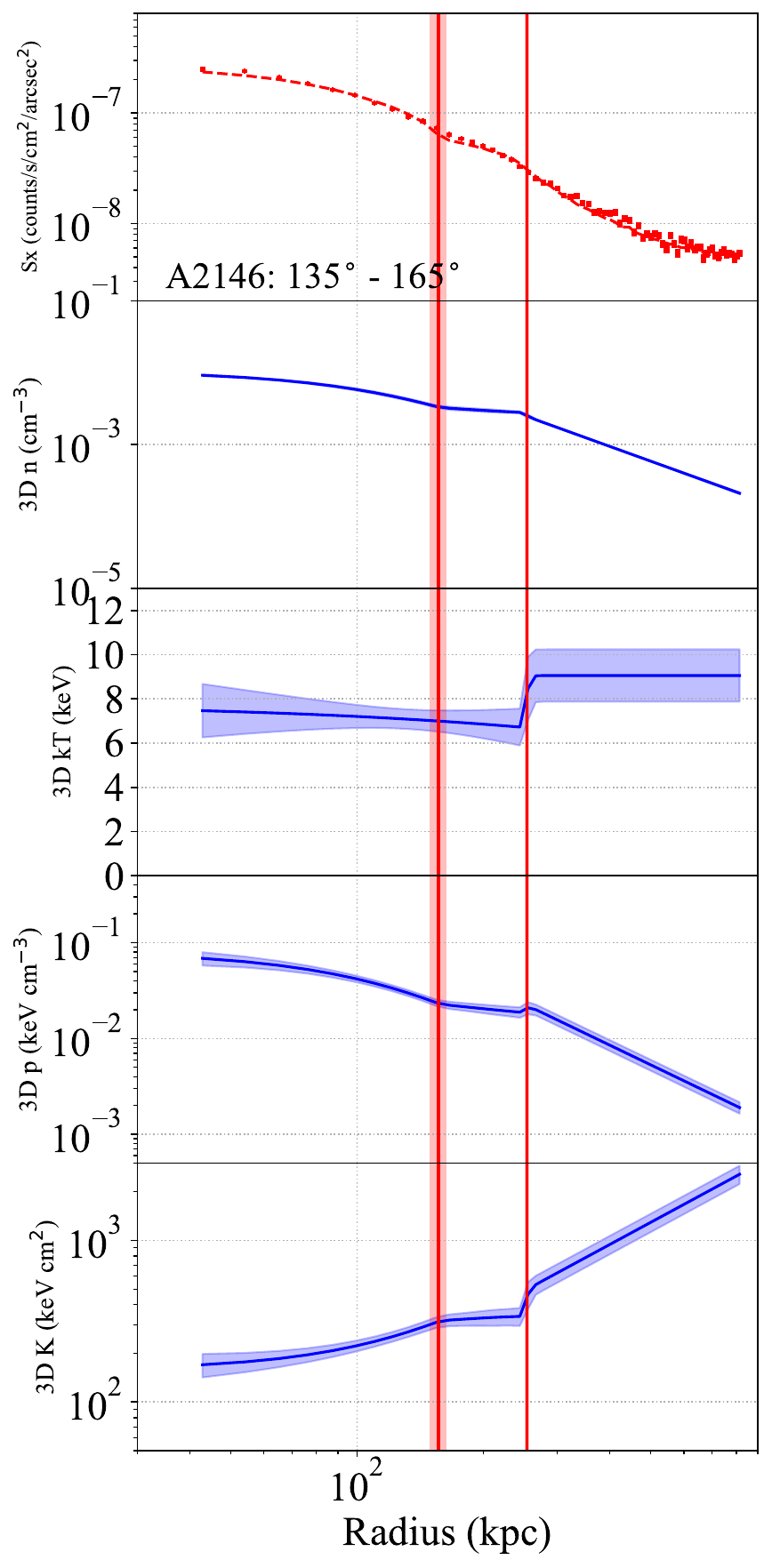}
  \includegraphics[width=5cm]{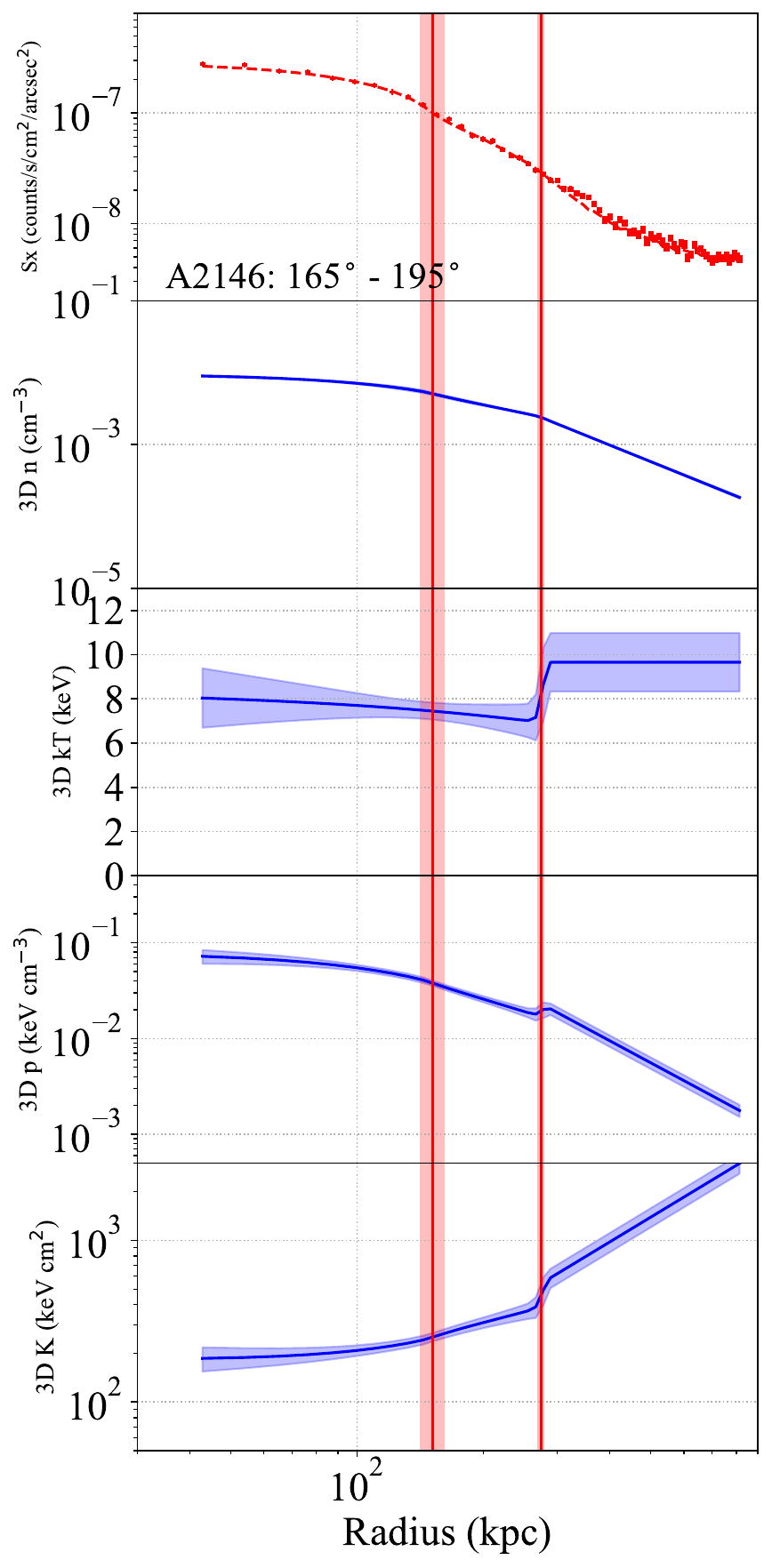}
  \includegraphics[width=5cm]{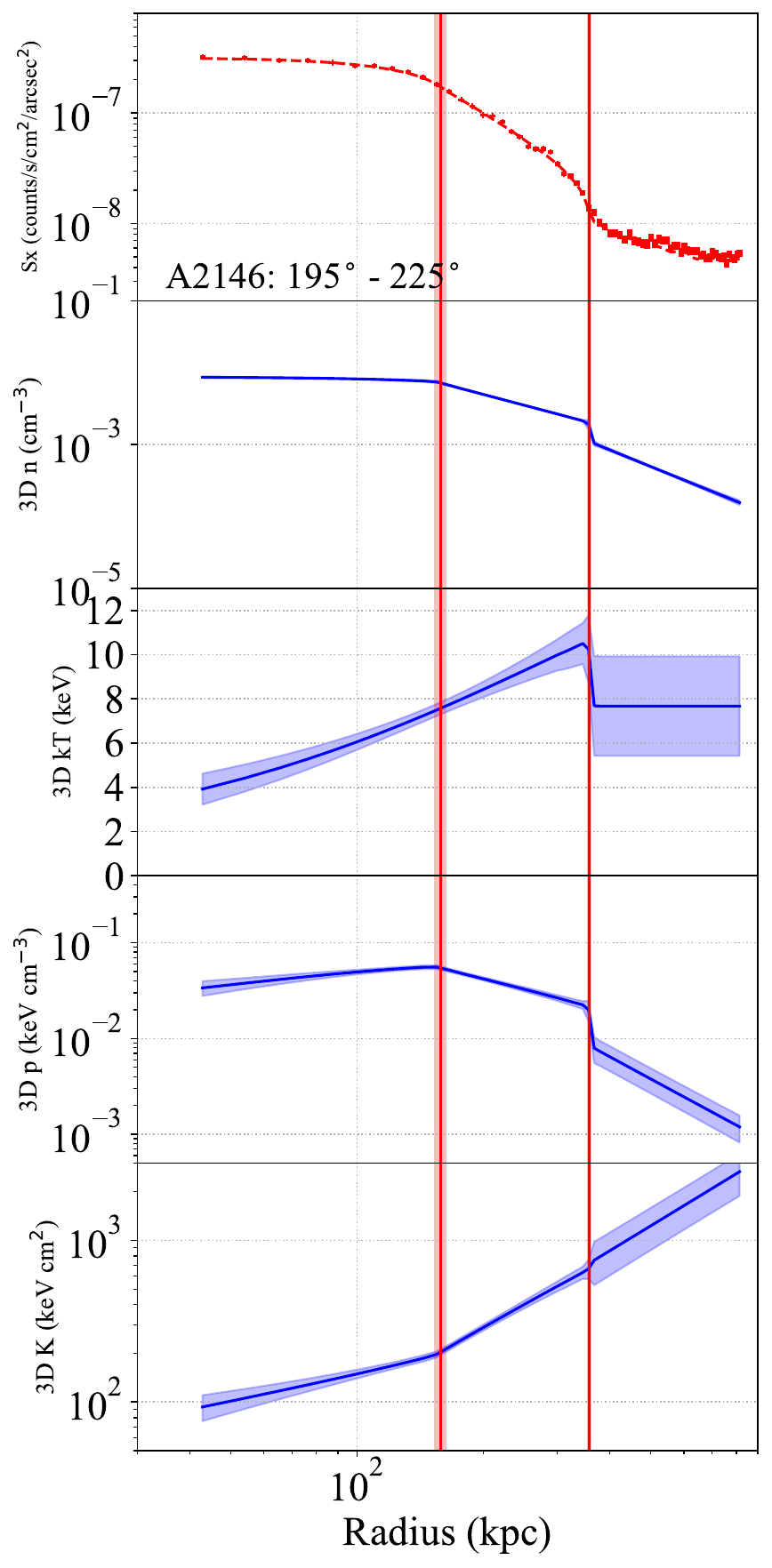}
  \includegraphics[width=5cm]{./A2146_225to255deg_all_profiles.pdf}
  \includegraphics[width=5cm]{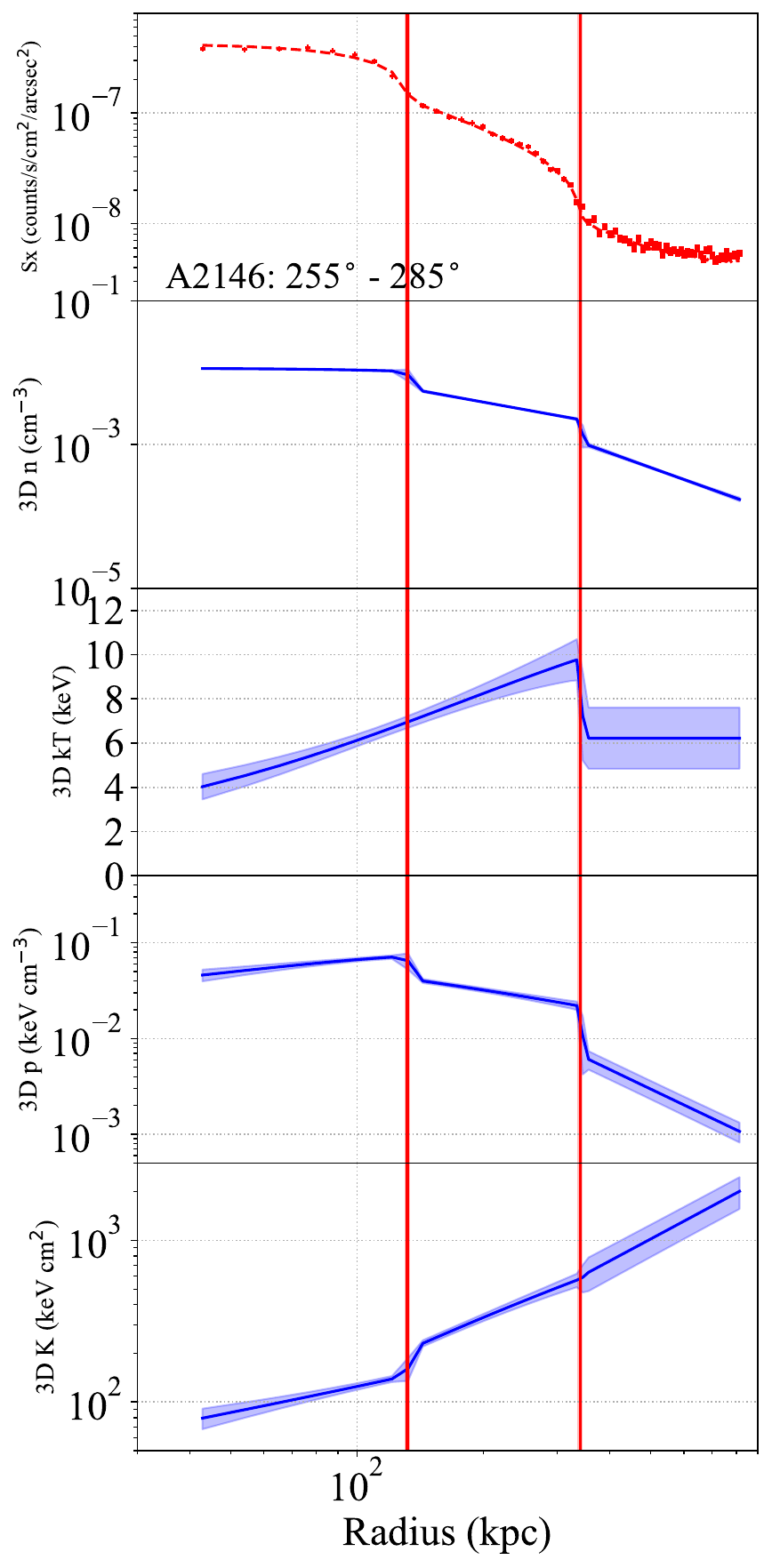}
  \includegraphics[width=5cm]{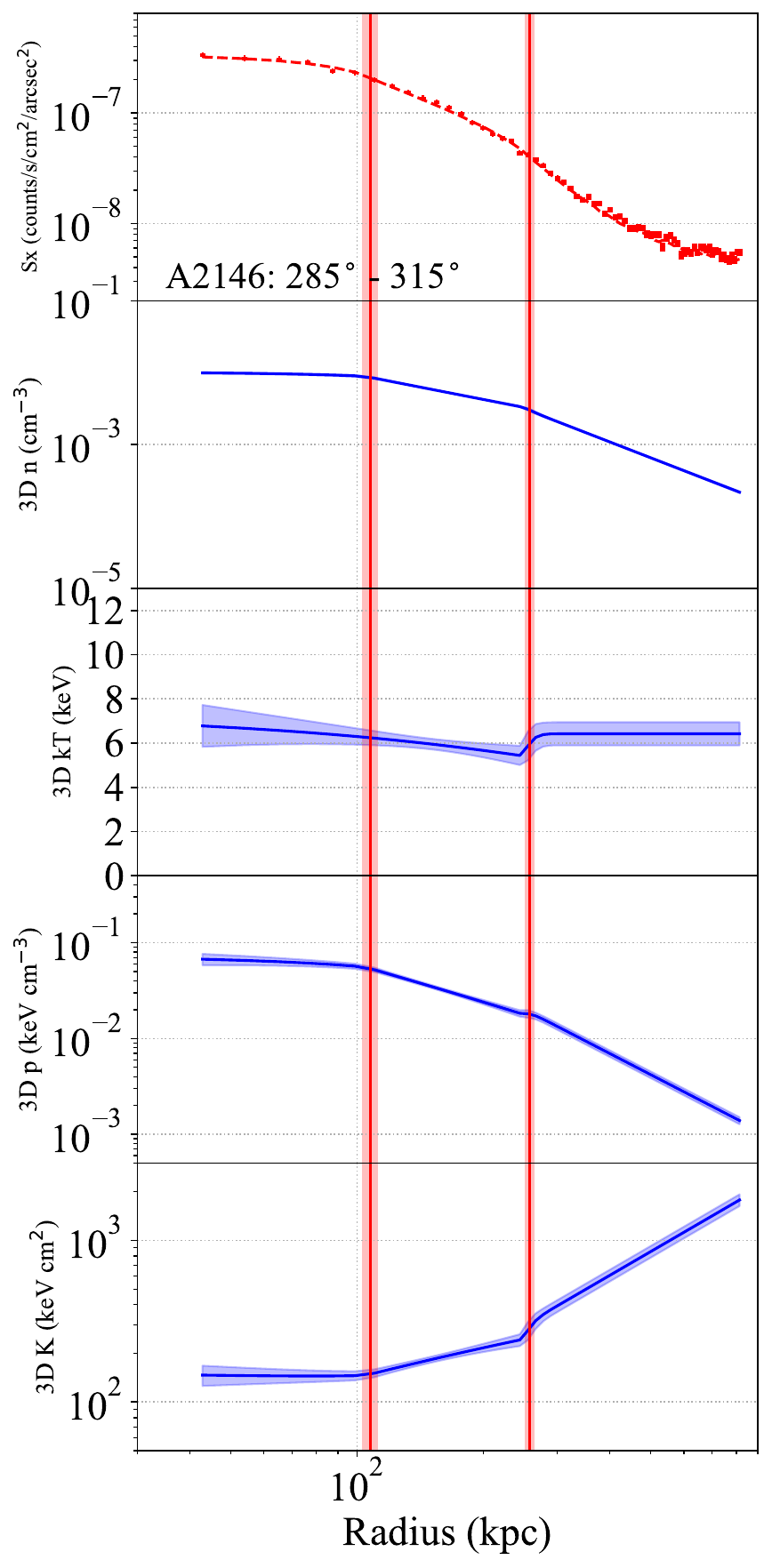}
 \end{center}
\caption{
Same as Figure~\ref{fig:A520_Appendix}, but for A2146. 
(Top left):  the $135^{\circ} - 165^{\circ}$ sector.
(Top middle): the $165^{\circ} - 195^{\circ}$ sector.
(Top right): the $195^{\circ} - 225^{\circ}$ sector.
(Bottom left): the $225^{\circ} - 255^{\circ}$ sector.
(Bottom middle): the $255^{\circ} - 285^{\circ}$ sector.
(Bottom right): the $285^{\circ} - 315^{\circ}$ sector.
The plots for the $225^{\circ} - 255^{\circ}$ sector is the same as the left panel of Figure~\ref{fig:SF_profile}.
}
\label{fig:A2146_Appendix}
\end{figure*}

Here, we show all radial profiles of each sector in our sample obtained from the forward modeling analysis. Figures~\ref{fig:A3667_Appendix}, \ref{fig:A2319_Appendix}, \ref{fig:A520_Appendix}, and \ref{fig:A2146_Appendix} show the observed and best-fit X-ray surface brightness profiles as well as the obtained 3D thermodynamic profiles of the ICM for A3667, A2319, A520, and A2146, respectively.

For A520, no significant temperature jump at the edge position of the shock front in the $295^{\circ} - 325^{\circ}$ sector is found, while this sector corresponds to the front of the shock front. As shown in the X-ray surface brightness image of A520 (bottom left panel of Figure~\ref{fig:img}) and the radial profiles of the $295^{\circ} - 325^{\circ}$ sector (top left panel of Figure~\ref{fig:Sx_SF}), this sector hosts the apparent substructures. \cite{Wang16b} measured the ICM temperature map in A520, and found that a lower-temperature ICM is co-spatial with the regions of the apparent substructures. In fact, \cite{Wang16b} measured the ICM temperature in the dense, compact substructure located at just below the shock front as $\sim 3$\,keV. We confirm the presence of such low-temperature gas by the spectral analysis (see the top right panel of Figure~\ref{fig:vs_A520}). Therefore, contamination from such low-temperature gas significantly impacts the forward modeling analysis of this sector, leading to the absence of an apparent temperature jump at the interface.


\section{Comparisons of the projected temperature profile between the forward modeling analysis and the X-ray spectral analysis}
\label{sec:comp_kT_p}

\begin{figure*}[ht]
 \begin{center}
  \includegraphics[width=5.6cm]{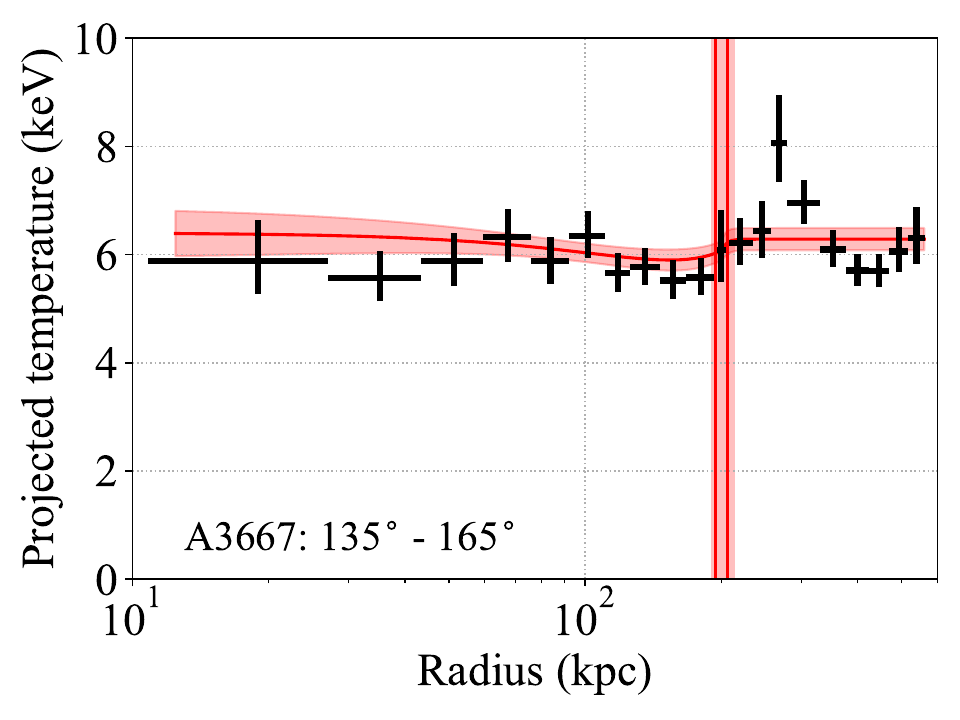}
  \includegraphics[width=5.6cm]{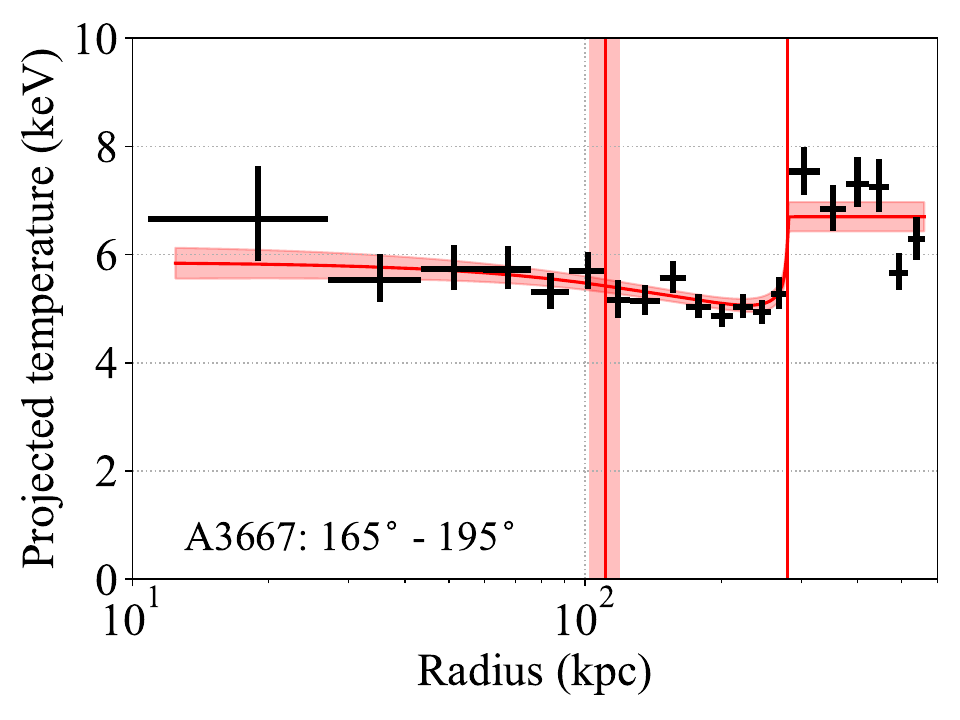}
  \includegraphics[width=5.6cm]{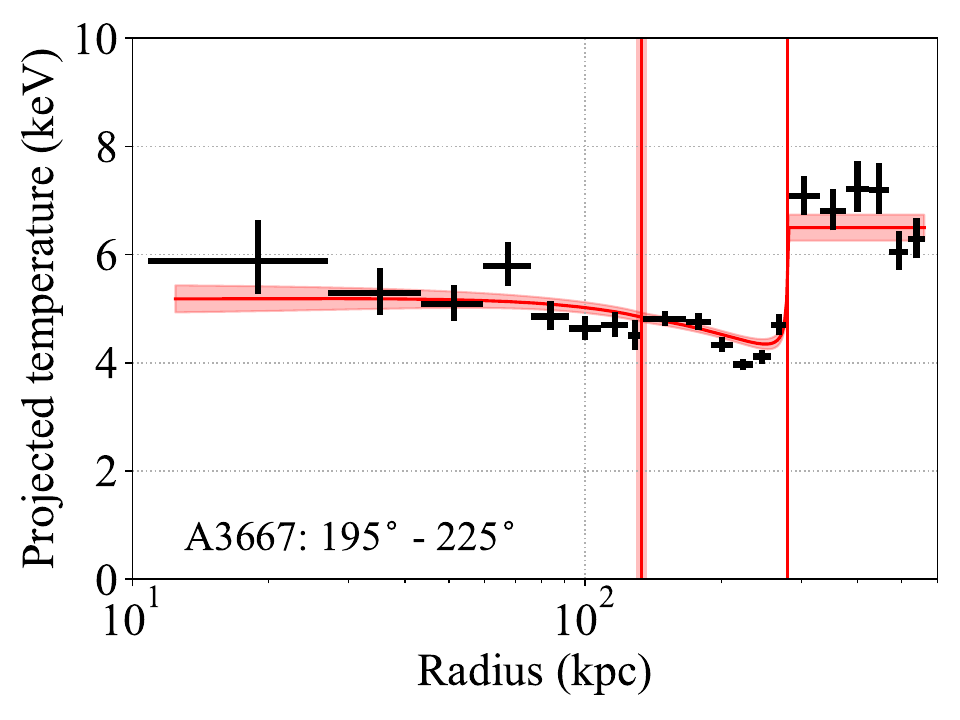}
  \includegraphics[width=5.6cm]{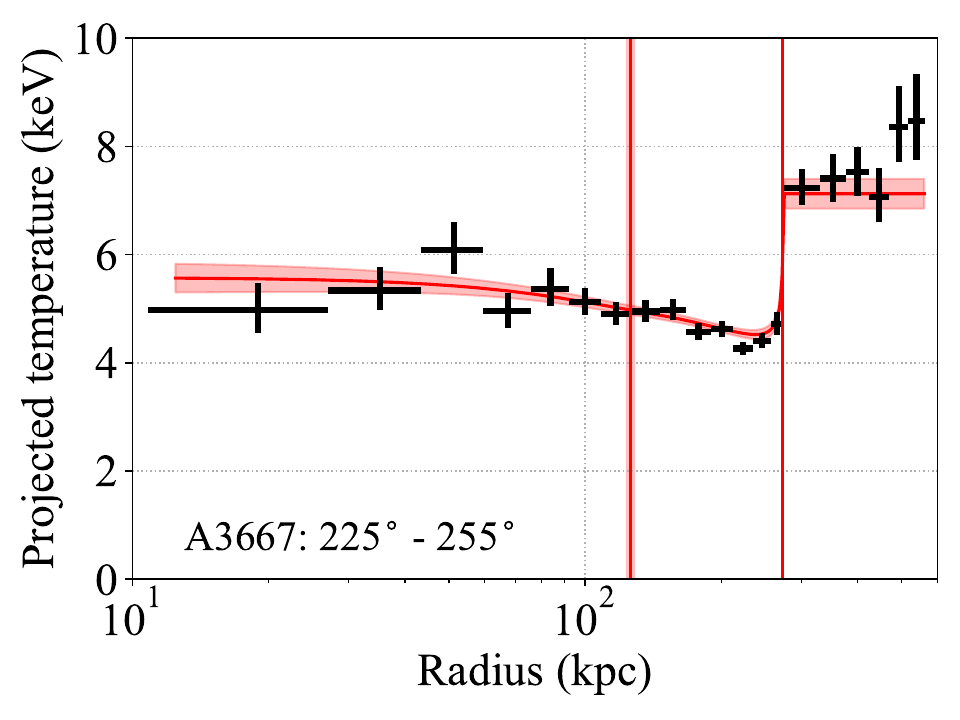}
  \includegraphics[width=5.6cm]{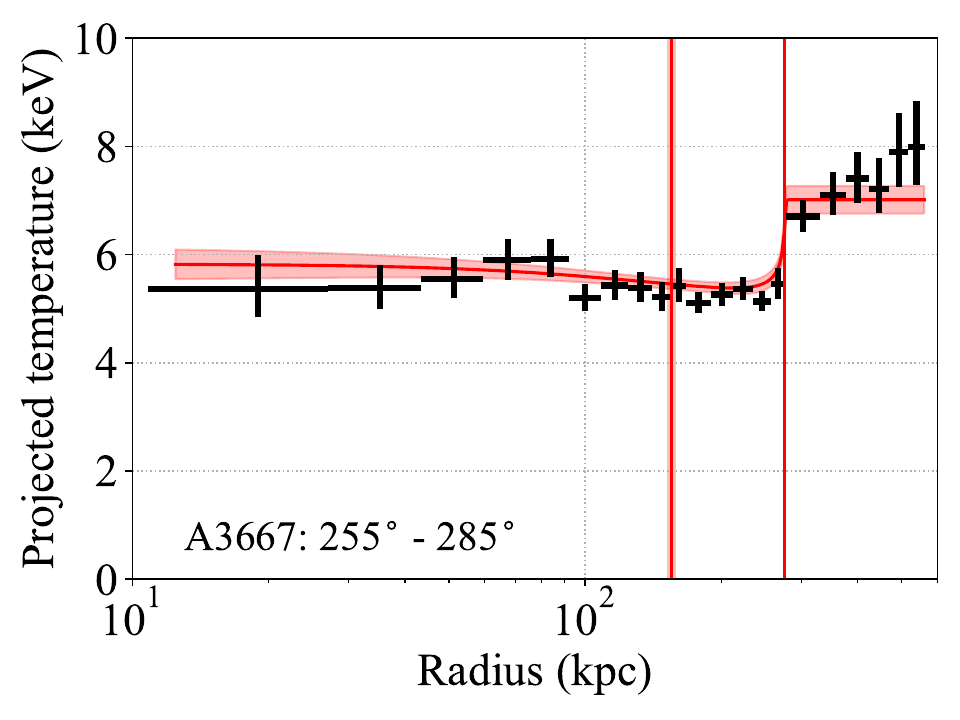}
  \includegraphics[width=5.6cm]{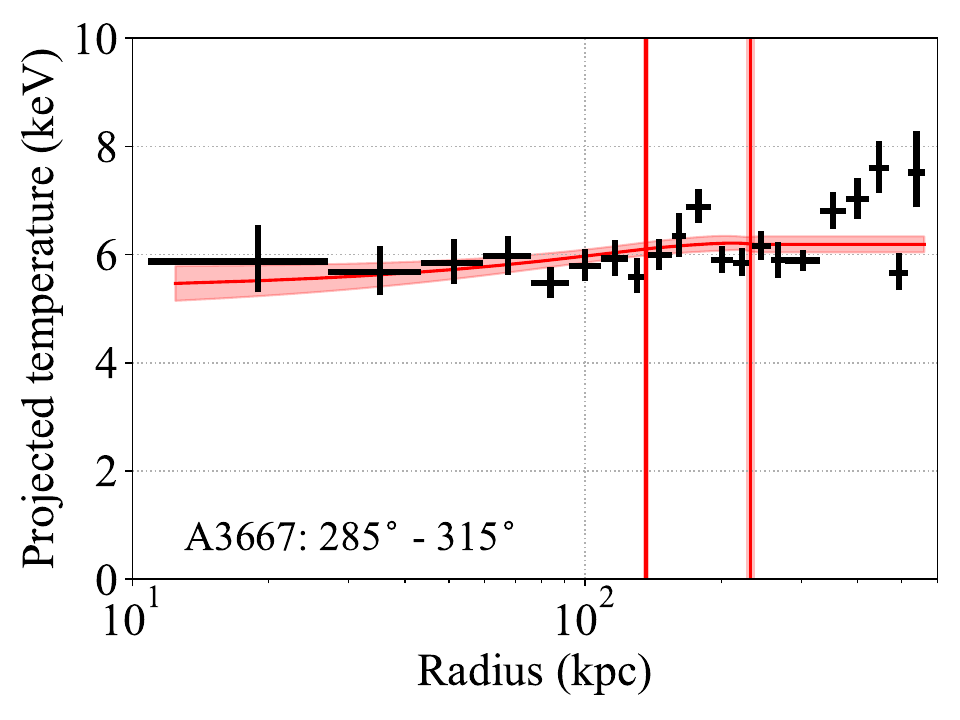}
 \end{center}
\caption{
Comparisons of the projected ICM temperature profiles obtained from the forward modeling analysis with those measured by the X-ray spectral analysis in A3667. The red solid line and the red shaded area show the best-fit projected ICM temperature and its $1\sigma$ confidence range. The black data points correspond to the projected ICM temperature in each radial bin measured by the X-ray spectral analysis with \texttt{XSPEC}. The two red vertical lines and the corresponding red shaded areas show the positions of $r_{12}$ and $r_{23}$ and their  $1\sigma$ confidence ranges, respectively.
(Top left): the $135^{\circ} - 165^{\circ}$ sector.
(Top middle): the $165^{\circ} - 195^{\circ}$ sector.
(Top right): the $195^{\circ} - 225^{\circ}$ sector.
(Bottom left): the $225^{\circ} - 255^{\circ}$ sector.
(Bottom middle): the $255^{\circ} - 285^{\circ}$ sector.
(Bottom right): the $285^{\circ} - 315^{\circ}$ sector.
}
\label{fig:vs_A3667}
\end{figure*}

\begin{figure*}[ht]
 \begin{center}
  \includegraphics[width=5.6cm]{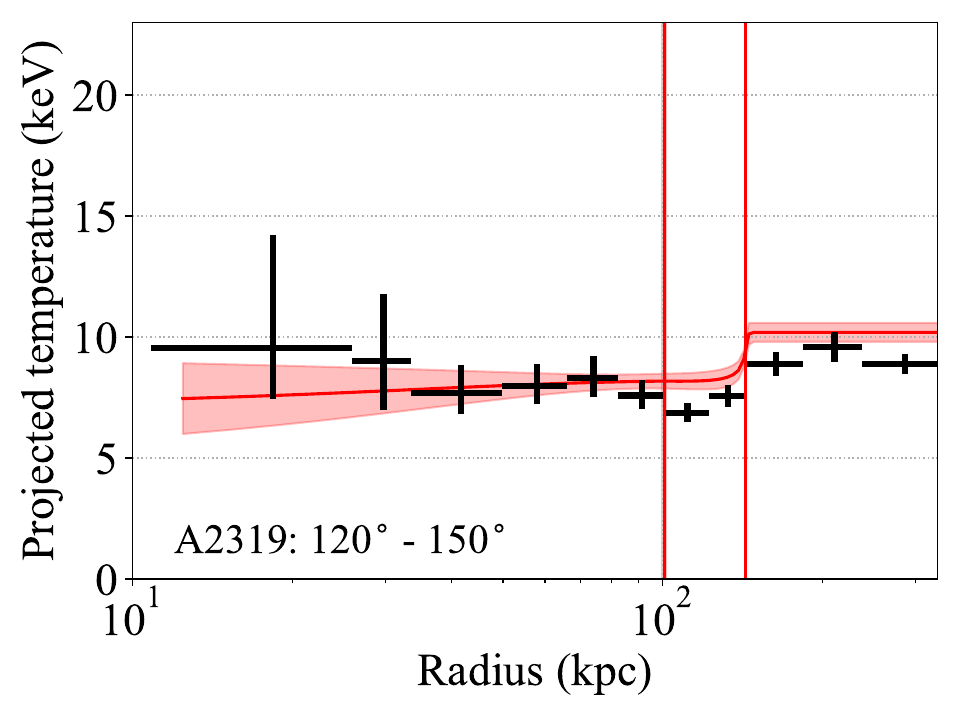}
  \includegraphics[width=5.6cm]{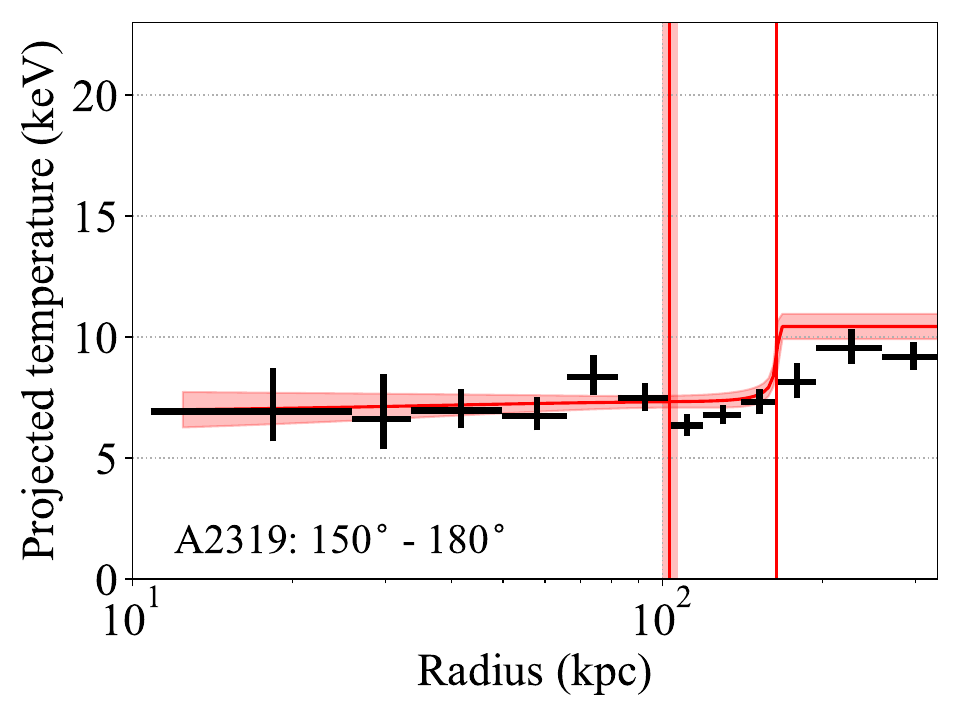}
  \includegraphics[width=5.6cm]{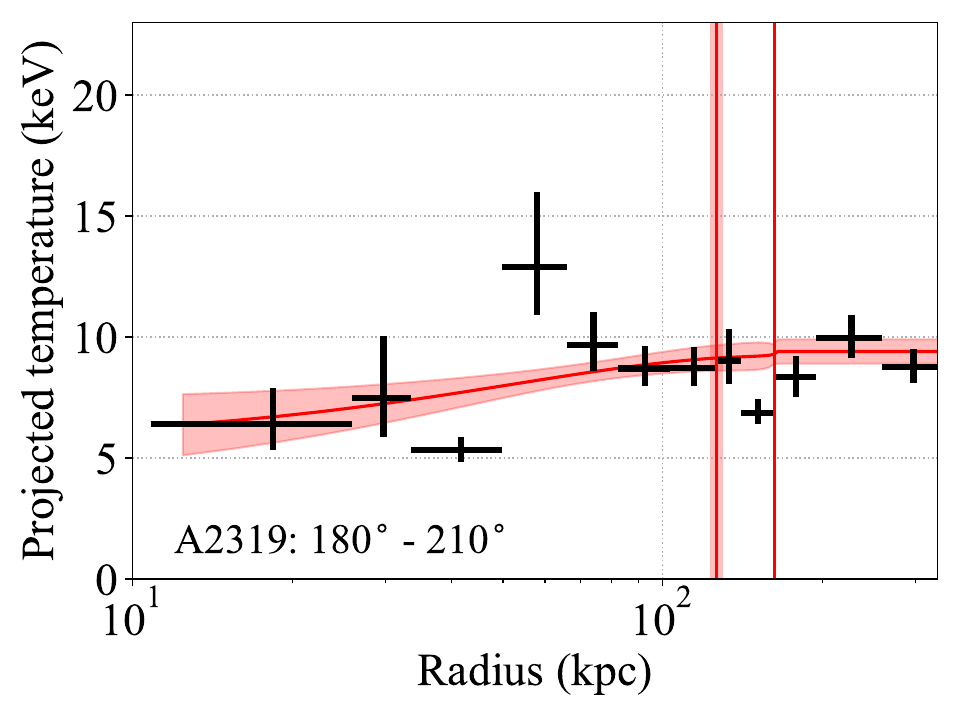}
  \includegraphics[width=5.6cm]{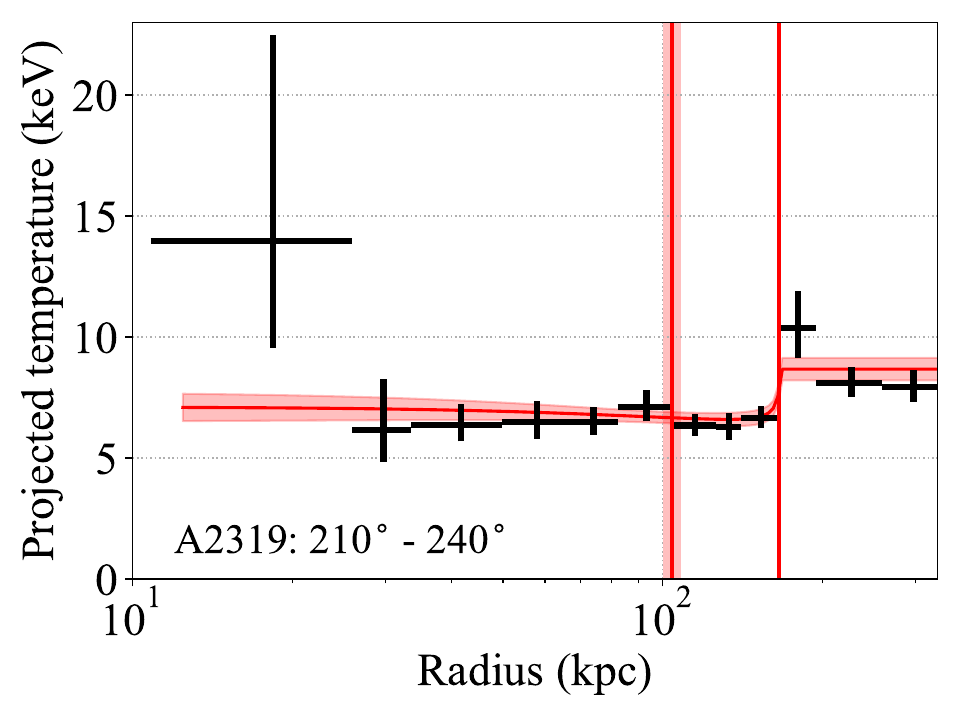}
  \includegraphics[width=5.6cm]{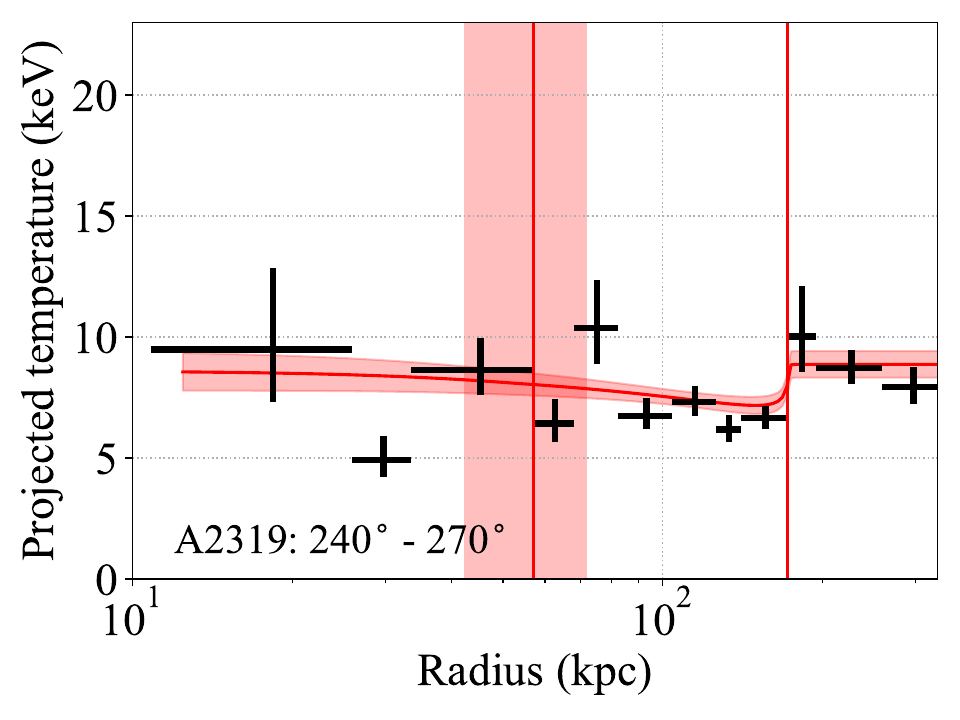}
  \includegraphics[width=5.6cm]{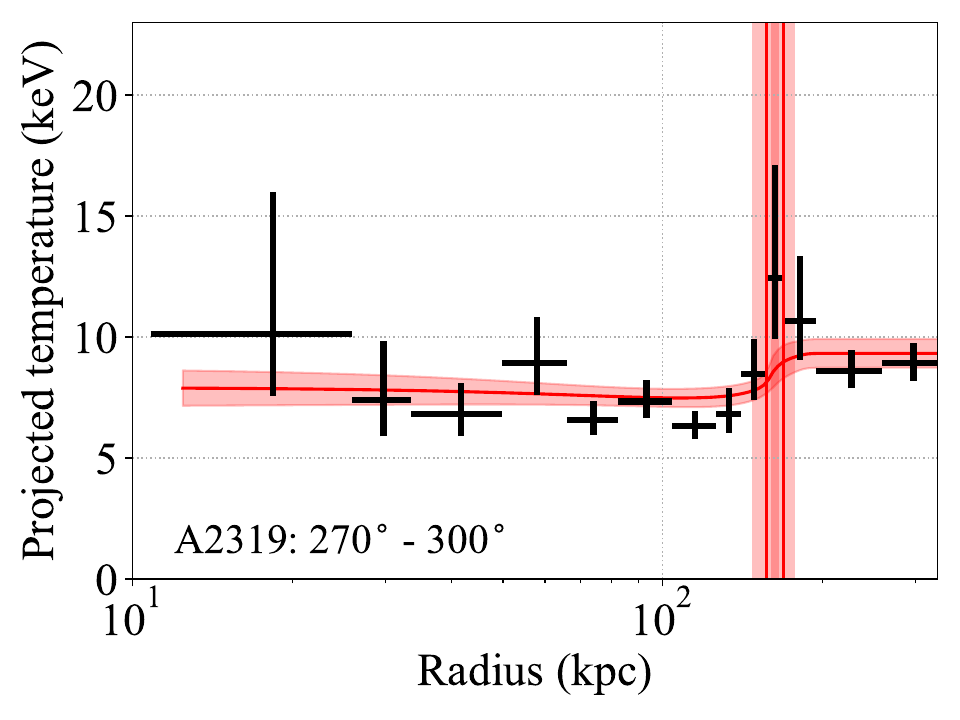}
 \end{center}
\caption{
Same as Figure~\ref{fig:vs_A3667}, but for A2319.
(Top left): the $120^{\circ} - 150^{\circ}$ sector.
(Top middle): the $150^{\circ} - 180^{\circ}$ sector.
(Top right): the $180^{\circ} - 210^{\circ}$ sector.
(Bottom left): the $210^{\circ} - 240^{\circ}$ sector.
(Bottom middle): the $240^{\circ} - 20^{\circ}$ sector.
(Bottom right): the $270^{\circ} - 300^{\circ}$ sector.
}
\label{fig:vs_A2319}
\end{figure*}

\begin{figure*}[ht]
 \begin{center}
  \includegraphics[width=5.6cm]{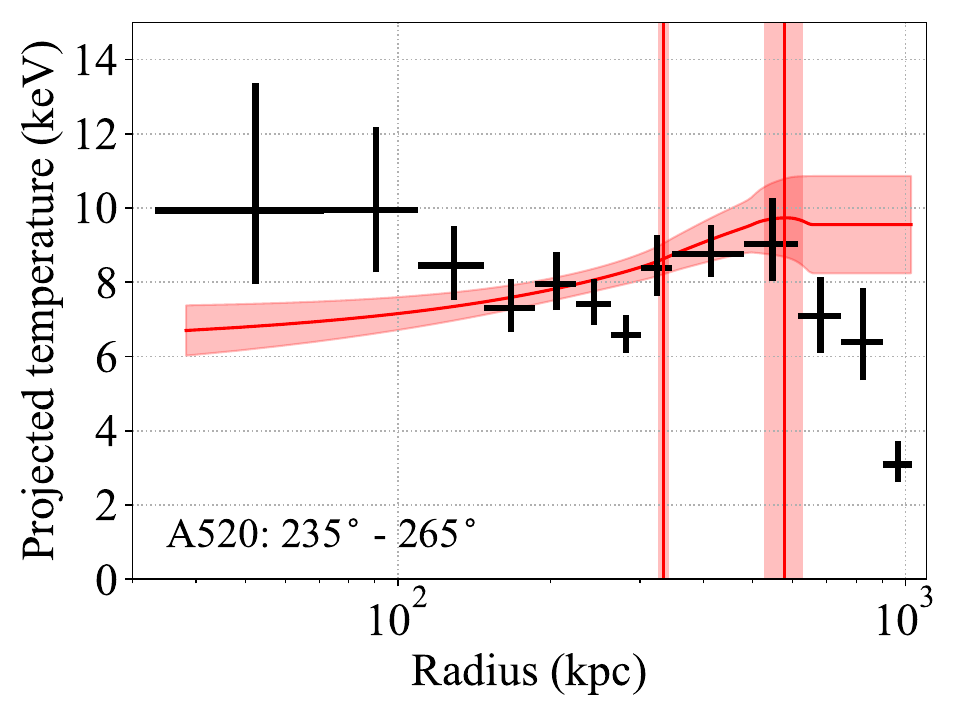}
  \includegraphics[width=5.6cm]{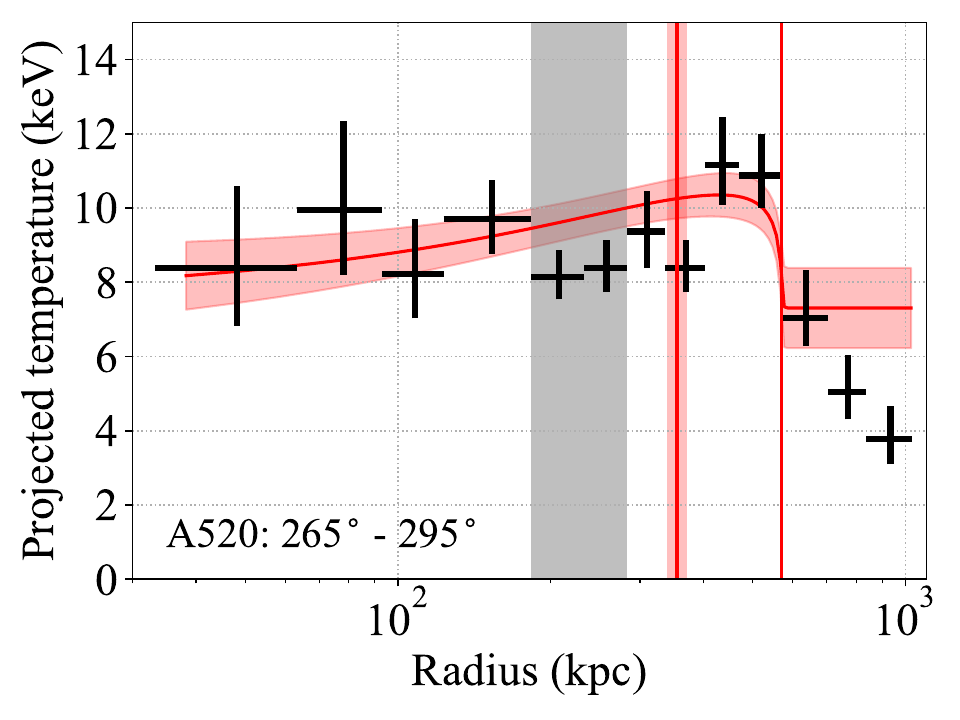}
  \includegraphics[width=5.6cm]{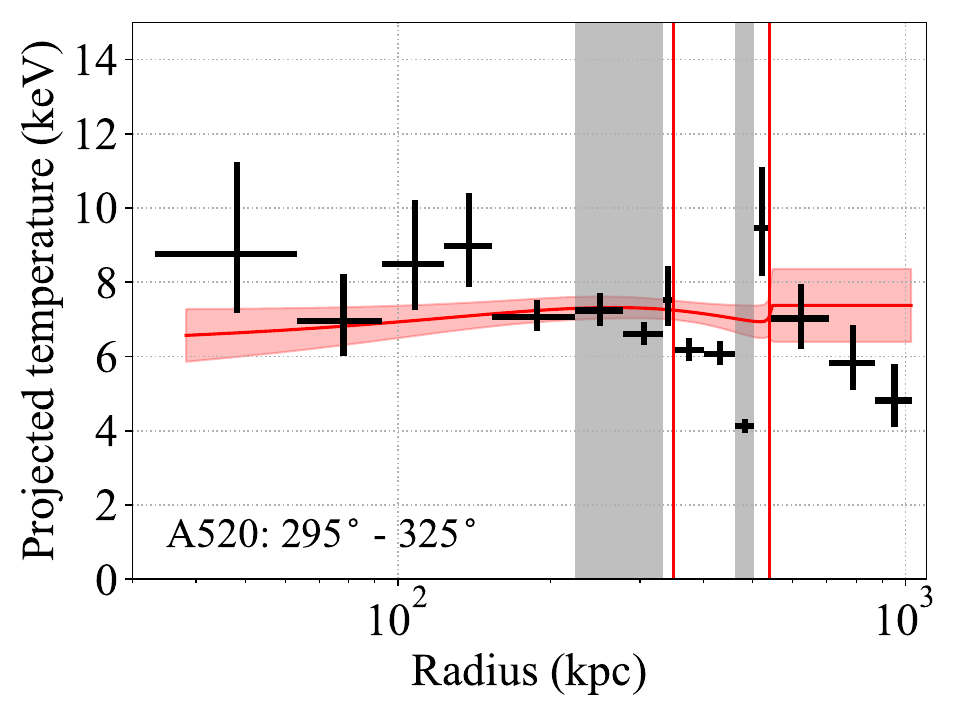}
  \includegraphics[width=5.6cm]{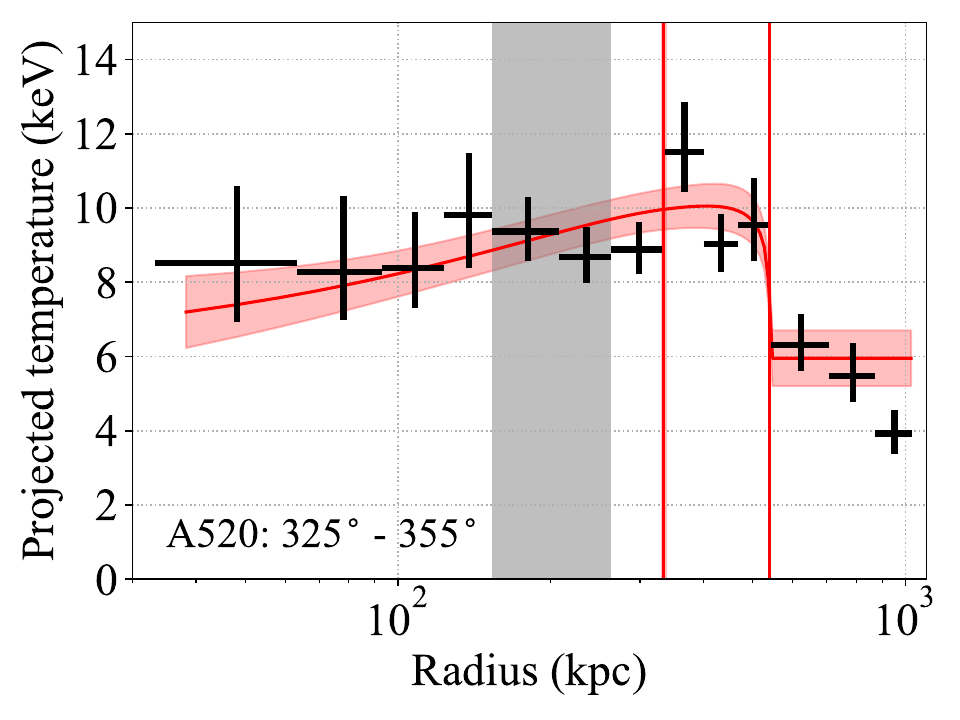}
  \includegraphics[width=5.6cm]{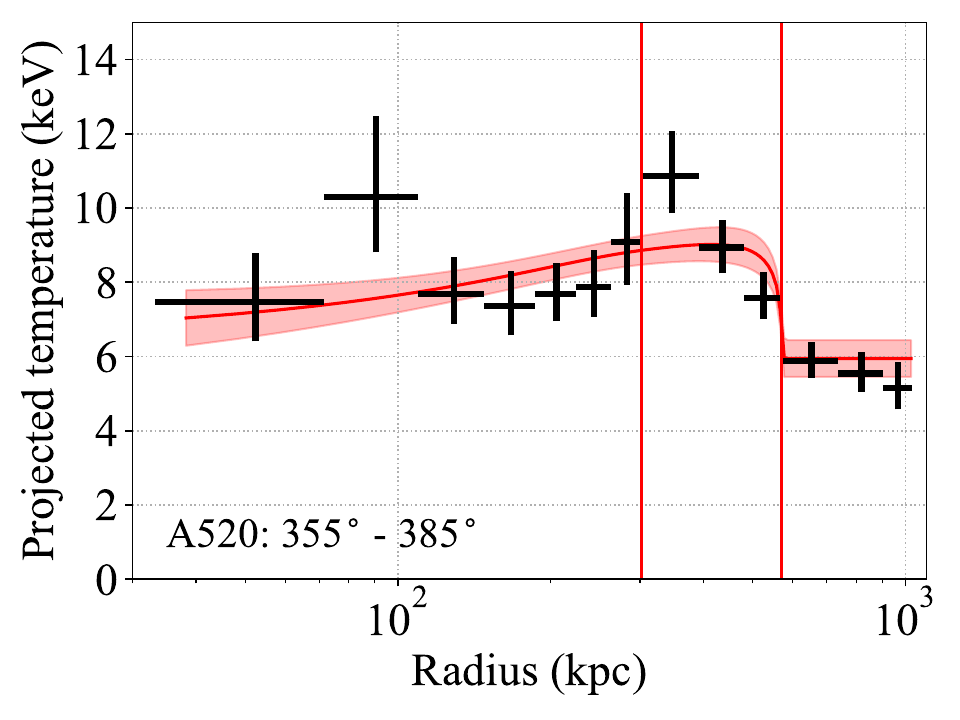}
  \includegraphics[width=5.6cm]{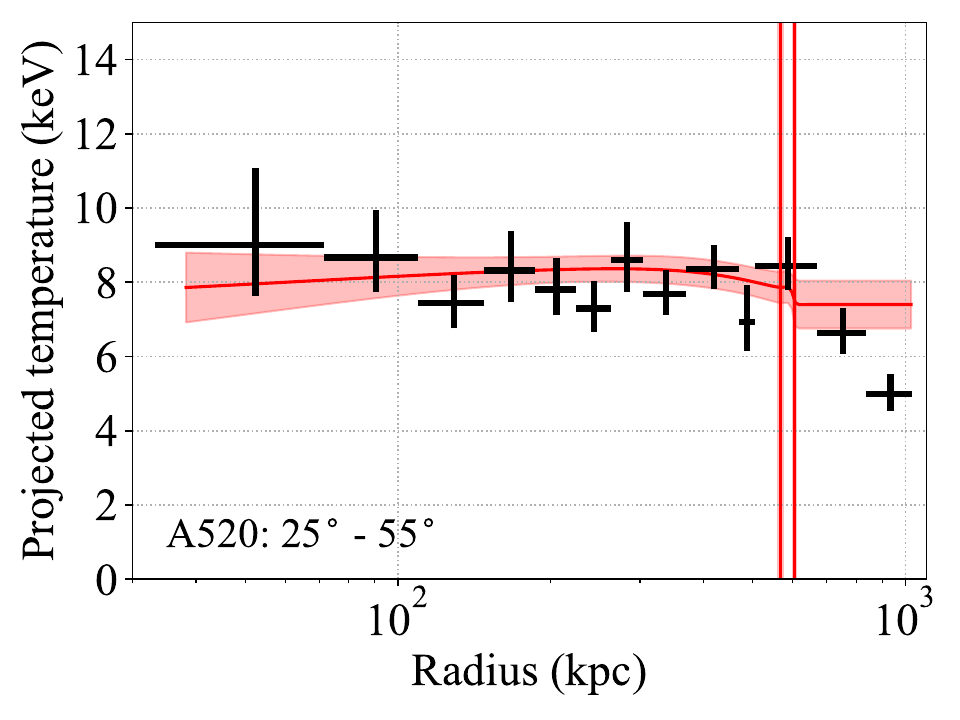}
 \end{center}
\caption{
Same as Figure~\ref{fig:vs_A2319}, but for A520.
The gray vertical shaded area shows a masked region for the forward modeling analysis, while the projected ICM temperature in the mask region is measured by the X-ray spectral analysis using \texttt{XSPEC}.
(Top left): the $235^{\circ} - 265^{\circ}$ sector.
(Top middle): the $265^{\circ} - 295^{\circ}$ sector.
(Top right): the $295^{\circ} - 325^{\circ}$ sector.
(Bottom left): the $325^{\circ} - 355^{\circ}$ sector.
(Bottom middle): the $355^{\circ} - 385^{\circ}$ sector.
(Bottom right): the $25^{\circ} - 55^{\circ}$ sector.
}
\label{fig:vs_A520}
\end{figure*}

\begin{figure*}[ht]
 \begin{center}
  \includegraphics[width=5.6cm]{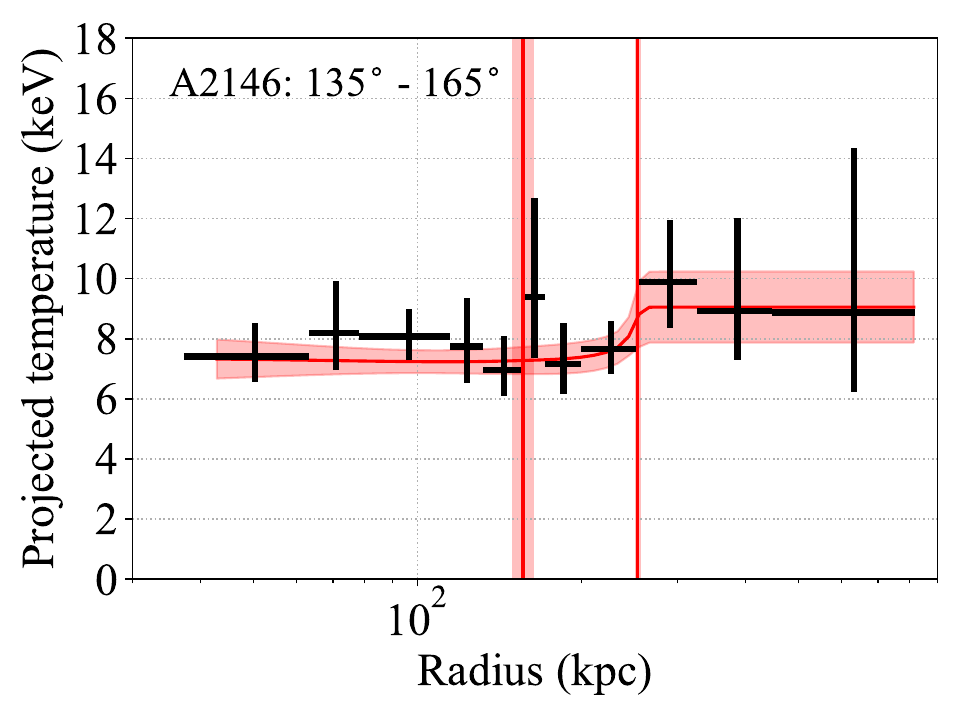}
  \includegraphics[width=5.6cm]{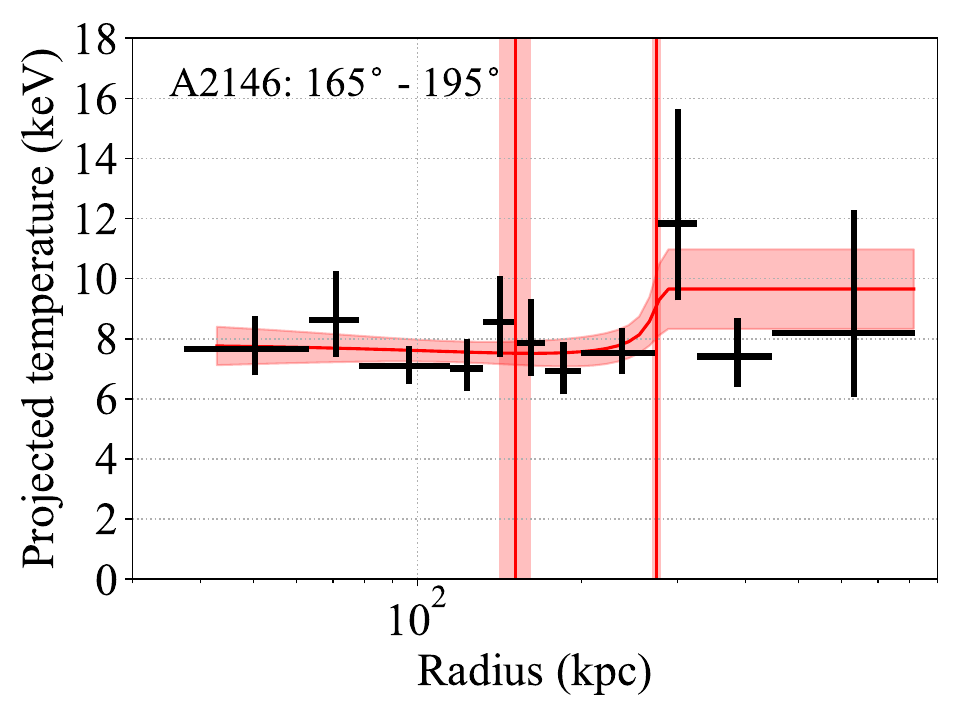}
  \includegraphics[width=5.6cm]{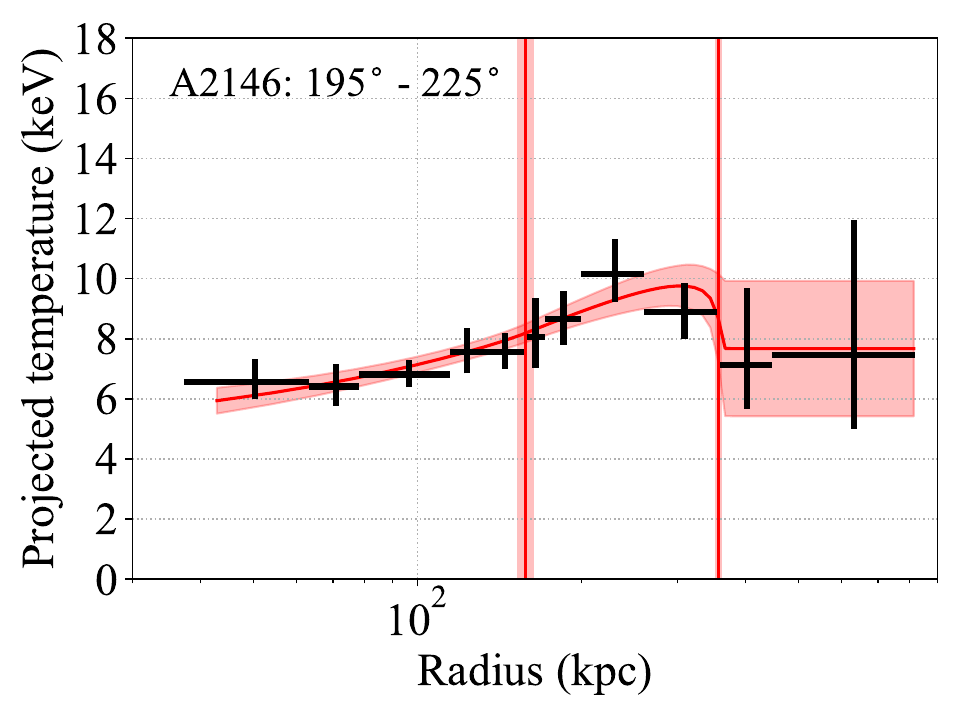}
  \includegraphics[width=5.6cm]{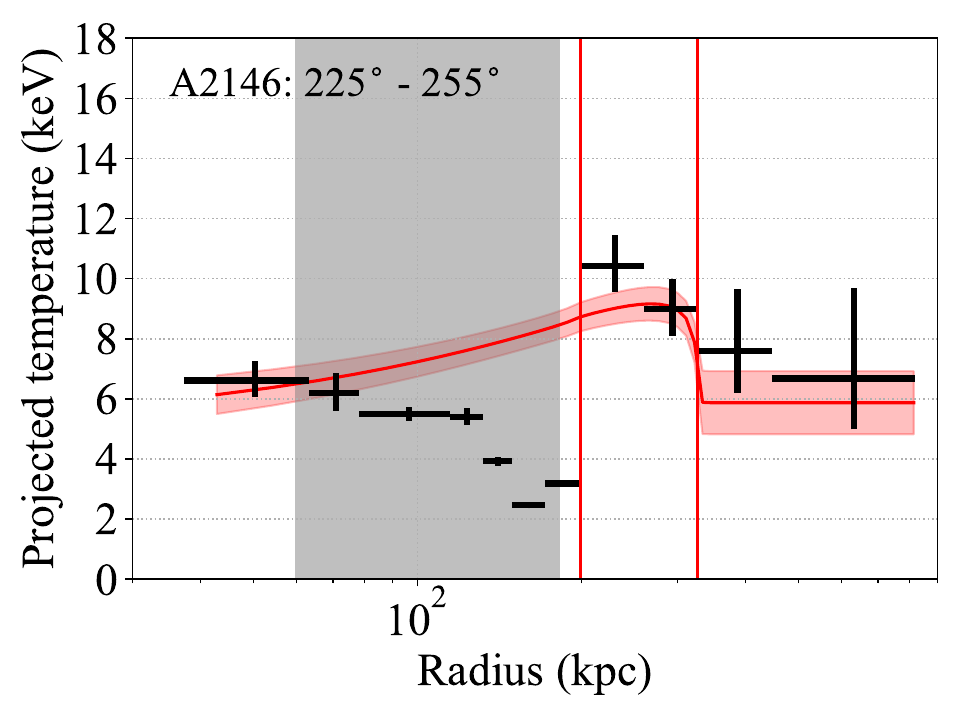}
  \includegraphics[width=5.6cm]{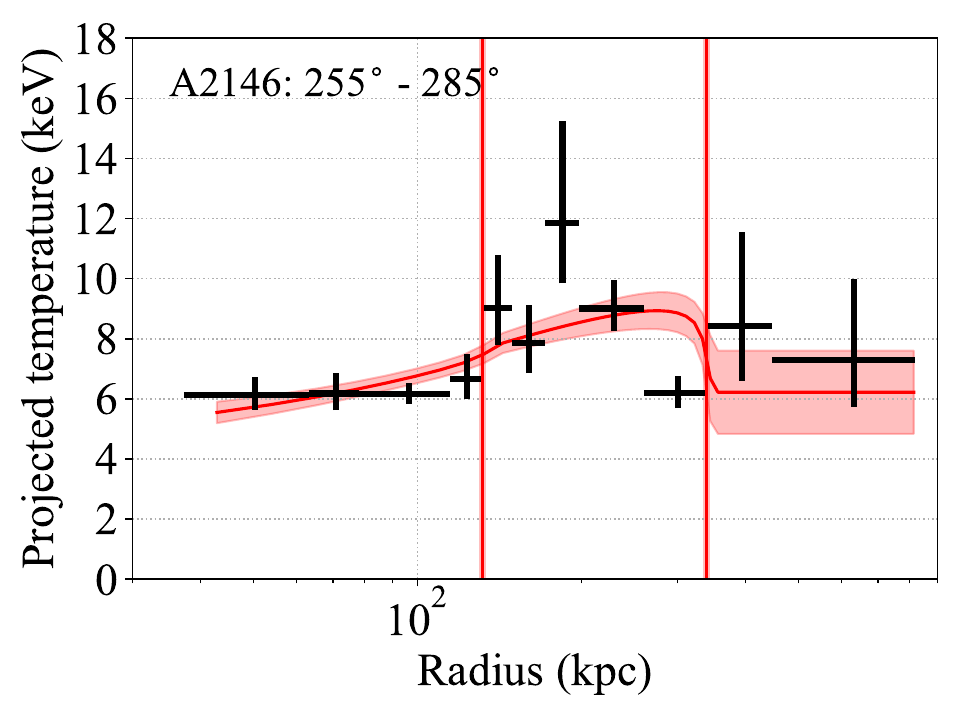}
  \includegraphics[width=5.6cm]{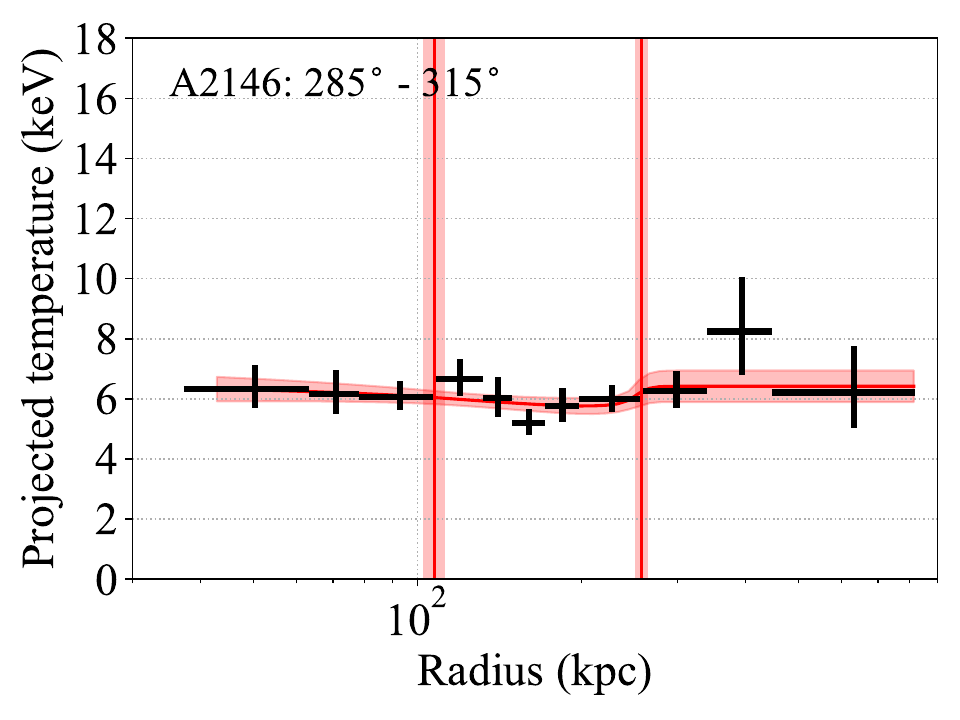}
 \end{center}
\caption{
Same as Figure~\ref{fig:vs_A520}, but for A2146.
(Top left): the $135^{\circ} - 165^{\circ}$ sector.
(Top middle): the $165^{\circ} - 195^{\circ}$ sector.
(Top right): the $195^{\circ} - 225^{\circ}$ sector.
(Bottom left): the $225^{\circ} - 255^{\circ}$ sector.
(Bottom middle): the $255^{\circ} - 285^{\circ}$ sector.
(Bottom right): the $285^{\circ} - 315^{\circ}$ sector.
}
\label{fig:vs_A2146}
\end{figure*}

Here, we present comparisons of the projected ICM temperature profiles obtained from the forward modeling analysis with those measured by the X-ray spectral analysis. Figures~\ref{fig:vs_A3667}, \ref{fig:vs_A2319}, \ref{fig:vs_A520}, and \ref{fig:vs_A2146} display these comparisons for A3667, A2319, A520, and A2146, respectively. We extracted and analyzed the X-ray spectra of the ICM from the masked region as well. The lower-temperature gas is found in the masked regions.


\newpage
\section{Forward modeling analysis of a sloshing cold front in the Perseus cluster}
\label{sec:Perseus}

\begin{figure*}[ht]
 \begin{center}
  \includegraphics[width=8.5cm]{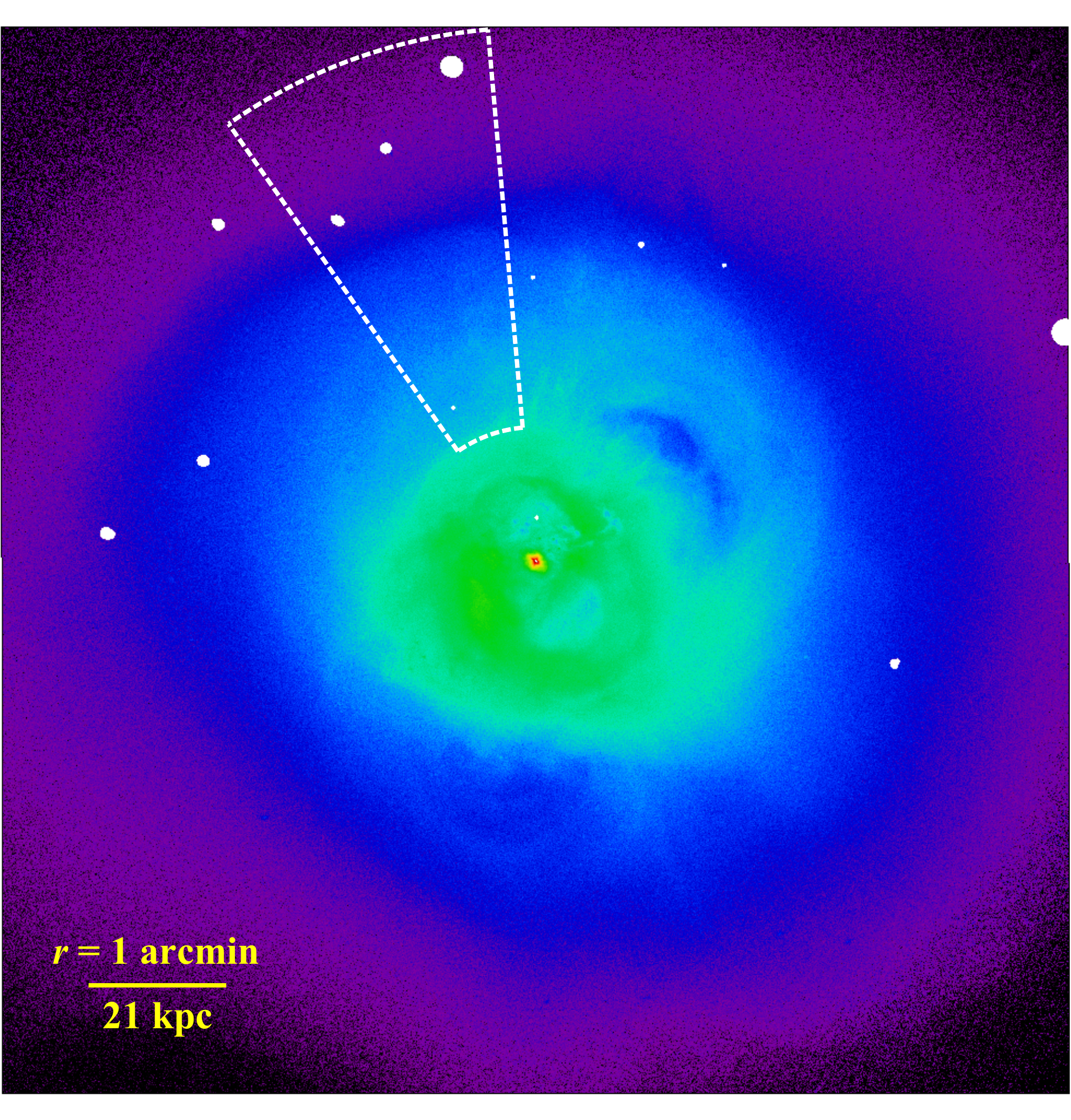}
  \includegraphics[width=8.5cm]{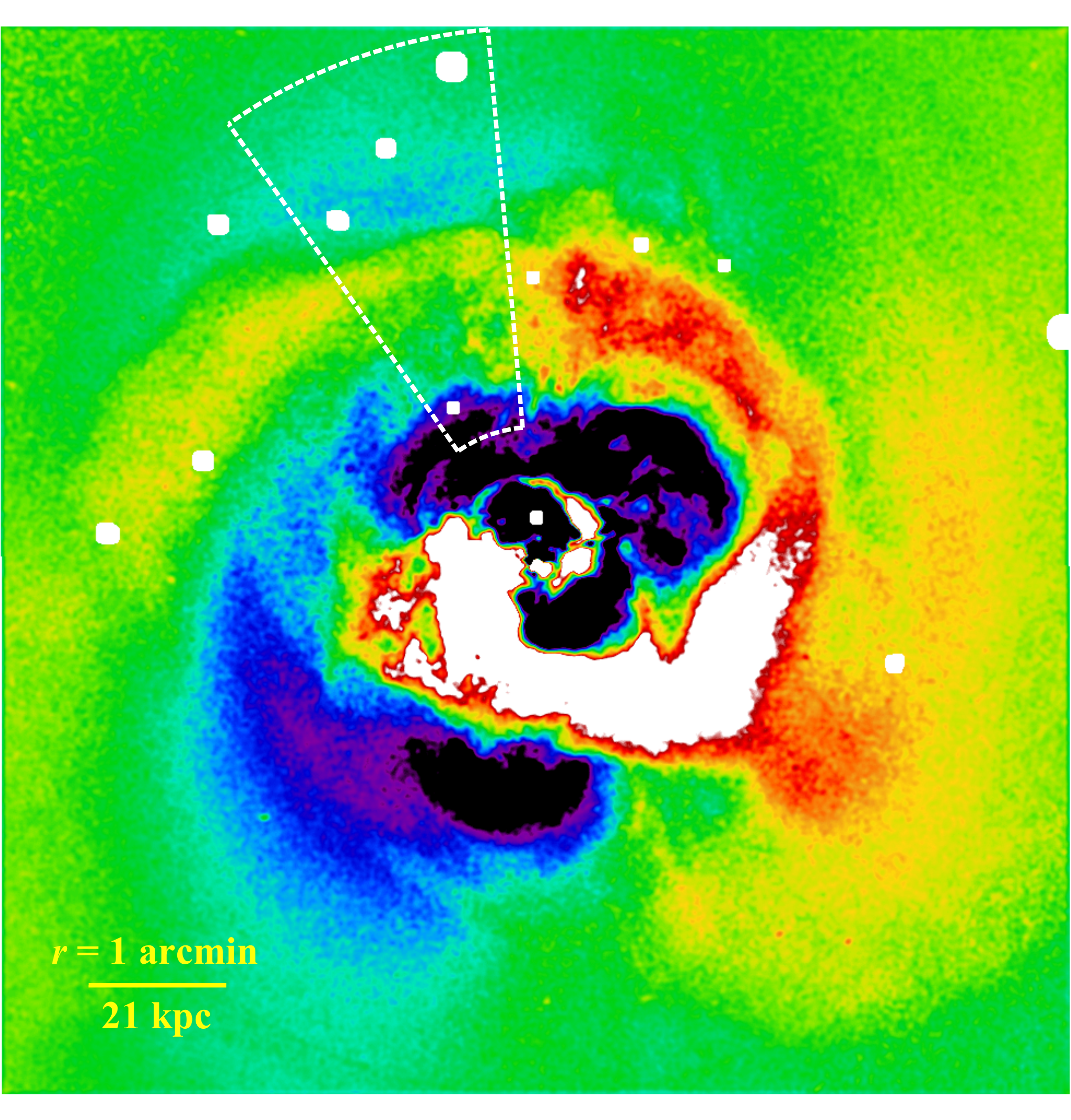}
 \end{center}
\caption{
Exposure corrected, background subtracted \Chandra\ X-ray surface brightness of the Perseus cluster in the $0.5 - 7.0$\,keV band (left) and its residual image (right). The point sources identified are masked with white ellipses. The overlaid, dashed white sector denotes the direction along where the X-ray surface brightness profiles are extracted ($\theta\in [60\arcsec, 240\arcsec]$) with an opening angle of $95^{\circ}$ to $125^{\circ}$.
(Left): the \Chandra\ X-ray image of the Perseus cluster in the $0.5 - 7.0$\,keV band.
(Right): the residual image of the Perseus cluster after subtracting the mean surface brightness. This image is smoothed by a Gaussian kernel with $3.5''$ FWHM.
}
\label{fig:Perseus}
\end{figure*}

To investigate the mechanisms to create the cold fronts in A3667 and A2319, we analyze a well-known prominent sloshing cold front in the Perseus cluster using the forward modeling analysis. The Perseus cluster is the X-ray brightest, very nearby galaxy cluster \citep[$z = 0.017284$, $1'' = 0.351$\,kpc;][]{Hitomi18d}. We analyze the datasets of the Perseus cluster taken with deep \Chandra\ observations. Data reduction is conducted with the same manner as Section~\ref{sec:obs}. The Perseus cluster exhibits a clear spiral pattern in its X-ray residual image \citep[][also see the right panel of Figure~\ref{fig:Perseus}]{Churazov03, Zhuravleva14}, which is considered to be generated by gas sloshing \citep[e.g.,][]{Churazov16}. To define a sector analyzed using the forward modeling analysis, we first make an X-ray residual image of the Perseus cluster. We apply the method presented in \cite{Ueda17} to compute the mean surface brightness. A circle model centered on the position of the central active galactic nucleus, (RA, Dec) = (3:19:48.1815, +41:30:42.521), is used. Then, we model the mean surface brightness using the concentric circle fitting algorithm of \cite{Ueda17}, by minimizing the variance of the X-ray surface brightness relative to the circle model. Figure~\ref{fig:Perseus} shows the X-ray surface brightness image of the central region of the Perseus cluster taken with \Chandra\ and its residual image. We define a sector to cross a striking edge of the sloshing cold front with an opening angle of $95^{\circ}$ to $125^{\circ}$.

We use the density profile provided from Equation~\ref{eq:ne} to analyze the observed X-ray surface brightness profile. Even though small-scale substructures with a spatial scale of a few to tens kpc are found in the X-ray surface brightness, this model still allows us to describe the global feature of the profile and determine the edge position accurately. Using the same approach as Section~\ref{sec:model}, we extract the radial profiles of X-ray surface brightness in the ten energy bands, and sample those in 90 linearly spaced radial bins in the ranges $\theta\in [60\arcsec, 240\arcsec]$ (i.e., $2''$ per bin) centered on the curvature center (RA, Dec) = (3:19:48.1815, +41:30:42.521). Then, we conduct the forward modeling analysis of this sector.

\begin{figure*}[ht]
 \begin{center}
  \includegraphics[width=8.5cm]{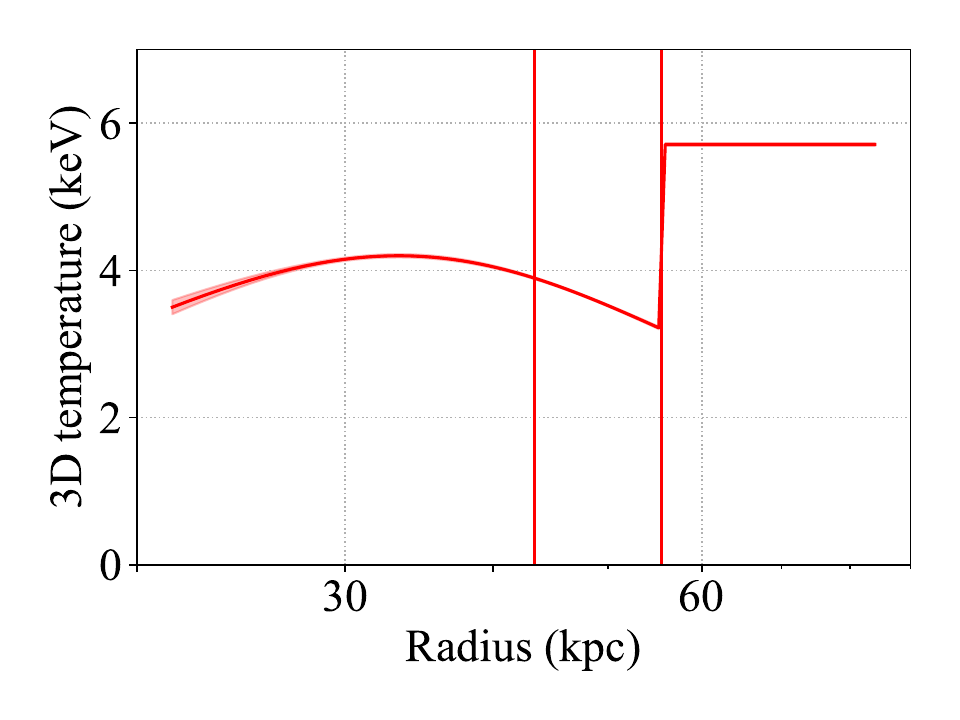}
  \includegraphics[width=8.5cm]{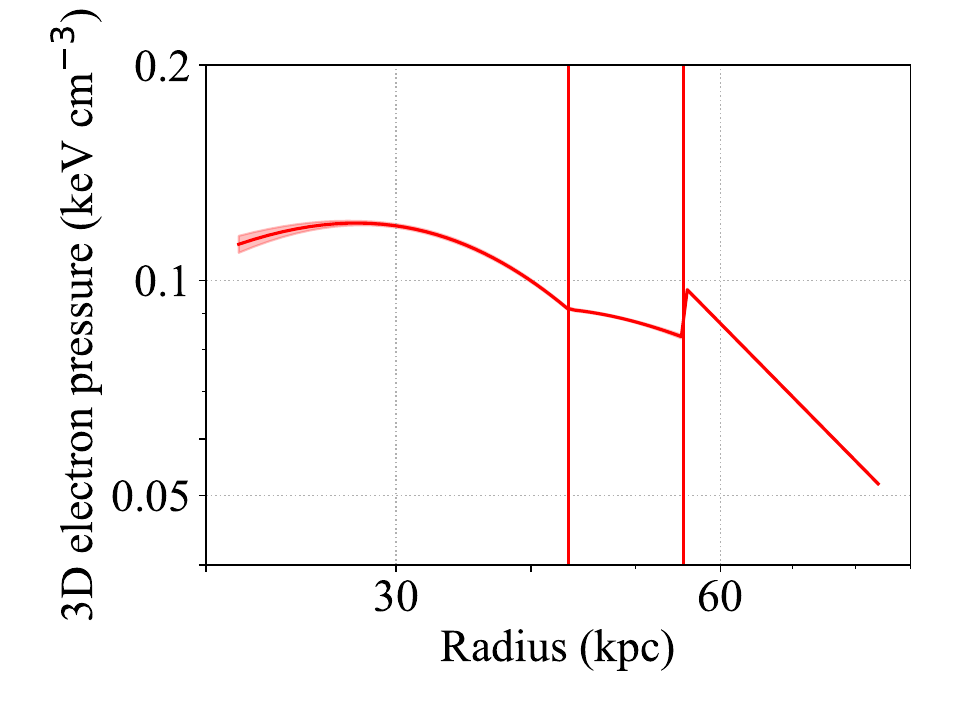}
 \end{center}
\caption{
Observed 3D ICM temperature (left) and pressure (right) profiles of the $95^{\circ} - 125^{\circ}$ sector with the forward modeling analysis. The red line and the red shaded area show the best-fit profile and its $1\sigma$ confidence range, respectively. The two red vertical lines correspond to the positions of $r_{12}$ and $r_{23}$, respectively. The red shaded vertical ares show the $1\sigma$ confidence ranges of $r_{12}$ and $r_{23}$, respectively.
}
\label{fig:Perseus_kT_P}
\end{figure*}

The edge position is measured at $55.44 \pm 0.02$\,kpc from the center of the curvature, corresponding to the interface between the sloshing cold front (i.e., the edge of the spiral pattern) and the ambient component of the ICM. Figure~\ref{fig:Perseus_kT_P} shows the inferred 3D ICM temperature and pressure profiles. The temperature jump is clearly detected. We also find a pressure jump at the interface, indicating that the sloshing cold front is also maintained by non-thermal support, e.g., magnetic fields and centrifugal force. However, the observed pressure jump is positive toward the outer region, which is distinct from the cold fronts in A3667 and A2319. Therefore, the observed features of the cold fronts in A3667 and A2319 are different from those of the sloshing cold front in the Perseus cluster, implying that the mechanisms to create cold fronts might not be similar to the Perseus cluster. Note that this positive pressure jump has been reported in \cite{Ichinohe19}.

\clearpage
\bibliography{00_BibTeX_library}{}
\bibliographystyle{aasjournal}



\end{document}